\documentclass[twocolumn,usenatbibs]{aa}
\usepackage{graphicx,natbib}
\usepackage{txfonts}

\def\ms{\,m\,s$^{-1}$}         
\def\kms{\,km\,s$^{-1}$}       
\def\vsini{$v$\,sin\,$i$}      
\def\Msol{\ {\mathrm M}_\odot}             
\def\Rsol{\ {\mathrm R}_\odot}

\def\MJ{{\mathrm M}_{\mathrm J}}
\def\RJ{{\mathrm R}_{\mathrm J}}

\begin{document}
\bibliographystyle{aa}

   \title{Doppler follow-up of OGLE planetary transit candidates in Carina\thanks{Based on observations 
	   collected with the UVES and FLAMES spectrographs at the VLT/UT2 Kueyen telescope 
	   (Paranal Observatory, ESO, Chile, Programme 72.C-0191)}
	   }

   \author{F. Pont\inst{1}, F. Bouchy\inst{2,3}, C. Melo\inst{4}, N.C. Santos\inst{5},
           M. Mayor\inst{1}, D. Queloz\inst{1} \and S. Udry\inst{1}
          }

   \offprints{\email{frederic.pont@obs.unige.ch}}

   \institute{Observatoire de Gen\`eve, 51 ch. des Maillettes, 1290 Sauverny, Switzerland
         \and
 Laboratoire d'Astrophysique de Marseille, 
               Traverse du Siphon, 13013 Marseille, France           
         \and
	Observatoire de Haute Provence, 04870 St Michel l'Observatoire, France 
	 \and
	      ESO, Casilla 19001, Santiago 19, Chile  
	  \and
	     Lisbon Observatory, Tapada da Ajuda, 1349-018 Lisboa, Portugal  
            }

   \date{Received ; accepted }

   \abstract{ We present the results of our high-resolution spectroscopic follow-up of 42 planetary transit candidates in Carina from the OGLE survey. This follow-up has already allowed the discovery of three new transiting exoplanets, OGLE-TR-111, 113 and 132, presented in earlier Letters \citep{Bou04, Pon04}. Here we analyse the data for the remaining 39 candidates. The radial velocity data show that most of them are eclipsing binaries, in very varied configurations. Precise radial velocity orbits were derived for 15 binaries, revealing 9 transits of small stars (generally M-dwarfs) in front of F-G dwarfs, 1 grazing equal-mass eclipsing binary, 4 triple and 1 quadruple systems. A remaining 14 systems appear binary, but the exact orbit is uncertain or was not determined. 2 objects do not show any radial velocity variations in phase with the transit signal, and 6 do not possess spectral lines strong enough for a reliable cross-correlation function to be measured. Among these last two categories, up to 6 objects are suspected false positives of the photometric transit detection. Finally 2 objects are unsolved cases that deserve further observations.

Our study illustrates the wide variety of cases that can mimic photometric planetary transits, and the importance of spectroscopic follow-up. Multi-fiber capacities and an optimized follow-up strategy, which we present here, can help deal with the high number of candidates that are likely to turn up in the near future.

An important by-product of this study is the determination of exact masses and radii for six very low-mass stars, including two at the very edge of the stellar domain, OGLE-TR-106 ($M=0.116\pm 0.021 \Msol$) and OGLE-TR-122 ($M=0.089 \pm 0.007 \Msol$). The radius of these objects is consistent with theoretical expectations. Two further objects, OGLE-TR-123 and OGLE-TR-129, may harbour transiting companions near the brown-dwarf/stellar limit ($M\simeq 0.07 \Msol$), whose confirmation requires further high-resolution spectroscopic monitoring.

No transiting massive planets ($M=2-10 \MJ$) were detected, confirming the rarity of such systems at short period as indicated by Doppler surveys. No light ($M<0.5 \MJ$), large ($R>\RJ$) planets were found either, indicating that "hot Saturns" generally have smaller radii than hot Jupiters. Three short period binaries with a M-dwarf companion show definite orbital eccentricities, with periods ranging from 5.3 to 9.2 days. This confirms theoretical indications that orbital circularisation in close binaries is less efficient for smaller companion masses.

We also discuss the implications of our results for the statistical interpretation of the OGLE planetary transit survey in Carina in terms of planet frequency and detection efficiency. We find that the actual transit detection threshold is considerably higher than expected from simple estimates, and very strongly favours the detection of planets with periods shorter than about 2 days. The apparent contradition between the results of the OGLE transit survey and Doppler surveys can be resolved when this detection bias is taken into account.

   }

   \authorrunning{F. Pont et al.}
   \titlerunning{Follow-up of OGLE transits in Carina}

   \maketitle

\section{Introduction}

Since the groundbreaking discovery of 51 Pegasi \citep{May95}, more that one hundred planets have been detected around other stars, the vast majority of them by radial velocity surveys. In the wake of the first exoplanet detection, \citet{Gil00} have monitored several thousand stars in the globular cluster 47 Tuc with the HST in search of photometric planetary transits, but none were found. A results that is now attributed to the paucity of short-period planets around metal-poor stars \citep{Sak05}. 

Transiting planets are especially valuable in the study of exoplanets, because the presence of a transit allows the determination of the exact mass and radius -- and therefore mean density -- of the planet. These are obviously very important parameters in the physical understanding of exoplanet structure and evolution. Data from transits are thus an important complement to radial velocity planet searches which provide only an estimate of $M_{pl} \sin i$ in addition to orbital parameters.

The first exoplanet transit was detected around HD 209458 by \citet{Cha00} and \citet{Hen00}, the planet having been previously discovered in radial velocity by \citet{Maz00}. This planet has turned out to have a radius much larger than Jupiter, $R=1.35\pm0.06$ R$_J$ \citep{Bro01}, and to be undergoing significant evaporation \citep{Cha02, Bro02, Vid03, Vid04}.

In recent years, increasing telescope automation and the capacity to process large amounts of CCD photometric data have make possible several ambitious ground-based searches for planetary transits: e.g. STARE \citep{Bro00}, WASP \citep{Kan01}, PLANET \citep{Sak04}. Up to now, the most successful of these searches has been the OGLE survey, which monitored 3 fields in the direction of the Galactic bulge and 3 fields in Carina during 2001/2002, annoucing 137 possible transiting candidates \citep{Uda02a, Uda02b, Uda02c, Uda03}. After these candidates were announced, several spectroscopic follow-up programmes were initiated \citep{Kon03b, Dre02, Gal05}, including our own follow-up of 17 candidates in the Galactic bulge field \citep[hereafter Paper I]{Bou05}. It was found that the vast majority of transiting candidates were eclipsing binaries, as predicted by the simulations of \citet{Bro03}.

In this context, \citet{Kon03} announced the first exoplanet discovered by photometric transit surveys, OGLE-TR-56b \citep[confirmed by]{Bou05}. This planet was found to orbit with the unexpectedly short period of 1.2-days, much shorter than the observed pile-up of periods above 3 days in radial velocity surveys. 

In March 2004, using the FLAMES/UVES multi-fiber spectrograph on the VLT, we acquired high-resolution spectroscopic information for 42 of the most promising candidates in the OGLE Carina fields, up to 8 spectra per object. Three planet were discovered in this way: OGLE-TR-111b \citep{Pon04}, OGLE-TR-113b  and OGLE-TR-132b \citep{Bou04}. Two of them are other instances of very short-period planets, revealing that the case of OGLE-TR-56b was not uncommon. OGLE-TR-113b was subsequently confirmed by independent radial velocities \citep{Kon04} and a high-accuracy photometric transit curve was obtained for OGLE-TR-132 by \citet{Mou04}. One more object in the OGLE survey, OGLE-TR-10, has gradually emerged as a solid transiting planet detection \citep{Kon03b, Bou05, Kon05}. 

The only other transiting planet discovered to date by transit searches is TrES-1 \citep{Alo04}. The total of known transiting planets now stand at seven, and have already spurred several interesting studies  \citep[e.g.][]{Bur04, Cha04, Lec04, Sas03, Maz05}.

In Paper I, we showed that most of the OGLE transiting candidates are in fact eclipsing binaries. These binaries can be interesting in their own right, especially since in many of them the eclisping body is a very small star. Because the mass and radius of the transiting body can be obtained from the transit and radial velocity data, an important by-product of the radial velocity follow-up is to provide constraints on the mass-radius relation for low-mass stars (see Paper I). These data augment those from brighter eclipsing binaries and from interferometric studies or nearby M-dwarfs.

The present paper exposes the results of our spectroscopic follow-up of OGLE transiting candidates in Carina  \citep[Candidates number 60-132 described in][]{Uda02c, Uda03}. Section~\ref{obs} presents the spectroscopic data and reduction, Section~\ref{analysis} explains the tools used in the analysis of the light curve and spectroscopic data, Section~\ref{results} presents the resolution of all cases by category, Section~\ref{indiv} includes individual notes on some objects, and Section~\ref{concl} discusses some implications of this study, notably on the stellar mass-radius relation for low-mass stars and the statistical implications of the planet detections.

\section{Observations and reductions}

\label{obs}

The observations were acquired in 8 half-nights (32 hours) on FLAMES on 13 to 21 March 2004 (Prog. 72.C-0191).
FLAMES is a muli-fiber link which allows to feed the high-resolution spectrograph UVES
with up to 7 targets on a field of view of 25 arcmin diameter, in addition to a simultaneous thorium calibration. In a previous run on this instrument (Paper I), we have shown that FLAMES was able to measure radial velocities with an accuracy of about 30~\ms\ on stars down to the 16th magnitude in $I$. Some trials with HARPS and UVES in slit mode (see Paper~I) led to the conclusion that FLAMES was a very efficient instrument for the radial velocity follow-up of OGLE transit candidates.

\subsection{Target selection and observing strategy}

\label{select}

For the 73 OGLE transit candidates in the Carina field (OGLE-TR-60 to TR-132), the most promising candidates in terms of planetary transits were selected according to three main criteria:

\begin{enumerate}
\item{ the radius of the eclipsing body indicated by the depth and duration of the transit.}

\item{ the shape of the transit. A U-shape (flat-bottom) transit indicates a central transit, while a V-shape indicates a grazing transit, therefore a probable eclipsing binary.}
\item{ the amplitude of the sine and double-sine modulations seen in the light  curve. Sirko \& Paczinski (2003) have analysed the OGLE candidate light curves and shown how the influence of a massive companion could be detected by modulations of the light curve outside the transits. }
\end{enumerate}

Using these three criteria and other occasional indications such as the presence of an anti-transit secondary signal, an initial list was built with 14 first-priority and 16 second-priority objects. 43 candidates were considered almost certainly binaries from these considerations alone. 

In order to optimize the use of telescope time, we used a real-time replacement strategy throughout the observing run: as soon as an object was detected as a binary (i.e. photometric transit signal not caused by a planet but by an eclipsing binary), the object was marked for possible re-allocation of the corresponding FLAMES fiber to another target. The objects were also marked if no signal was detected in the cross-correlation function (indicating a fast-rotating star, an early-type star or a heavily blended system). In the following night, the fiber corresponding to the objects marked was re-allocated to a new target, if allowed by the distribution of the targets in the sky. In that way, we could observe and characterise 42 objects with four fields of 7 fibers and make a highly efficient use of the multiplex facility. We could observe all first-priority targets, and 11 of the 16 second-priority targets (OGLE-TR-70, 71, 73, 74 and 115 were left unobserved because of their inconvenient position in the field).

\begin{figure}[ht!]
\resizebox{8cm}{!}{\includegraphics[angle=-90]{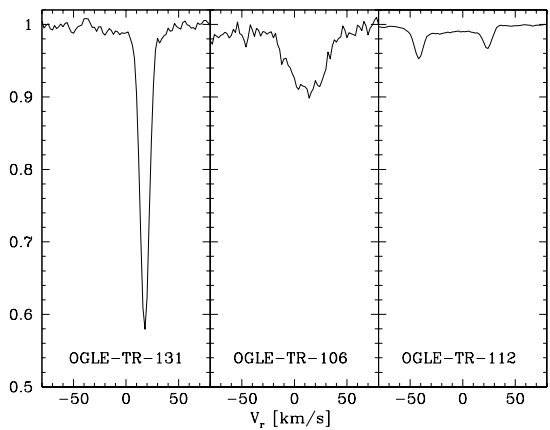}}
\caption{Example of cross-correlation functions from our programme: for an unrotating star (left), for a rotating star synchronised with the eclipsing companion (middle) and for a triple-lined spectroscopic binary (right). In the last case, the third component is the wide depression between the two deeper dips. } 
\label{ccf}
\end{figure}

As reminded in the introduction, most OGLE candidates are eclipsing binaries rather than planets. We used three criteria to identify eclipsing binaries from the spectroscopic measurements: (i) presence of more than one set of lines in the spectrum (double- or triple-lined binaries). (ii) rotational broadening of the lines corresponding to tidal synchronisation. (iii) large variations of the radial velocity. Figure~\ref{ccf} shows examples of spectral cross-correlation function (CCF) illustrating the different cases.

Note that criteria (i) and (ii) require only one spectroscopic measurement. Criteria (ii) is based on the fact that binaries with periods smaller than about 10 days are expected to be tidally locked to their companions in synchronous rotation.  The increased rotation velocity is observed as a broadening of the correlation dip in the cross-correlation function. See Section~\ref{analysis} for more details.



\subsection{Radial velocities}

The spectra obtained from FLAMES/UVES  were extracted using 
the standard ESO-pipeline with bias, flat-field and background correction. Wavelength 
calibration was performed with the ThAr spectra. The radial velocities were obtained by 
 cross-correlation with a numerical template constructed from the Sun spectrum atlas. 
The ThAr spectrum was used in order to compute 
the instrumental drift by cross-correlation with a Thorium template (see Paper~I for details).
 Radial velocity uncertainties were computed  as in Paper~I.
The most precise of our measurements ($\sigma<50$~\ms) are not photon-noise limited and 
we added quadratically an uncertainty of 30 {\ms} in order to take into the account 
systematic errors probably due to wavelength calibration errors, fiber-to-fiber 
contamination, and residual cosmic rays. This value was ajusted on 
the O-C residuals of the non-rotating star without significant radial velocity variations OGLE-TR-131 \citep{Bou04}. 

In the case of double-lined and triple-lines spectra, the radial velocities of each component were calculated with standard techniques for spectroscopic binaries. Iterative solutions were often necessary to disentangle the components when the multiple line systems are blended with eachother. We note that the radial velocity uncertainties were calculated assuming single-lined spectra and in the case of blended spectra they may be underestimated.

Our radial velocity measurements and cross-correlation function parameters are listed 
in Table~\ref{tablevr} and analysed in Section~\ref{results}.

\subsection{Rotation velocities}

Rotational velocities were computed from the observed cross-correlation function  using rotationally broadened line profiles convolved with a Gaussian instrumental profile of width $4.0$ {\kms} (found suitable for our instrumental setup in Paper I). The profiles were fitted to the CCF simultaneously with the radial velocity to determine the projected rotation velocity  {\vsini} of the target objects.
A  quadratic limb-darkening with coefficients  $u1\! + \! u2\! =\! 0.6$ was assumed 
 \citep[The computations of][find that such a coefficient is a suitable approximation for a wide range of spectral types in wavelengths corresponding to the V filter]{Bar03}. 

\subsection{Stellar spectroscopic parameters}

For the slowly-rotating stars in our sample, the stellar parameters
(temperatures, gravities and metallicities) were obtained from an analysis 
of a set of \ion{Fe}{i} and \ion{Fe}{ii} lines, following the procedure used 
in \citet{San04}. 
The precision of the derived atmospheric parameters 
is limited by the relatively low S/N of the combined spectra (30-50), 
together with some possible contamination coming from the ThAr spectrum. 

For the stars in our sample rotating with \vsini $\geq$20 {\kms}, or multiple-lined spectra,
the method described in \citet{San04} is not applicable,
because the measurement of individual equivalent widths is
not accurate enough due to line blending. The majority of the objects in our sample are actually in this case.
When necessary for the resolution of the case,  we have determined very rough estimates of the spectral type by visual comparison of the observed spectra with a grid of synthetic spectra.

\section{Analysis}
\label{analysis}
A detailed description of our analysis tools for photometric and spectroscopic data of transiting candidate is given in Paper I. 
We briefly sketch the analysis "toolbox" below and refer to Paper I for a more detailed description.

\subsection{Synchronised rotation of eclipsing binaries}

For close binaries, with rotation periods of the order of a few days, we expect the 
rotation axis to be aligned with the orbital axis and the system 
to be tidally locked \citep[e.g.][]{Lev76,Hut81}.  For known close binaries, the alignment of the axes and the tidal locking is fast. In that case, $P_{rot}=P_{transit}$ and the rotation velocity is 
directly related to the radius of the primary. In Paper I, we could indeed verify this hypothesis in all the target observed, and we shall use it here both to estimate the primary radius $R$ and as an indication that the companion is massive when we have only one spectroscopic measurement. 

\begin{figure}[ht!]
\resizebox{8cm}{!}{\includegraphics{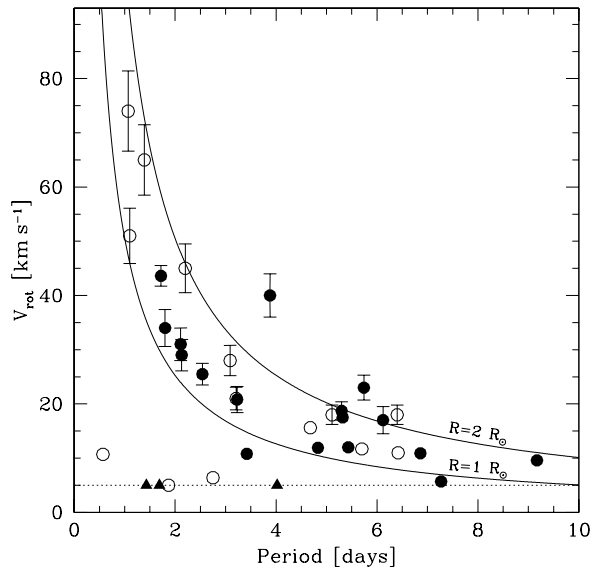}}
\caption{Rotation velocity of the primary vs period for our sample. The lines show the expected rotation velocity for tidally locked systems with radii of 1 $\Rsol$ and 2 $\Rsol$ for the primary. For objects at \vsini=5 \kms (dotted line), the value given is an upper limit. Uncertainties are indicated only when larger than the symbols (an uncertainty of 10\% is used for values based on single spectra). Black dots show objects with a precise radial velocity orbits, triangles the detected planets.  The three open circles isolated in the lower left are OGLE-TR-97 (triple system), 124 and 131 (suspected false transit detections). For all other objects the rotation velocity is compatible with synchronous rotation.} 
\label{rotvit}
\end{figure}

Figure~\ref{rotvit} shows, for the Carina sample, the rotation velocity of the primary target as a function of the period of the transit signal (for transiting objects, $\sin i\simeq 1$ and therefore $V_{rot}\simeq v \sin i$). Most objects have a rotation velocity compatible with synchronisation for reasonable values of the primary radius (1-2 $\Rsol$), indicating that eclipsing binaries dominate the sample. The only exceptions are the three planet hosts OGLE-TR-111, 113 and 132, and the objects OGLE-TR-97, 124 and 131. The first of these has a triple-lined spectrum and the last two are suspected false transit detections (see Sect.~\ref{snovar}).

\subsection{Revised period}

The radial velocity data, obtained two years after the photetric data, generally allows a significant improvement of the determination of the orbital periods. Some systems turn out to be grazing, equal-mass eclipsing binaries where both the eclipse and anti-eclipse were visible in the light curve. In these cases the real orbital period of the system is double the period in \citet{Uda02c, Uda03}.

Contrarily to Paper~I, there were no cases where the radial velocity indicated a completely different period than the photometry. This reflects the fact that much less  candidates with only two or three measured transits were accepted in the OGLE Carina sample than in the Galactic bulge sample.


\subsection{Analysis of the transit shape}
\label{agol}

The depth, width and general shape of the transit signal depend on a combination of 
physical variables, mainly the radius ratio $q_R$ (noted $\overline{r}$ in Paper~I), the primary radius $R$ and the impact 
parameter $b$ (or, equivalently for circular orbits, the angle $i$ of the normal of the 
orbital plane with the line-of-sight) and the orbital eccentricity. It is also more weakly 
dependent on the total mass $(m+M)$ -- via the orbital period and semi-major axis for a 
Keplerian orbit -- and the limb darkening coefficients. The parameter $q_R$ is mainly 
constrained by the transit depth, $b$ by the transit shape and the factor
$R (m+M)^{-1/3}$ by the transit duration. 

The light curves were fitted by non-linear least square fitting with analytic transit curves 
computed according to \citet{Man02}, using a 
quadratic limb darkening model with $u1\! + \! u2\! =\! 0.3$. Note that this is 
different from the coefficients used for the determination of the rotational velocity, 
because the wavelengths are different. The OGLE data was 
obtained with an $I$ filter while the spectra are centered on the visible. The fitted 
parameters were $q_R$, $V_T/R$ and $b$, where $V_T$ is the transversal orbital velocity 
at the time of the transit.  The uncertainties on these parameters were estimated using a method that 
takes into account the covariance of the photometric residuals (see Paper I for details).


\subsection{Synthesis of the spectroscopic and photometric constraints}

 For convenience, the six relations 
used to infer the system parameters from the observables are repeated below from Paper~I:

 \begin{eqnarray}
\frac{v \sin i}{\sin i} \cdot P &=& 50.6 \cdot R
\\
K&=&214 \cdot \frac{m}{(m+M)^{2/3}} \cdot P^{-1/3}
\\
i&=& acos (\frac{b \cdot R}{a})  
\\
V_T&=&2 \pi \cdot  \frac{a}{P} 
\\
q_R&= &r/R
\\
(\log T_{eff}, \log g, [Fe/H], R)&=& f(M, age, Z)
\end{eqnarray}

with $a = 4.20 \cdot P^{2/3}  \cdot (m+M)^{1/3}$ \\

where \vsini\ is the projected rotation velocity in \kms, $i$ the orbital inclination, $P$ the period in days, $R, r, M$ and $m$ the radii and masses of the eclipsed and eclipsing bodies in solar units, $K$ the primary radial-velocity semi-amplitude in \kms, and $a$ the orbital semi-major axis in units of $\Rsol$. The function $f$ in the last equation is the relation given by the theoretical stellar evolution models of \citet{Gir02}. 


We assumed that all orbits were circular ($e=0$) unless the radial velocity data clearly indicated otherwise.
 Very close binaries are expected to be circularized on a timescale shorter than typical stellar ages. However, the circularisation timescale
increases with decreasing companion mass \citep{Zah92}, so that some of the low-mass companions in our sample may still have eccentric orbits.
Four objects show definite indications of $e\neq 0$ in the velocity curve.  For eccentric orbits, Equ. (2), (3) and (4) above must be slightly modified. The distribution of eccentricities are discussed in Section~\ref{ecc}.

\section{Results}
\label{results}

The analysis of the 42 objects measured in spectroscopy is presented below, divided for convenience according to categories reflecting the treatment of the data and the nature of the systems: single-lined spectroscopic binaires (eclipsing binaries with small companion), multiple-lined spectroscopic binaries (grazing eclipsing binaries, multiple systems, line-of-sight blends), transiting planets, suspected false transit detections, and objects without CCF signal.

\subsection{Single-lined spectroscopic binaries}
\label{ssb1}

\subsubsection{Eclipsing binaries with resolved orbits}
\label{ssb1_1}

\begin{figure*}[ht!]
\resizebox{18cm}{!}{\includegraphics[angle=-90]{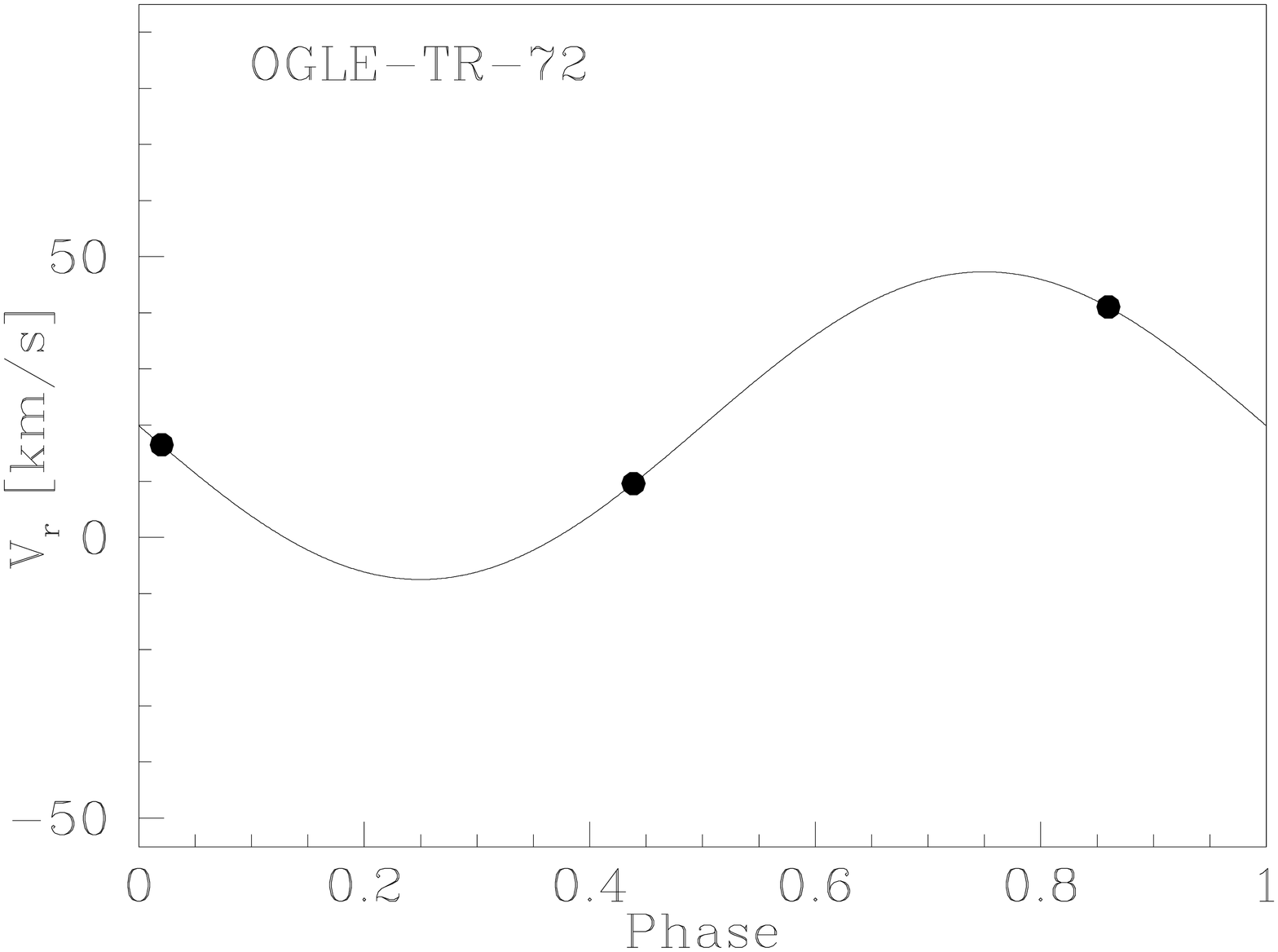}\includegraphics[angle=-90]{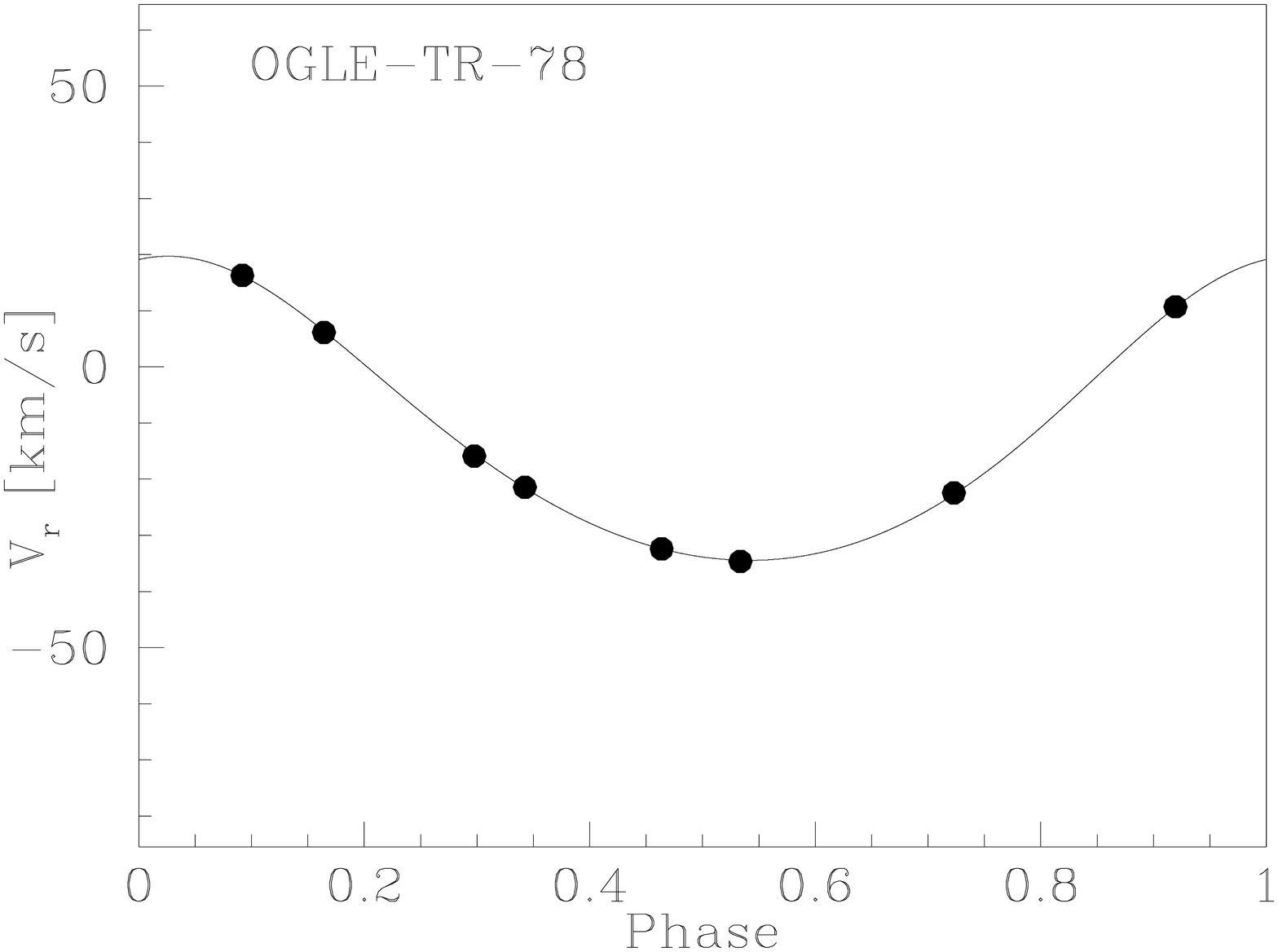}\includegraphics[angle=-90]{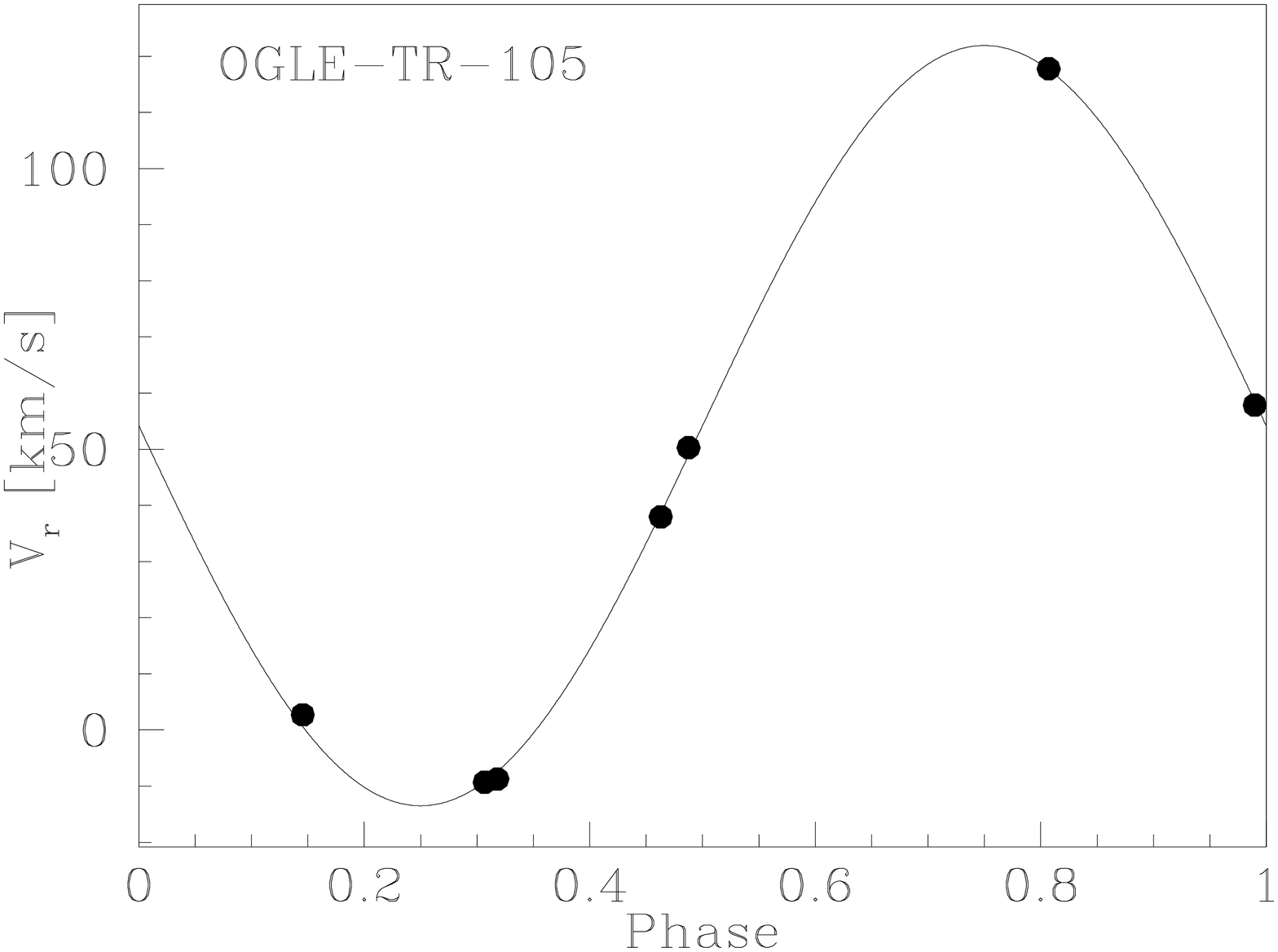}}
\resizebox{18cm}{!}{\includegraphics[angle=-90]{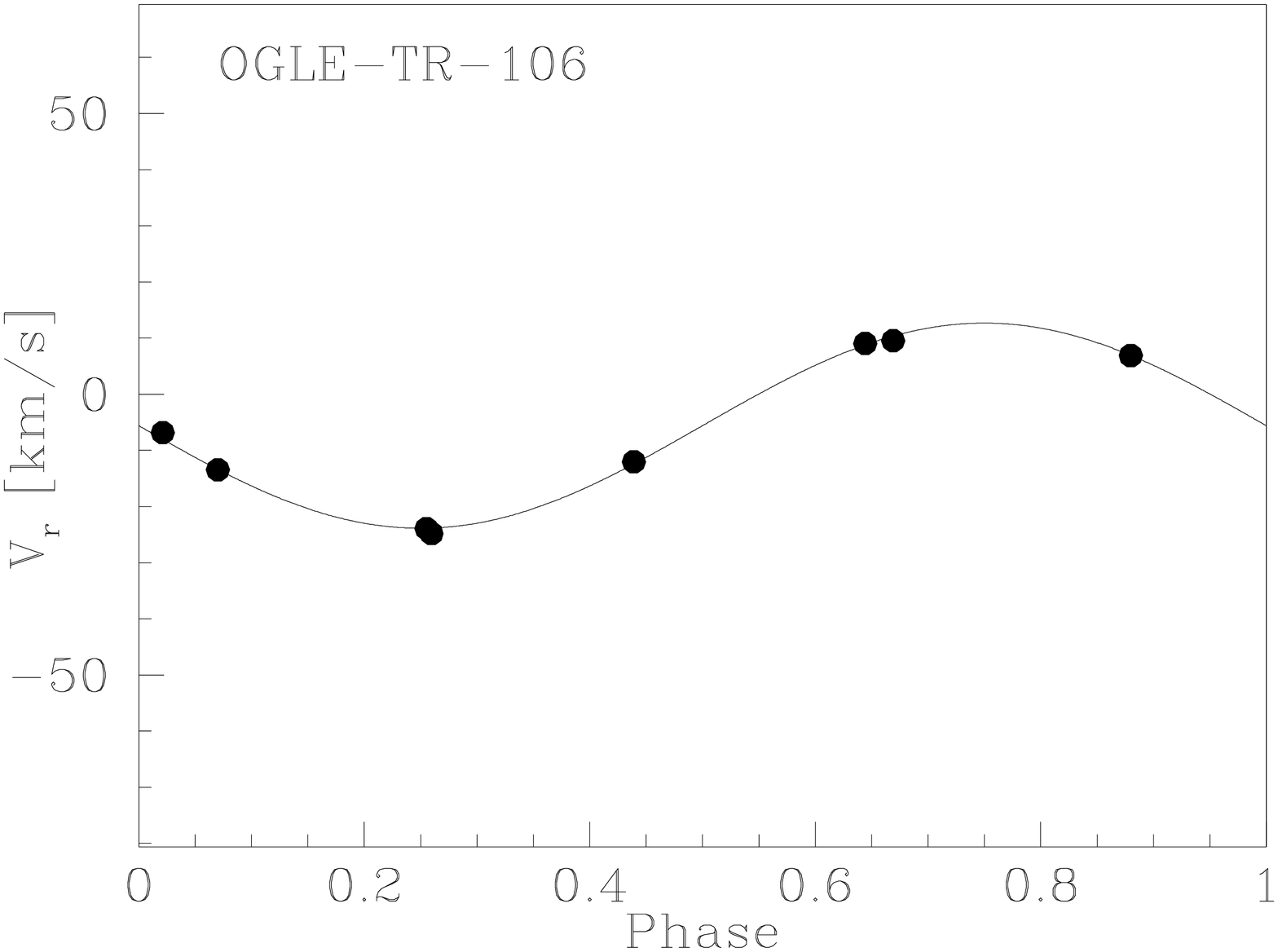}\includegraphics[angle=-90]{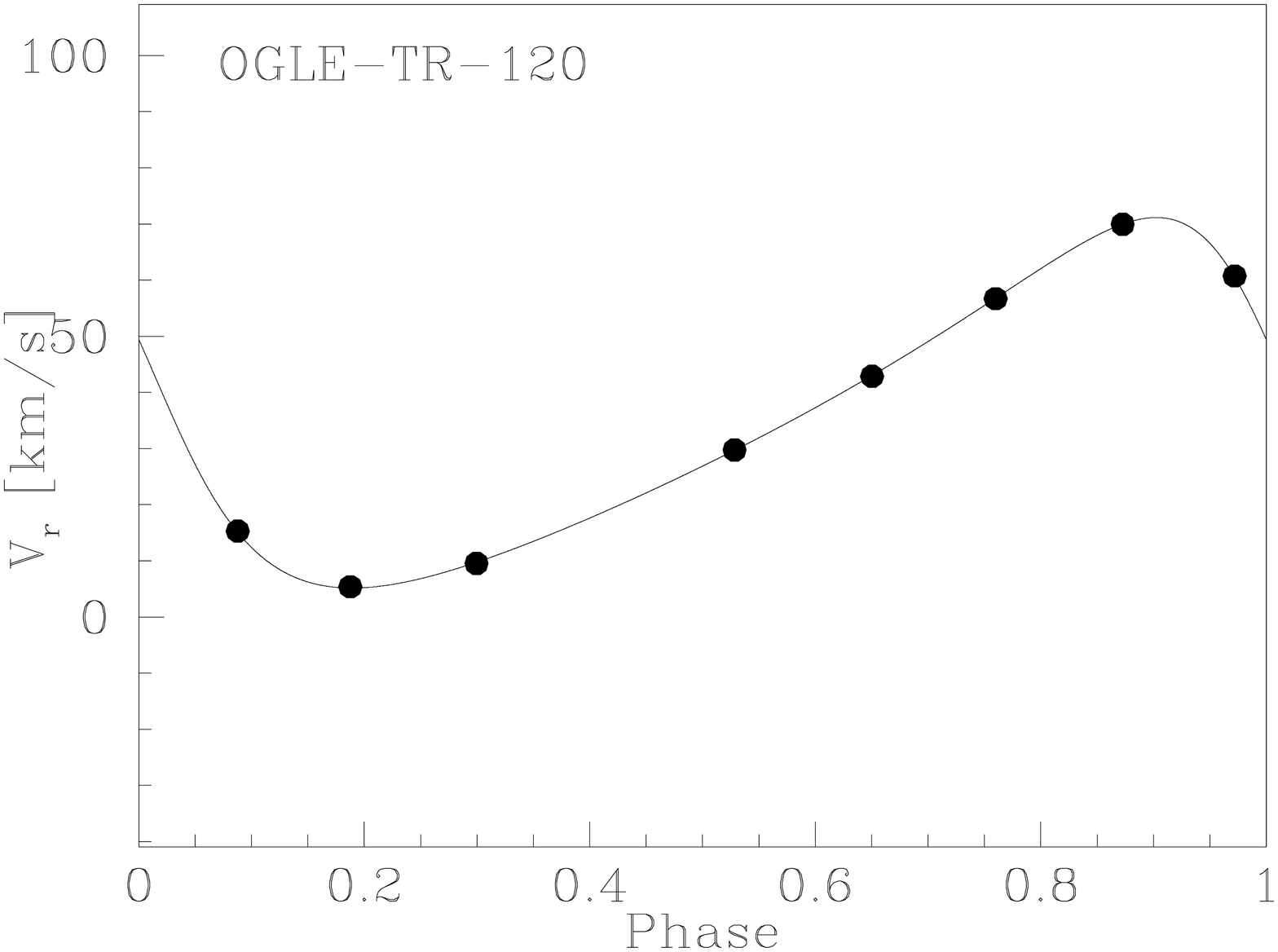}\includegraphics[angle=-90]{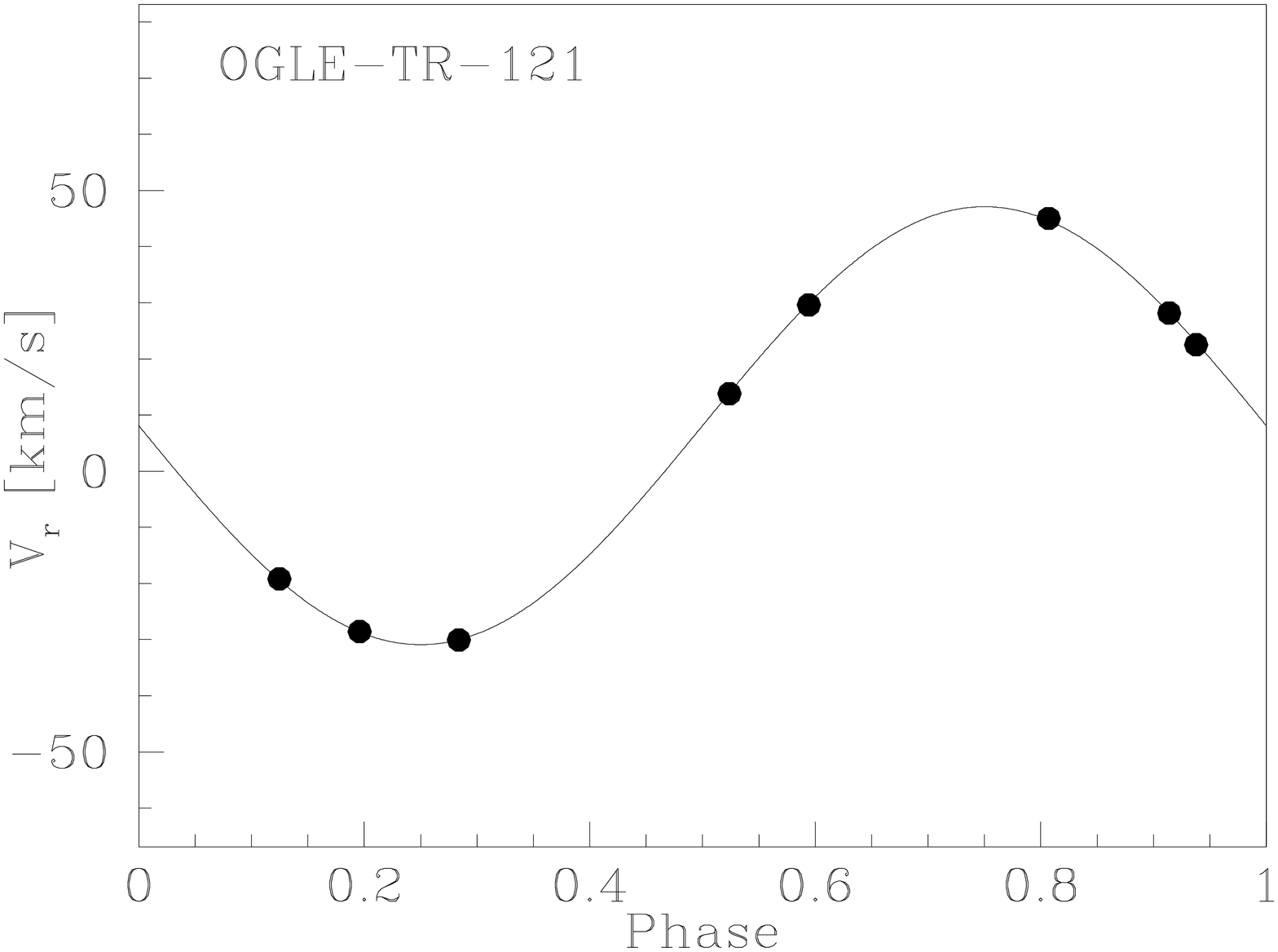}}
\resizebox{18cm}{!}{\includegraphics[angle=-90]{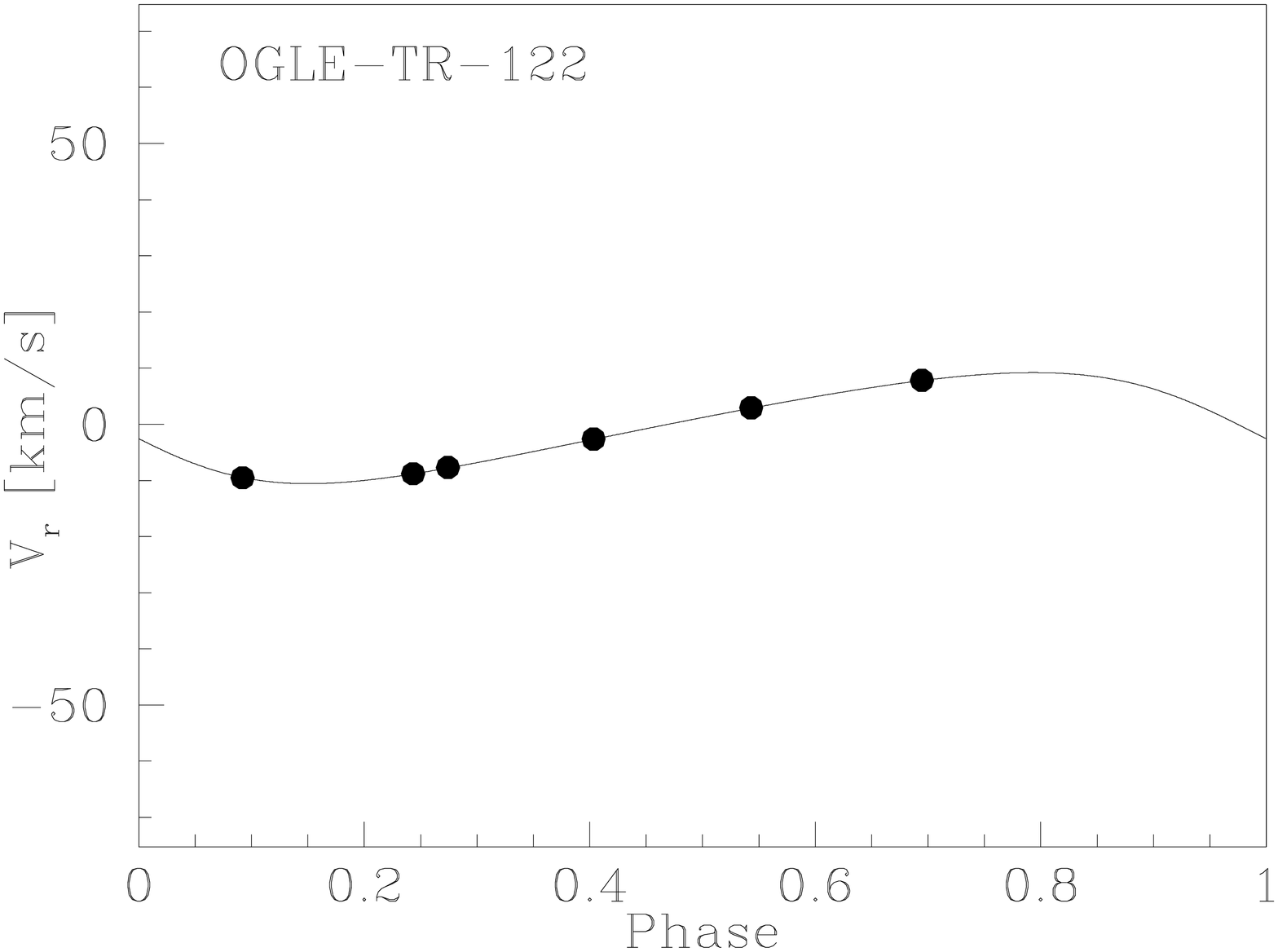}\includegraphics[angle=-90]{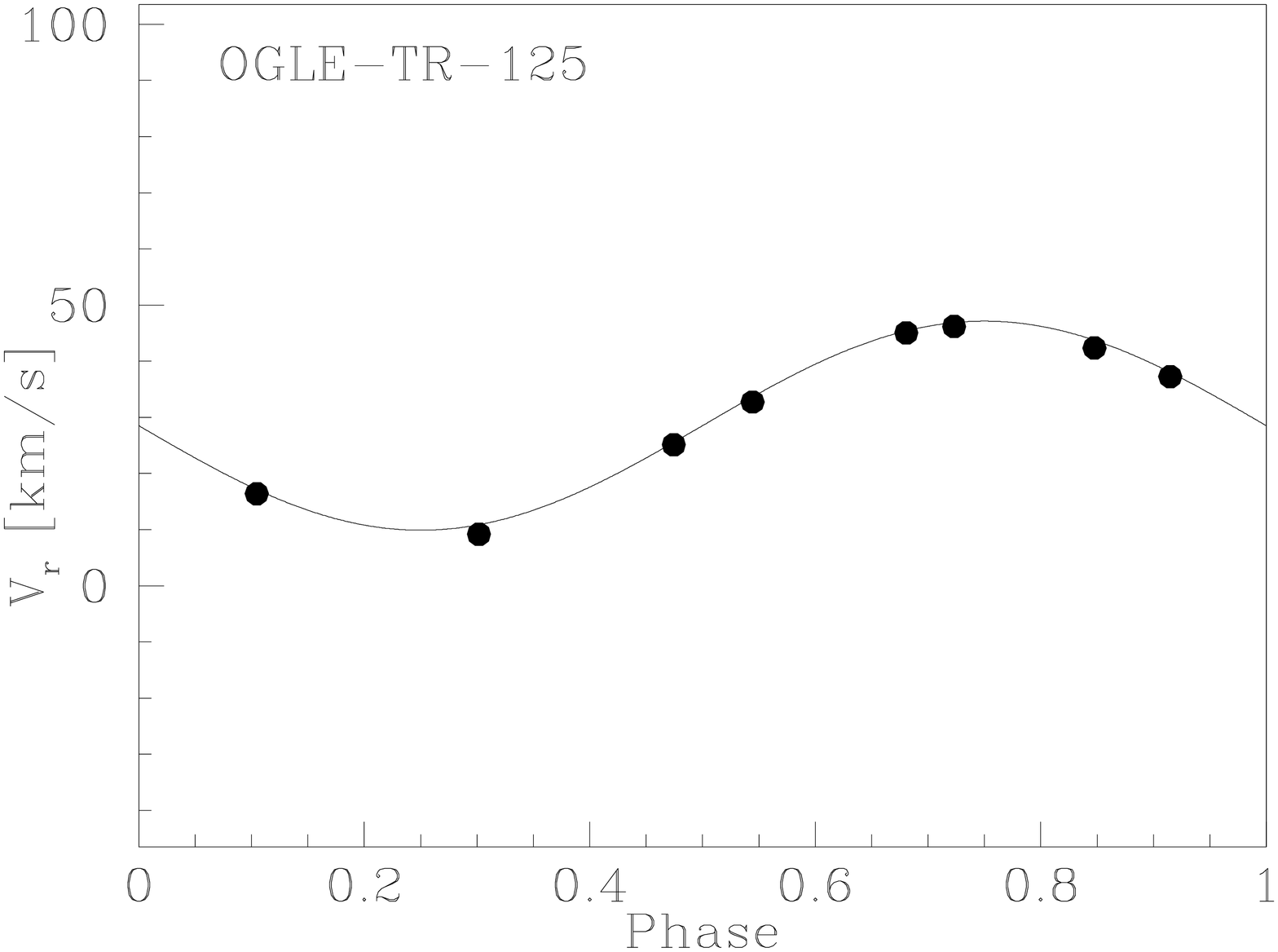}\includegraphics[angle=-90]{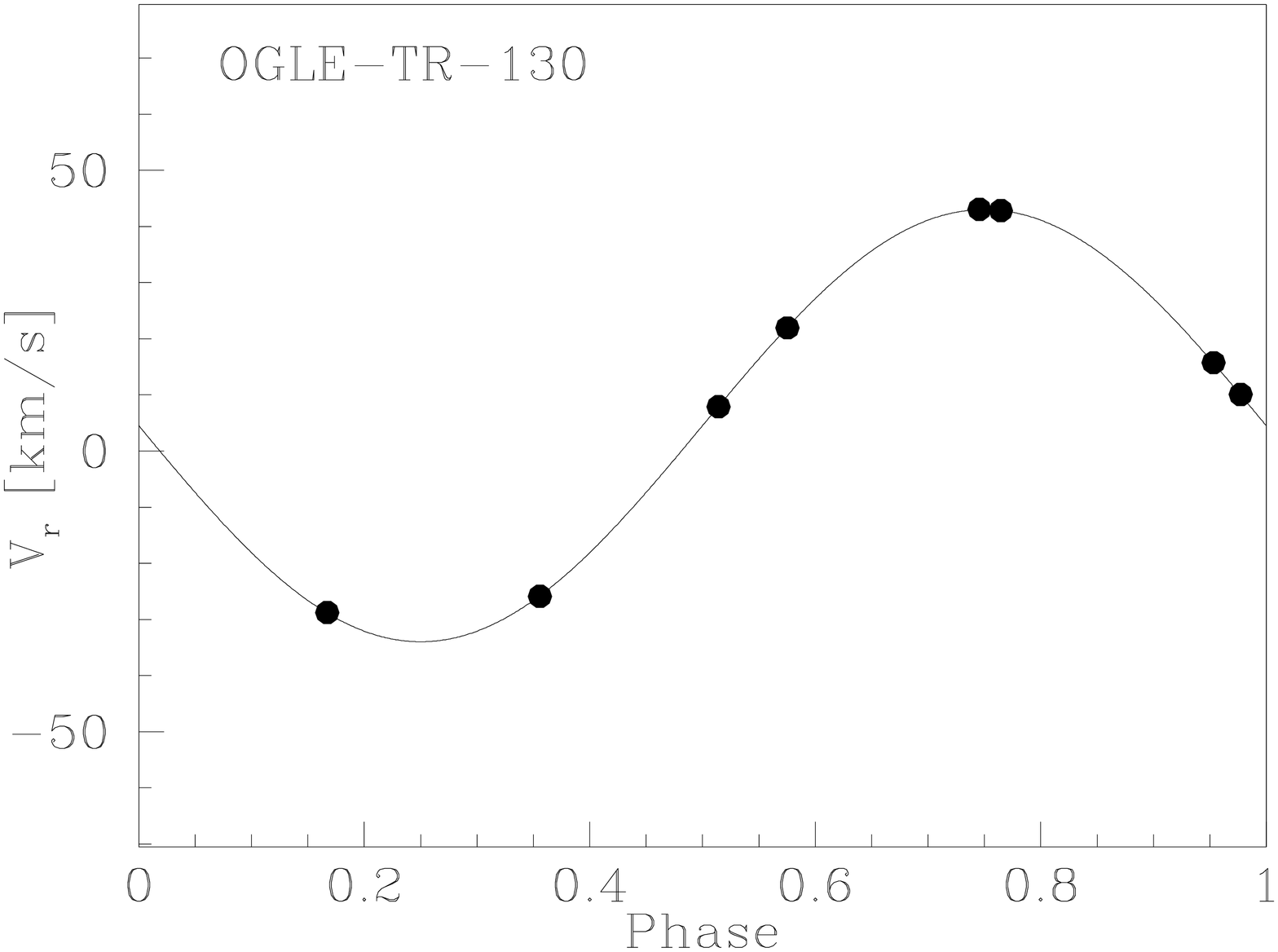}}
\caption{Radial velocity data and orbits for single-lined spectroscopic binaries. The orbital periods and epochs are constrained in combination with the photometric signal. The corresponding parameters are given in Table~\ref{table_sb1}. The measurements uncertainties are smaller than the symbols. The solutions for OGLE-TR-123 and OGLE-TR-129 are tentative (see Text). } 
\label{sb1}
\end{figure*}

Objects OGLE-TR-72, 78, 105, 106, 120, 121, 122, 125 and 130 show single-lined spectra with radial velocity variations compatible with orbital motion at the period given by the photometric transits. They are eclipsing binaries, with a fainter, unseen companion causing the observed photometric signal. The radial velocity data were fitted with Keplerian orbits. Figure~\ref{sb1} shows the resulting radial velocity orbits. The spectroscopic and light-curve data were then combined as described briefly above and more fully in Paper I to determine the mass and radius of both components. The target stars are assumed to be tidally locked to their companion in synchronous rotation, which implies that the rotation velocity derived from the broadening of the cross-correlation function is directly related to the star's radius (Equ.~1). 
Contrarily to Paper I, for most of the objects in this study we have no precise spectroscopic estimate of the temperature, because of the rotational line broadening and the low signal-to-noise of the spectra. When necessary to constrain the solution, the existence of a signal in the cross-correlation function was used to give a broad upper limit to the temperature, $T_{eff}<7000$~K. This improves the solution for $m$ by eliminating very high values of $M$ in cases when $R$ is large (namely for OGLE-TR-78 and 125).

The results for the radial velocity orbit and the rotation  velocity are given in Table~\ref{table_sb1}. The parameters obtained from the combination of the light curve and spectroscopy are given in Table~\ref{param}.
Two different methods were used to combine the constraints (1)-(6): if the impact parameter was low, the solution was obtained by $\chi^2$ minimisation as in Paper I. For higher values of the impact parameter, the radius ratio is partly degenerate with the orbital angle. In these cases (OGLE-TR-72, 105, 121), we first obtained an estimate of the primary mass using the approximate relation $M \sim \sqrt{R}$, then fitted the light curve with $V_T/R$ fixed and only the radius ratio and impact parameters as free parameters. Obviously, in that case some values obtained can have higher uncertainties, as are indicated in Table~\ref{param} with columns. 

For OGLE-TR-105, the radial velocity data show that the period needs to be doubled, and that both transit and anti-transit were visible in the photometric data.

OGLE-TR-122 is the smallest stellar object in our sample, with a planet-like radius, and as such is highly interesting. It is analysed in more details in a separate Letter \citep{Mel05}.

\subsubsection{Eclipsing binaries with tentative orbits}
\label{ssb1_2}

\begin{figure*}[ht!]
\resizebox{12cm}{!}{\includegraphics[angle=-90]{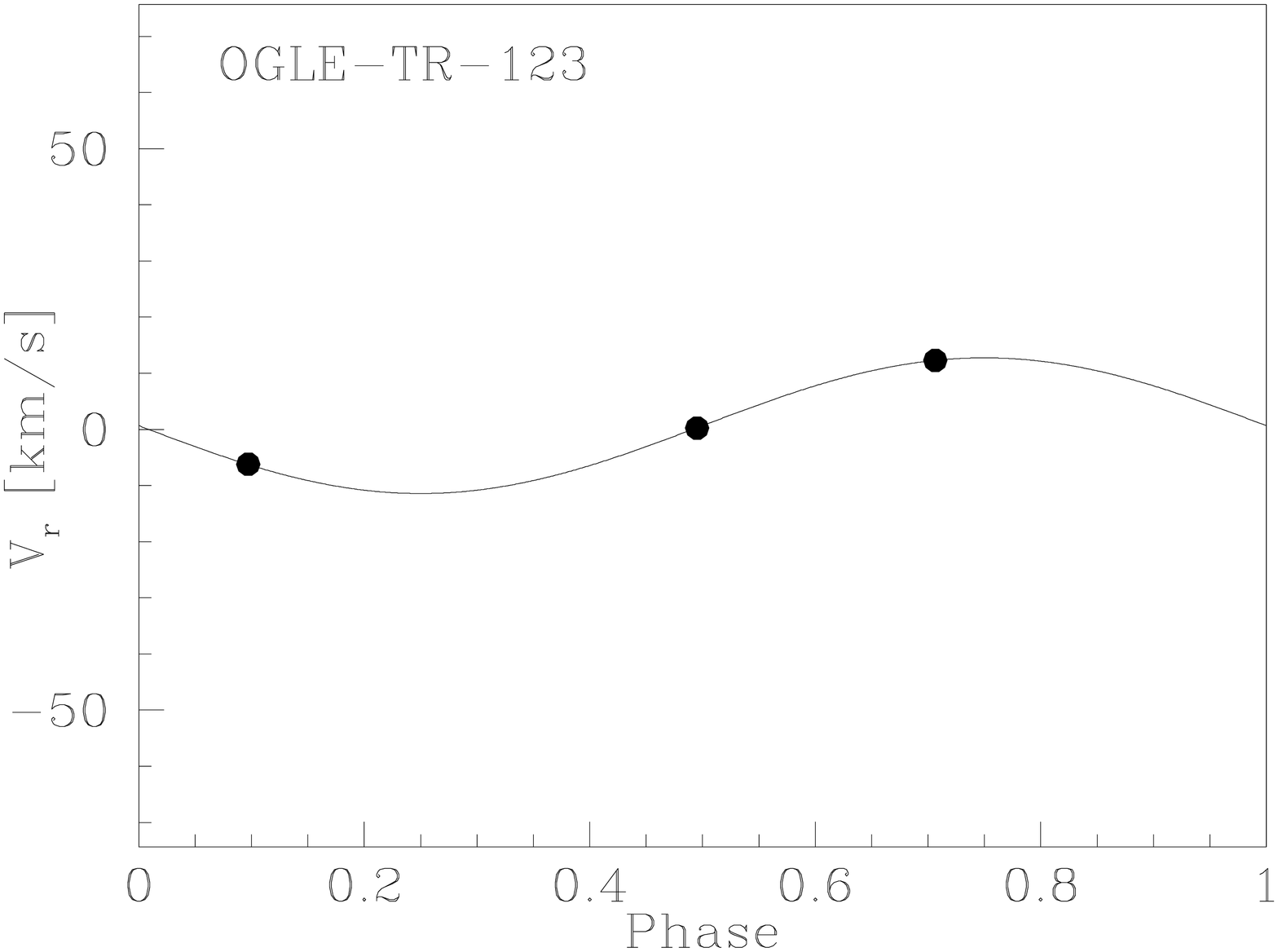}\includegraphics[angle=-90]{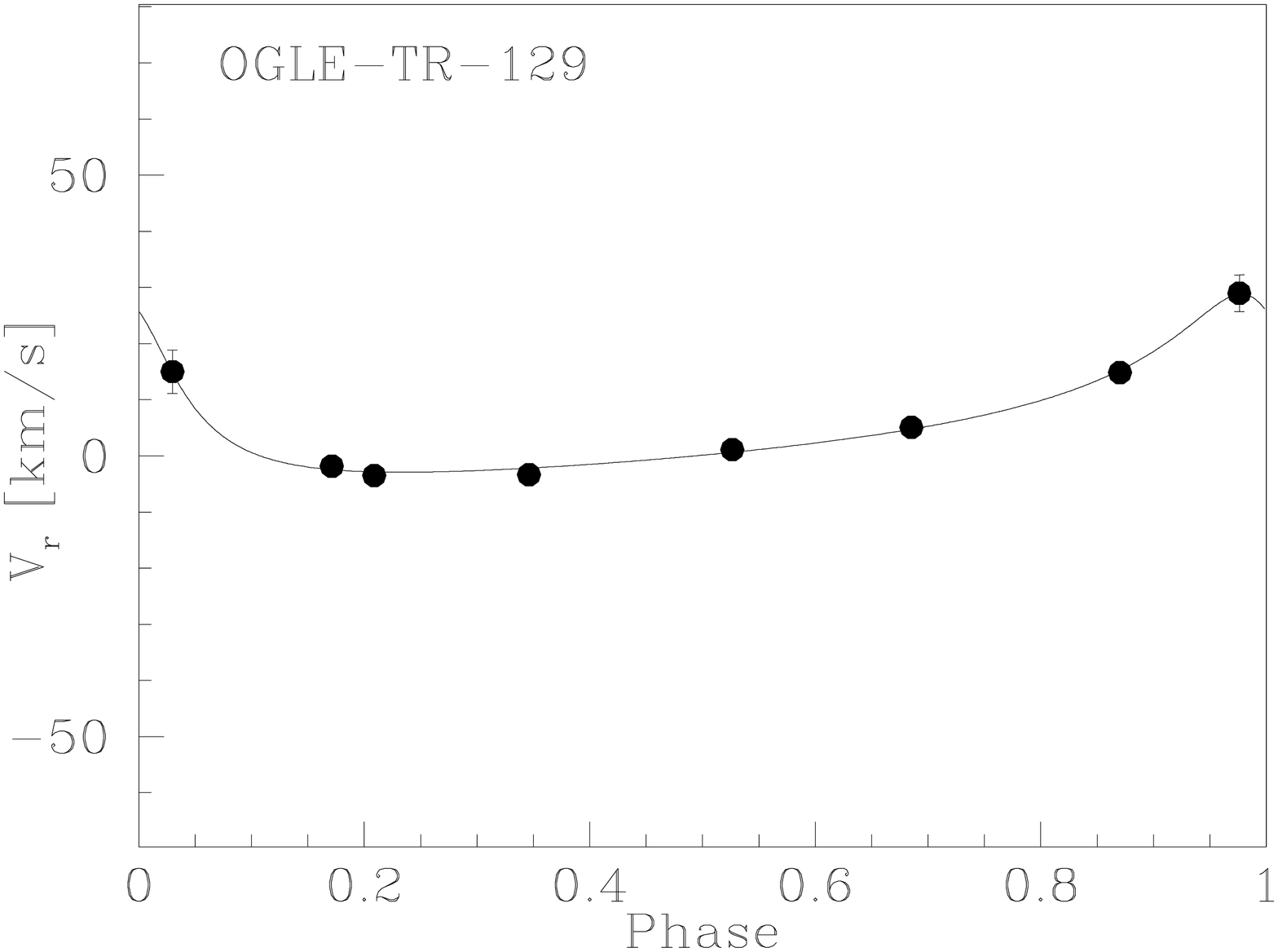}}
\caption{Radial velocity data and tentative orbital solution for OGLE-TR-123 and OGLE-TR-129. The orbital periods and epochs are constrained in combination with the photometric signal. The corresponding parameters are given in Table~\ref{table_sb1}. The measurements uncertainties are plotted only when larger than the symbols.  } 
\label{tentative}
\end{figure*}

OGLE-TR-123 was measured only three times and the solution proposed in Table~\ref{table_sb1} is only tentative. Our three spectra for this objects show a wide CCF signal, with \vsini $ \simeq 34.5$ \kms, and small velocity variations to the level of a few \kms. The rotation is compatible with synchronisation, and would indicate $R \sim 1.2 \Rsol$. A circular orbit fitted on the velocity points with the  epoch of the transit signal gives $K=12.08$ \kms, $V_0= 0.66$ \kms and $P=1.8038$ days. These are tentative values because the number of free parameters is equal to the number of points. If the period is fixed to the OGLE value as well, the orbit fit yields $K=11.32 \pm 0.6$ \kms and $V_0=0.77\pm 0.40$ \kms. This implies $r=0.097 \pm 0.006 \Rsol$ and $m\simeq 0.070 \Msol$ for the transiting body, making OGLE-TR-123 an extremely interesting candidate of transiting brown dwarf or low-mass star near the Hydrogen-burning limit (see Fig.~\ref{mrr2}). However, the transit duration is longer than would be predicted by this scenario. More radial velocity data would be needed to confirm the period and exclude more complex blend scenarios. 


OGLE-TR-129 has a wide and very shallow signal in the CCF, and the radial velocities have large uncertainties. The projected rotation velocity is \vsini $\sim 23$ \kms, which implies $R\simeq 2.6 \Rsol$ in case of synchronised rotation. The spectrum indicates a high temperature, $T_{eff}\geq 6500$, compatible with this radius value. The radial velocity data phased on the period of the photometric transit signal are compatible with a markedly excentric orbit. There is some degeneracy between the excentricity and $K$, and orbits with lower eccentricity and lower values of $K$ are also marginally compatible with the data. Possible values for $K$ would imply $m=0.07 - 0.13 \Msol$ for the transiting body. A transiting M-dwarf scenario is not coherent, however, with the transit shape and depth if the primary is synchronised: the amplitude of the radial velocity variation indicates $m\leq 0.2 \Msol$, but the transit depth indicates $r/R\sim 0.2$, therefore with a synchronised primary $r\sim 0.5 \Rsol$ for the secondary. Such a mass-radius relation for a M-dwarf is not expected. Moreover, the transit is clearly V-shaped, indicating a grazing transit with $b\sim 1$, which would imply an even larger secondary. The position of OGLE-TR-129 in Fig.~\ref{conf} indicates that it lies near the transit detection threshold, with $S_d=20$. Comparison with other similar cases (see Section~\ref{snovar}) suggest that it may be a false positive. In that case, the radial velocity variation may be unrelated to the photometric transit signal.  Alternatively, the radial velocity variations may be explaind by a blend of more than one set of lines. The radial velocities are not precise enough to distinguish a Keplerian orbit from other types of variations. In summary, a possible scenario for OGLE-TR-129 is that of a $0.07-0.015 \Msol$ transiting M-dwarf with a hot, fast-rotating primary that is not yet synchronised, but other scenarios cannot be entirely excluded.

The proposed orbits for these two objects are displayed in Fig.~\ref{tentative}.

\begin{table*}[ht!]

\begin{tabular}{l l l l l l l l l l }

Name &$P$  & $T_{tr}${\tiny (OGLE)} & $P_{\rm OGLE}$  & $T_p$       & w  & K & $V_0$ & e & \vsini    \\ 			       
     & [days]         & [-2452000]      &[days] & [-2452000] & [deg]  & [\kms] & [\kms] &  & [\kms]\\ \hline \hline
72   & 6.8581 & 77.399 &  (6.854)   & - 	& -	& 27.4            & 19.9	    & 0		          &  10.9$\pm$0.2  \\
78   & 5.3187 & 328.812&  (5.32038) & 327.849	& 348	& 27.08$\pm$0.19  & -10.47$\pm$0.12 & 0.117$\pm$0.007     &  17.5$\pm$0.8  \\
105  & 6.1161 & 324.380&  (3.0581)  & - 	& -	& 67.73$\pm$0.93  & 54.23$\pm$0.61  & 0.0$\pm$0.01	  &  17$\pm$2.5    \\
106  & 2.5359 & 324.783&  (2.53585) & - 	& -	& 18.27$\pm$0.38  & -5.60$\pm$0.27  & 0.0$\pm$0.02	  &  25.5$\pm$2.0  \\ 
120  & 9.1662 & 331.498&  (9.16590) & 331.268	& 70	& 32.99$\pm$0.08  & 34.072$\pm$0.06 & 0.361$\pm$0.002     &  9.6$\pm$0.4   \\
121  & 3.2321 & 325.689&  (3.2321)  & - 	& -	& 39.00$\pm$0.12  & 8.100$\pm$0.086 & 0.0$\pm$0.006	 &  20.8$\pm$2.4  \\
122  & 7.2695 & 342.283&  (7.26867) & 335.152	& 101	& 9.887$\pm$0.065 & -0.252$\pm$0.058& 0.231$\pm$0.006     &  5.7$\pm$0.6       \\
123  & 1.8039 & 324.979&  (1.8038)  & - 	& -	& 12.08           & 0.66            & 0		        	  & 34$\pm$3	 \\  
125  & 5.3039 & 343.825&  (5.30382) & - 	& -	& 18.83$\pm$0.25  & 27.60$\pm$0.18  & 0.0$\pm$0.01	&  18.7$\pm$1.7   \\
129  & 5.7339 & 327.368&  (5.74073) & 327.5:	& 36:	& 16:            & 5:                & 0.6:      &  23$\pm$3  \\
130  & 4.83103& 327.281&  (4.83027) & - 	& -	& 38.482$\pm$0.037& 4.516$\pm$0.026 & 0.0$\pm$0.001     &  11.9$\pm$0.4  \\    
\end{tabular}
\caption{Orbital parameters for single-lined systems.
Columns 2-9 : Parameters of the orbital solution. $P$: revised period, $T_{tr}${\tiny (OGLE)}: epoch of the transit (fixed), $P_{\rm OGLE}$: original OGLE period, $T_p$: epoch of periastron, $w$: omega angle, , $K$: orbital semi-amplitude, $V_0$: systemic velocity,  $e$: eccentricity. Column 10: projected rotation velocity from the cross-correlation function.}

\label{table_sb1}
\end{table*}

\subsubsection{Probable eclipsing binaries without resolved orbits}
\label{ssb1_3}

OGLE-TR-63 was measured three times in spectroscopy. Its rotation velocity is compatible with orbital synchronisation. Because the rotational broadening is very large (\vsini $\sim$ 74 \kms), the radial velocities have high uncertainties. Moreover, the period near 1 day causes an unfavourable phase sampling. As a consequence, the radial velocity data is compatible with both a small M-dwarf transiting companion and with a constant velocity. This case remains unsolved.

OGLE-TR-94, 98, 99 and 126 were measured only once in spectroscopy.  Their rotational line broadening is compatible with tidal synchronisation at the period of the observed transit signal (see Table~\ref{table_sb1} and Fig.~\ref{rotvit}), so that these objects are very likely to be eclipsing binaries. Note that even if by chance the high rotation velocity were not due to orbital synchronisation, it would make the detection of a planet orbital motion very challenging anyway, because the high width of the spectral lines significantly increases the uncertainty on the derived radial velocities. The resulting radial velocity uncertainties would be too high to reveal a planetary orbit with our FLAMES/VLT observational setup and reasonable exposure times. Moreover, if the rapid rotation were not due to synchronisation, it would most probably indicate an early-type primary. In that case, the large radius of the primary would imply a secondary radius too large for a planet. Therefore, we do not expect to miss any planet detection by rejecting these fast rotators.

For these objects, a tentative value of the primary radius can be obtained from the rotation velocity and Equ.~(1), and the secondary radius can be estimated through the radius ratio obtained by a fit of the lightcurve. No mass estimate can be derived. Table~\ref{unres} gives \vsini, $R$ and $r$ for these objects. These values are tentative estimates and no uncertainties are derived. 

In the case of OGLE-TR-94, the lightcurve indicates a grazing transit, so that the radius ratio is degenerate with the impact parameter. Only a lower limit can be assigned to $r$. It is also possible that both the eclipse and anti-eclipse are seen, and in that case the period has to be doubled and the secondary radius is comparable to the primary radius, so that the value of $r$ given in the table is really a lower limit.

\begin{table}[h!]
\begin{tabular}{l l l  l l l}
Name & $N$ & \vsini & $R$ & $r$  & $P_{\rm OGLE}$\\ 
 & & [ \kms  ] & [$\Rsol$] & [$\Rsol$] & [days] \\ \hline \hline
OGLE-TR-63 &3 &74 &1.6 &0.15 & 1.06698\\
OGLE-TR-94 &1 &28 &1.7 &$>$0.2&  3.09222\\
OGLE-TR-98 &1 &18 &2.2 &0.34 &  6.39800\\
OGLE-TR-99 &1 &51 &1.1 &0.19& 1.10280\\
OGLE-TR-126 &1 &18 &1.8 &0.25& 5.11080\\
\hline
\end{tabular}
\caption{Rotation velocity and radius estimates for primary and secondary components of single-lined suspected binaries without orbits. $N$: number of spectra; \vsini: projected rotational velocity of the primary;  $R$ and $r$: radius of the eclipsed and eclipsing bodies respectively.}
\label{unres}
\end{table}

\subsection{Double-lined or triple-lined spectroscopic binaries}

\label{ssb2}
\label{ssb}

\begin{figure*}[ht!]
\resizebox{15cm}{!}{\includegraphics[angle=-90]{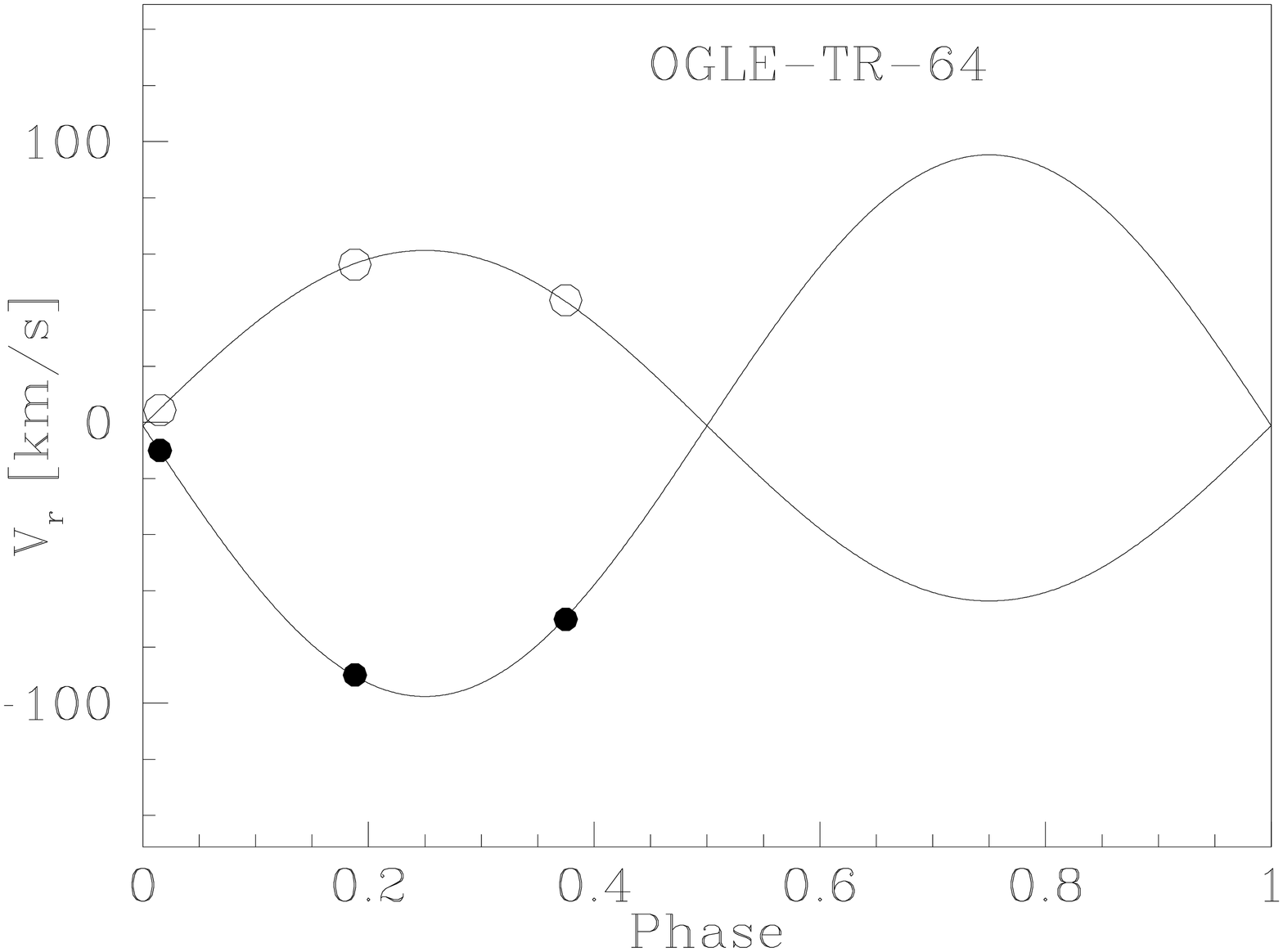}\includegraphics[angle=-90]{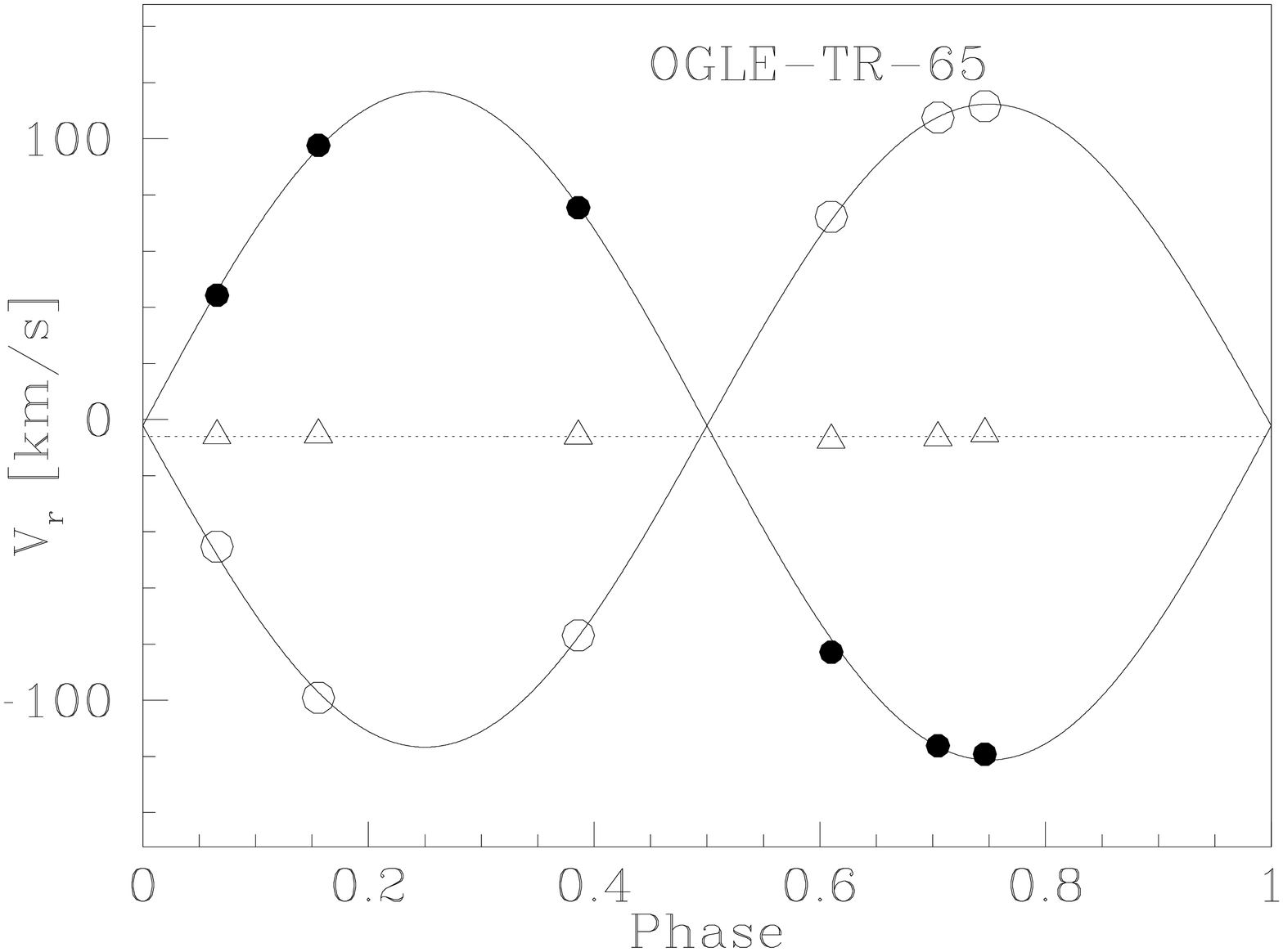}\includegraphics[angle=-90]{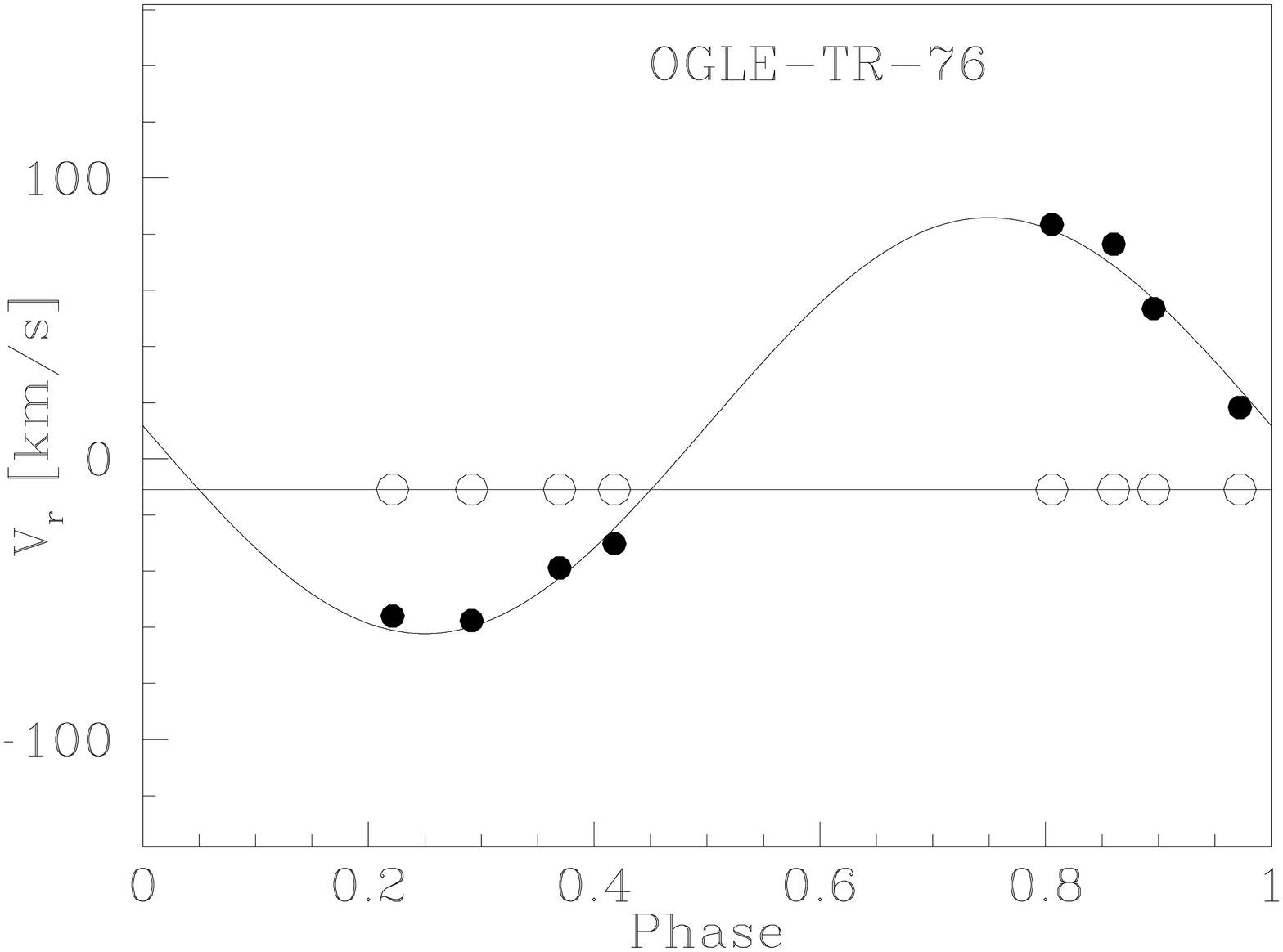}}
\resizebox{15cm}{!}{\includegraphics[angle=-90]{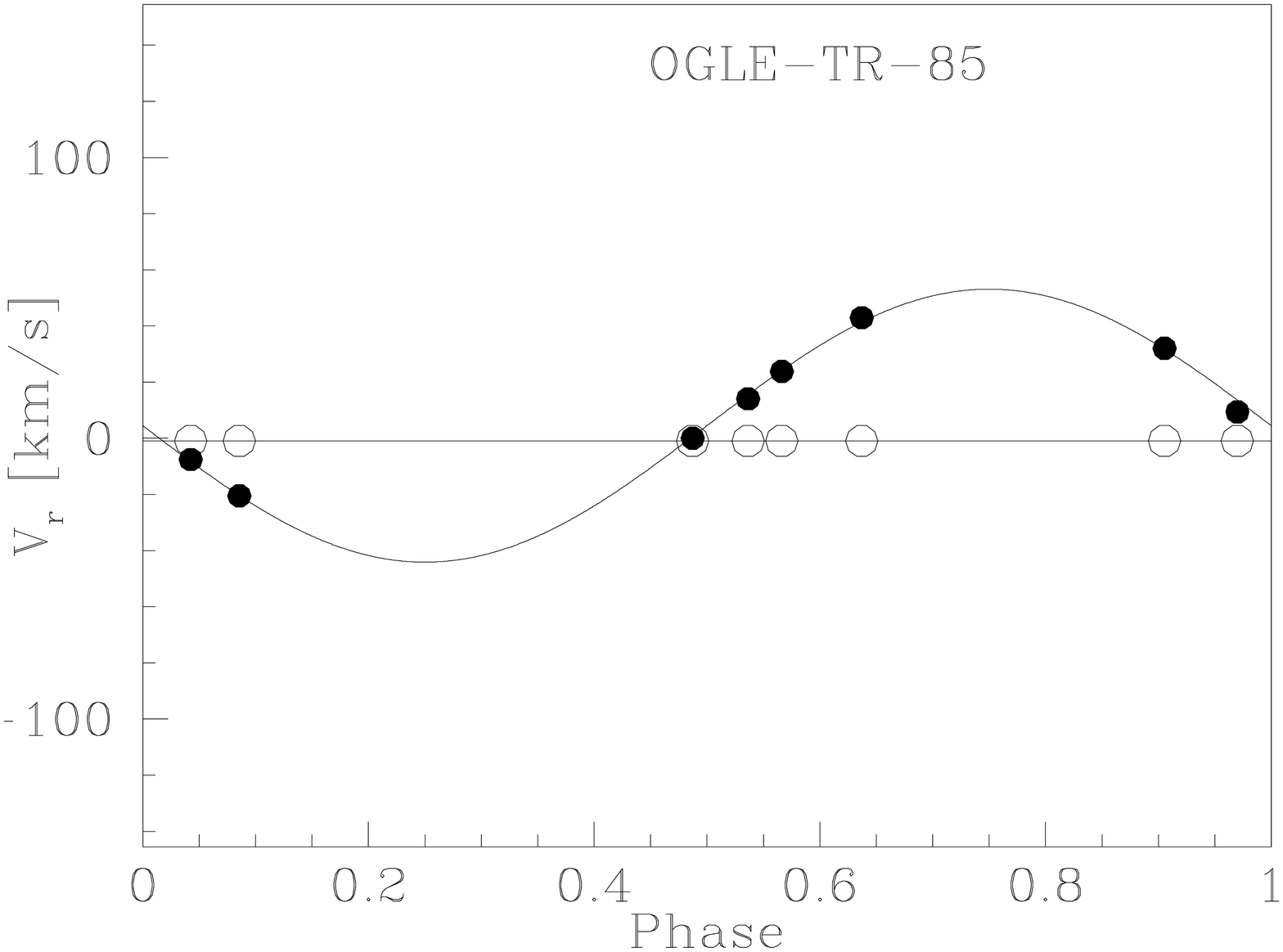}\includegraphics[angle=-90]{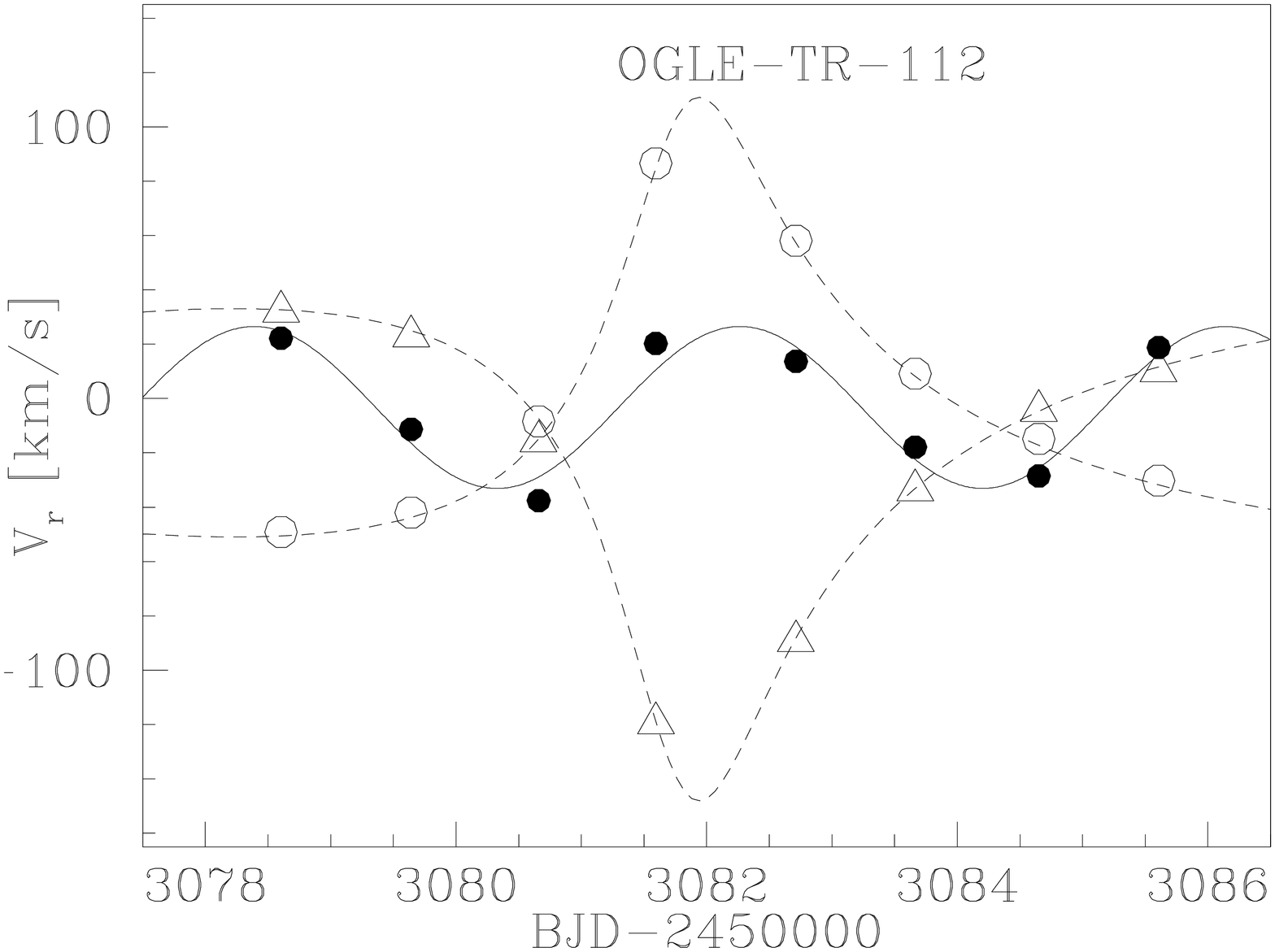}\includegraphics[angle=-90]{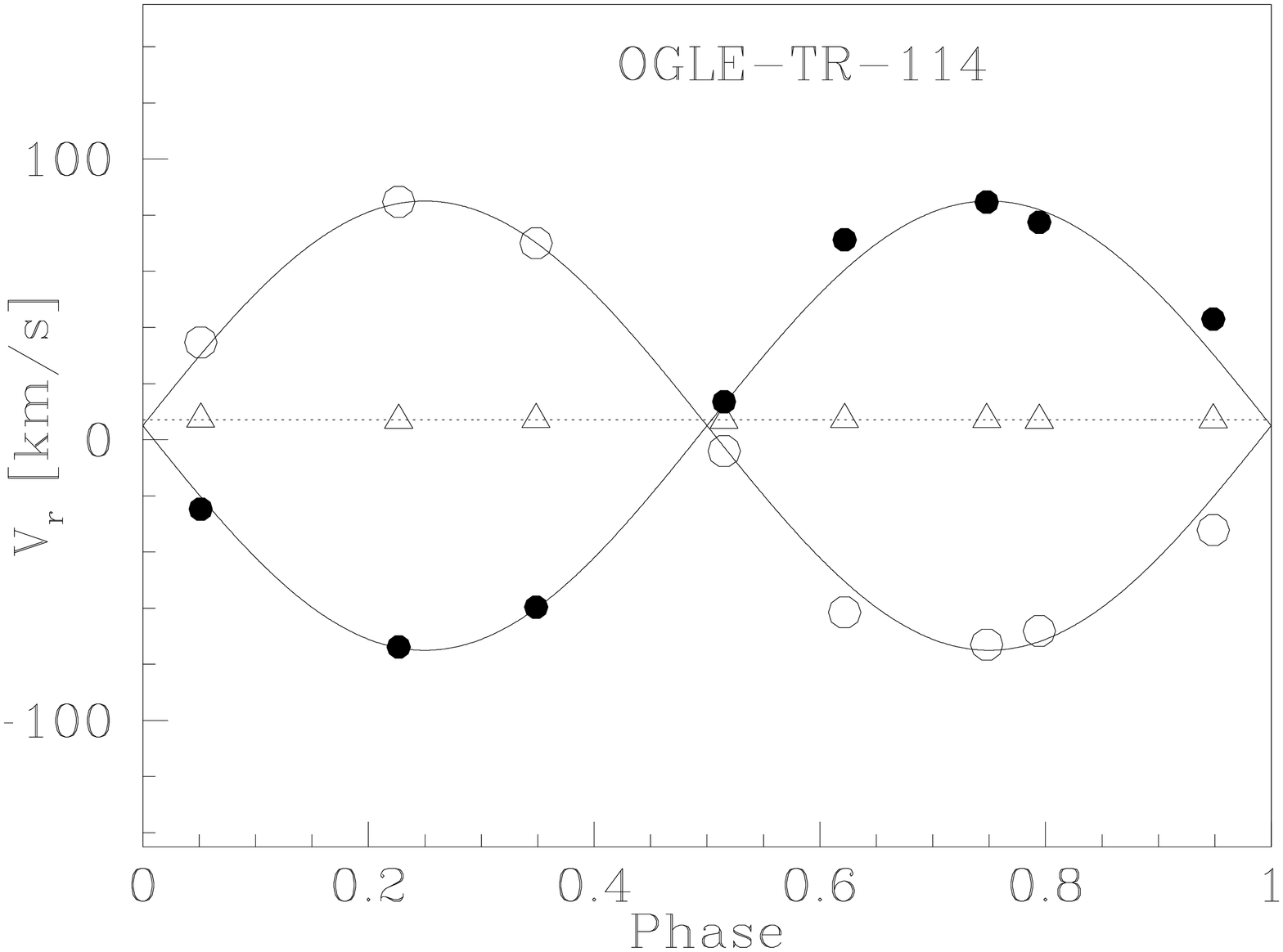}}

\caption{Radial velocity data and resulting orbits for double-lined and triple-lined spectroscopic binaries. The transit epoch is derived from the photometric data. The corresponding parameters are given in Table~\ref{table_sb2}. The different components identified in the spectra are identified with different symbols. In all plots the black dots indicate the object undergoing the eclipse at $T_{tr}${\tiny (OGLE)}. Error bars are comparable to the size of the symbols or smaller. The radial velocity of components witout significant radial velocity changes was fixed to a constant values to allow a better solution for the other components. Note that OGLE-TR-112 is plotted in date instead of phase because of the two different periods. }
\label{sb2}
\end{figure*}

Such systems range across a wide variety of cases: grazing eclipsing binaries, triple and quadruple systems. The resolution of the cases in this section necessitates the full arsenal of spectroscopic binaries analysis, that we will not repeat in detail here. The CCF components are often blended with eachother and their separation requires a global, iterative treatment of all the measurements. When an orbit could be determined, the orbital solution is given in  
Table~\ref{table_sb2} and illustrated in Figure~\ref{sb2}. Estimates of masses and radii are given in Table~
\ref{param}. Other cases are presented in Table~\ref{table_sb2_2}. The radius of probably synchronised component is estimated from the rotation velocity.

\subsubsection{Grazing equal-mass eclipsing binaries} 
\label{ssb2_1}

OGLE-TR-64 shows two dips in the cross-correlation function varying in anti-phase. The mass of both components can be determined precisely from the orbit in the usual way of double-lined spectroscopic binaries. Both radii can be determined from the rotation velocities derived from the line broadening in the CCF. The $\sin i$ factor can be calculated from the a posteriori fit of the light curve with both radii fixed. The period is the double of that given by \citet{Uda02c}, because both the eclipse and anti-eclipse are seen in the light curve. 

OGLE-TR-69  and 110, with only one or two measurements showing two sets of lines with a large radial velocity difference and rotational broadening compatible with orbital synchronisation, are also probable grazing eclipsing binaries. The masses of the components could be estimated from the velocities using the period and epoch of the photometric transits, but the time interval between the two sets of measurements is too large to provide reliable values.

\subsubsection{Triple systems with faint eclipsing companion} 
\label{ssb2_2}

OGLE-TR-76 and 85 are triple systems with two sets of lines in the spectrum, an eclipsing binary blended with a third body.   
In thes cases the CCF shows one set of lines with orbital motion, broadened by synchronous rotation, and the other without radial velocity change. Therefore the body causing the eclipse is an unseen third body in orbit around the first. The second body seen in the CCF is either gravitationally bound to the two others in a triple system, or an unrelated star along the same line-of-sight.  The treatment of these cases is the same as for single-lined spectroscopic binaries, except that the lightcurve also contains an unknown contamination from the third body, so that the radius ratio cannot be determined from the transit signal. OGLE-TR-81, 93 and 95, with one measurement only, show one broad and one narrow component in the CCF and are probably similar systems.

\subsubsection{Triple systems with equal-mass grazing binary}
\label{ssb2_3}

OGLE-TR-65 and 114 show three sets of lines in the CCF. These targets are equal-mass eclipsing binaries in triple systems. OGLE-TR-96, with only one measurement, appears to be a similar system with two synchronised components and a wider component (a fast-rotating F-dwarf). 

\subsubsection{Quadruple system}
\label{ssb2_4}

OGLE-TR-112 is a truly involved case with three dips in the CCF, all varying in radial velocity on short timescales. It is a quadruply system, with three components visible in the spectra. Two of them describe an excentric orbit around each other with a period unrelated to the photometric signal ($P\simeq 10.63$ days). The third has an orbit with the period of the transit signal, revealing that it is eclipsed by a fourth, unseen companion. Therefore OGLE-TR-112 is a system consiting in two close binaries. 

Note that for all triple systems, the radial velocity of the third component is near enough to the systemic velocity of the eclipsing binary for the systems to be gravitationally bound multiple systems rather than line-of-sight contamination.

\begin{table*}[ht!]
\begin{tabular}{l l l l l l l  l }

Name &Comp &$P$  & $T_{tr}${\tiny (OGLE)} & $P_{\rm OGLE}$  & $K$ & $V_0$  & \vsini    \\ 	

     &   & [days]         & [-2452000]      &[days] & [\kms] & [\kms] &  [\kms]\\ \hline \hline

OGLE-TR-64  & a & 5.434691 & 78.569 & (2.71740) & 62.43 $\pm$0.95  &$-$1.17 $\pm$0.75 &12$\pm$1  	 \\			
            & b &         &        &           &96.46 $\pm$1.02  &                  & 9: \\

OGLE-TR-65 &  a &1.720390 & 76.319 & (0.86013) & 114.53$\pm$0.84  &$-$2.18$\pm$0.70  &43.6$\pm$1.9   \\		
           &  b &         &        &    & 119.09$\pm$0.97  &                & 43.9$\pm$1.3 \\
           &  c &        &        &    &                &  $-$6:  &$<$5 \\

OGLE-TR-76 & a &2.12725& 323.545 & (2.12678) &74.1$\pm$2.4 &11.8$\pm$1.8 &29: \\					
           & c  &       &         &    &             &$-$11: & 20:: \\

OGLE-TR-85 & a &2.11481   & 324.440& (2.11460) &48.62$\pm$0.75 &4.48$\pm$0.30  &31$\pm$3   \\						
           & c &         &        &    &             &$-$1:  &  60:  \\

OGLE-TR-112 & a &3.8754  & 327.532 & (3.87900) &29.8 $\pm$2.5  &$-$3.3$\pm$1.8  &40::   \\
           & c  & $\sim$10.6&  - & - & 80.9$\pm$1.2 &  $-$11.32$\pm$ 0.04 & $<$5  \\
           & d  &&    &   & 90.6$\pm$1.2 &  & $<$5  \\

OGLE-TR-114 & a &3.4218  & 323.249 & (1.71213) &80$\pm$3  &5$\pm$2  &10.8$\pm$1.4 \\		
            &   b &    &             &  &80$\pm$3 &           &10.6$\pm$1.3  \\		
           & c    &    &             &    &    &7 (drifting)  & $<$5 \\

\hline

\end{tabular}

\caption{Orbital elements and rotation velocities for double-lined and triple-lined systems. Column 2: system components. By convention the eclipsing components are always "a" and "b".  Columns 3-7 : Parameters of the orbital solution. $P$: revised period, $T_{tr}$: epoch of the transits (fixed), $P_{\rm OGLE}$: original OGLE period, $K$: orbital semi-amplitude, $V_0$: systemic velocity. The excentricity was fixed to zero in all cases except for the second system ("cd") of OGLE-TR-112, where $e=0.55\pm 0.02$.
Column~8: rotation velocity from the cross-correlation function.}
\label{table_sb2}
\label{myst}
\end{table*}

 \begin{table}[ht!]
\begin{tabular}{l l l l l l}

Name & Comp &$N$ & \vsini & $R$ & $P_{\rm OGLE}$ \\
  &   & & [\kms] & $\Rsol$ & [days] \\ \hline \hline

OGLE-TR-69& a & 2 &  15.6   &	1.44  & 2 $\times$ 2.33708 \\							
          & b &   &   11.5 & 	1.06  &	 \\

OGLE-TR-81 & a  & 1  &  21: & 1.4 & 3.21650\\
           & b  & 1  &  $<$ 5 & -  &\\

OGLE-TR-93 & a & 1 & 45: &  2.0	&2.20674\\										
           & b &  & $<$5 & -	 & \\										

OGLE-TR-95 & a &1&  65:: &1.8 & 1.39358\\
           & b &  & 12	&-  & \\

OGLE-TR-96 &a &1 & 11:  &   1.4	& 2 $\times$ 3.20820 \\										
           & b &  & 9:	&1.2 &\\										
           & c &  & 71:	& - & \\										
	
OGLE-TR-97& a &1& 10.7 & -  & 2 $\times$ 0.56765\\
           & b &  & 9.4	& - &\\										
           & c &  & $<$5 & - &\\

OGLE-TR-110 & a &1 &11.7 &	1.32 & 2 $\times$ 2.84857\\											
           & b &  & 11.4	& 1.28 & \\										

\hline
\end{tabular}

\caption{Number of measurements and rotation velocity for double-lined and triple-lined systems without orbital solution. Column~5 indicates the implied radius in case of synchronised rotation. Uncertainties on the \vsini\ are of the order of 10 percent. When indicated, the orbital period is twice the OGLE period because the system contains an equal-mass eclipsing binary.}
\label{table_sb2_2}
\end{table}

\subsubsection{Probable binaries with ambiguity in the configuration}
\label{ssb2_5}

For some double-lined objects with only one spectroscopic measurement, the broadening of some of the spectral components are compatible with synchronous rotation with a massive eclipsing companion, but there is not enough information to determine the exact parameters of the system. Table~\ref{table_sb2_2} gives, for these objects, the rotation velocities indicated by the CCF components, and the corresponding radius if the rotation velocity is compatible with the period of the photometric transit signal, or with double that period for suspected grazing binaries.

For OGLE-TR-69, 81, 93, 96 and 110, the rotation velocities allows a tentative scenario to be proposed.
OGLE-TR-69, 96 and 110 appear to be grazing equal-mass binaries with double the OGLE period  (Sect.~\ref{ssb2_1}).  OGLE-TR-81 and 93 appear to be spectroscopic binaries blended with a non-rotating third body (Sect.~\ref{ssb2_2}).

For OGLE-TR-97, several scenarios are possible and none is clearly favoured. Details are given in the comments on individual objects in Section~\ref{indiv}.  
The CCF shows three clear dips, but none of the calculated rotation velocities are compatible with synchronisation with the (very short) period indicated with the photometry.
More measurements would be needed to determine the nature of this system. However, since the presence of  three sets of lines in its spectrum show that it is a blended system, it loses its interest in the context of planetary transit search, which was the main objective of our study.

\subsection{Transiting planets}

\label{planets}
\begin{figure}[ht!]
\resizebox{7cm}{!}{\includegraphics{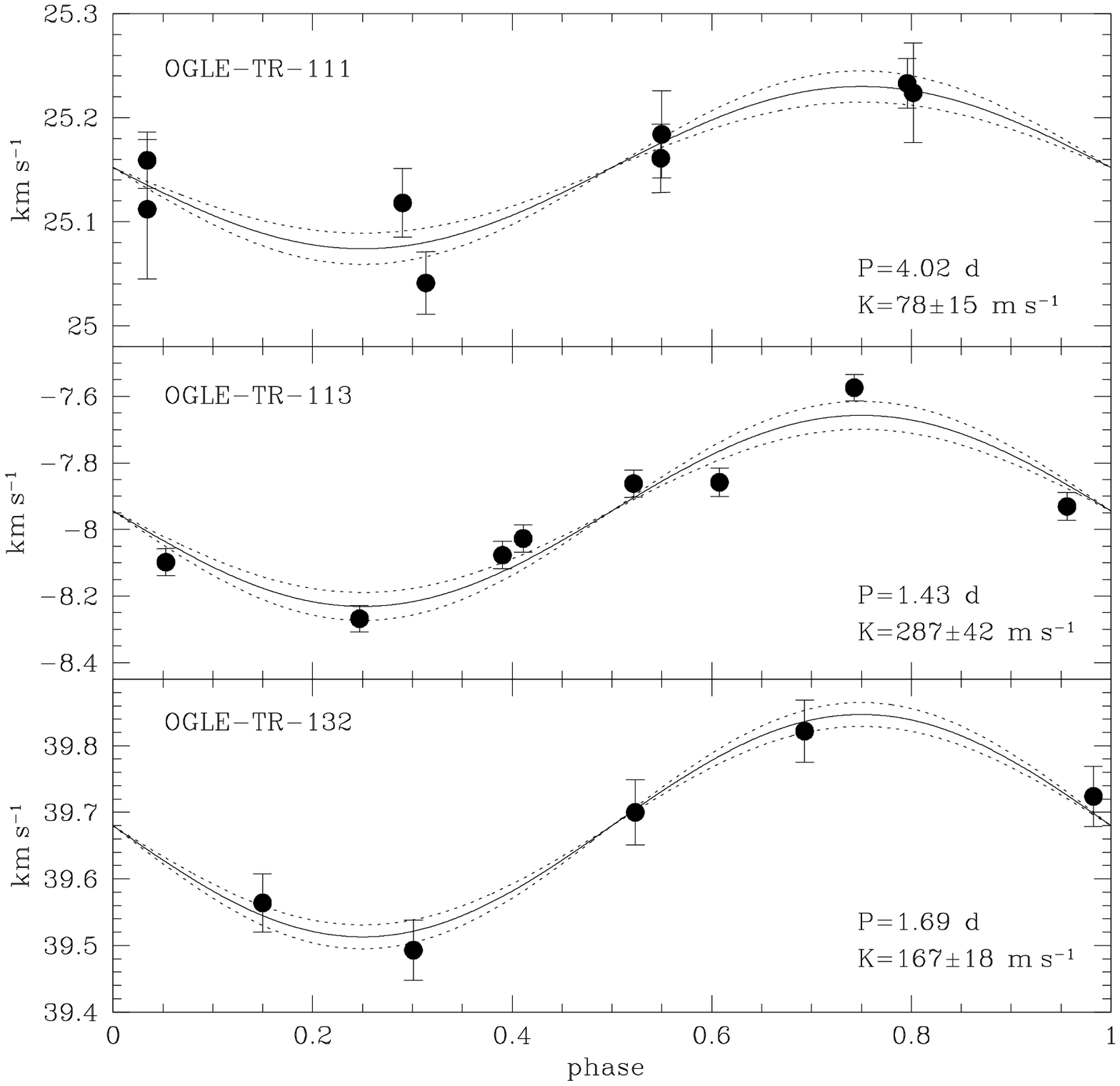}}
\caption{Radial velocity data and orbit for the three detected transiting planets. Adapted from \citet{Pon04, Bou04, Mou04}.} 
\label{threevr}
\end{figure}

Transiting exoplanets were securely detected around three of the objects: OGLE-TR-111, OGLE-TR-113 and OGLE-TR-132. A detailed account has been previously published in \citet{Bou04} and \citet{Pon04}.  The orbits are included in Figure~\ref{threevr} for completeness. 

\subsection{Stars without short-period radial velocity orbit}

\label{snovar}

\begin{figure*}[ht!]
\resizebox{15cm}{!}{\includegraphics[angle=-90]{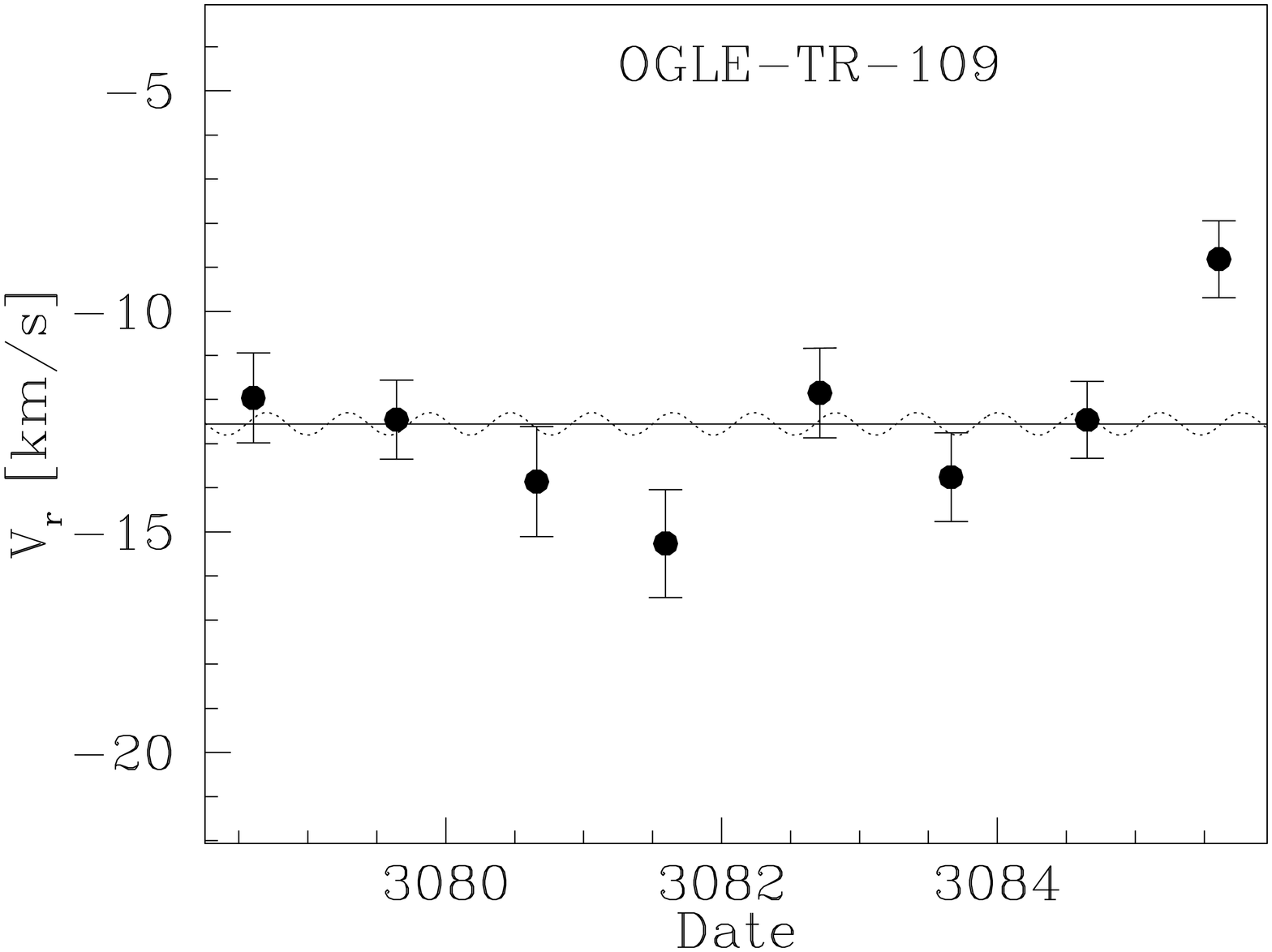}\includegraphics[angle=-90]{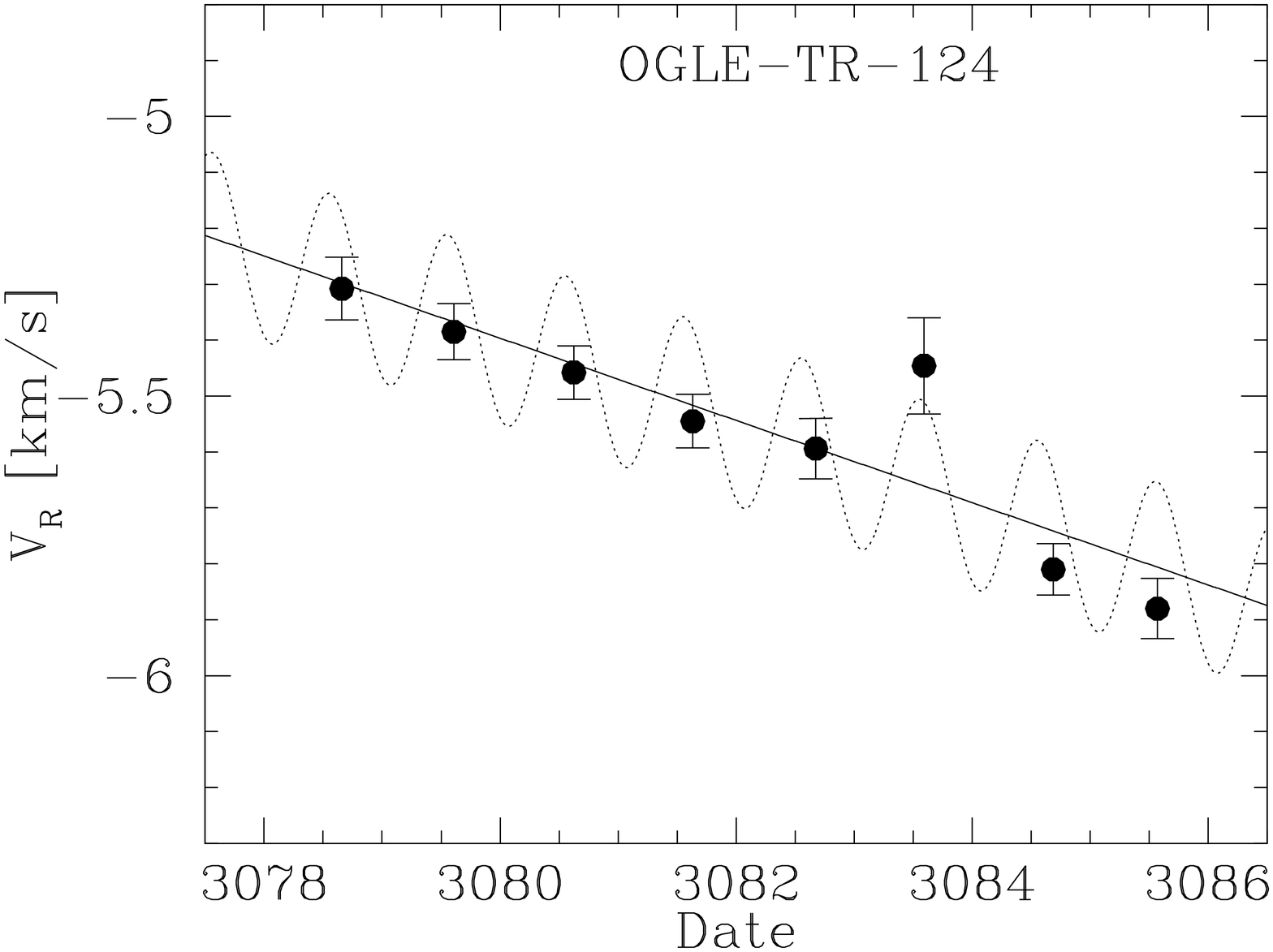}\includegraphics[angle=-90]{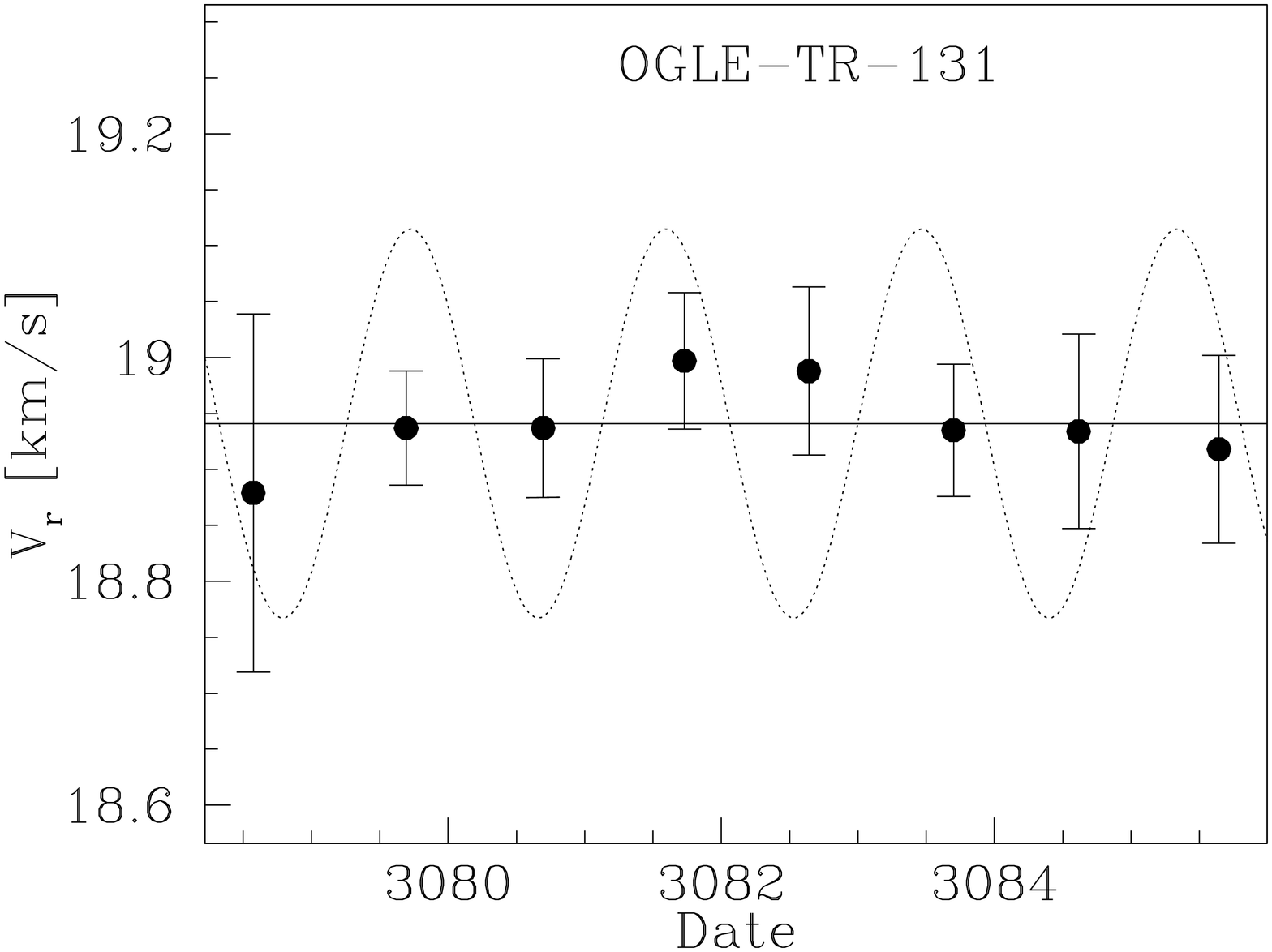}}
\caption{Radial velocity data for the objects without radial velocity variations in phase with the transit signal. The dotted line indicates the orbit corresponding to a transiting 1 M$_{\mathrm J}$ planet. } 
\label{novar}
\end{figure*}

Two objects, measured 8 times each, with very high-accuracy radial velocity data, show no radial velocity variation in phase with the transit signals to the level of less than 50 \ms (see Fig.~\ref{novar}). OGLE-TR-124 shows a constant drift of the radial velocity throughout the observation period, and OGLE-TR-131 shows radial velocity residuals consistent with the noise around a constant velocity. A third object, OGLE-TR-109 shows no significant radial velocity variations either, but with larger uncertainties due to a large rotational broadening. These objects are discussed in some detail here. They are interesting in the context of planet search because the absence of detected short-period radial velocity orbit is compatible with the explanation of the photometric data in terms of planetary transit if the planet mass is smaller than some upper limit. Spectroscopic information for these objects is given in Table~\ref{table_novar}. For the reasons exposed below however, we estimate that these objects are probably false positives of the transit detection procedure.

\ \\

OGLE-TR-109:
The spectral lines of OGLE-TR-109 show a large rotational broadening, indicating \vsini $=35.4 \pm 1.8$ \kms. The radial velocity uncertainties are correspondingly higher, and our 8 measurements show no significant variation within these uncertainties. The radial velocity data is displayed in Fig.~\ref{novar}. Fitting a circular orbit at the period of the transit gives $K=-1.5 \pm 0.9$ \kms. The temperature of the target is $T\sim 7000$ K according to our estimation, $T=7580 \pm 370$ K according to \citet{Gal05}. Synchronisation with the transiting companion would imply $R\simeq 0.4 \Rsol$, which is incompatible with the temperature observed. Even if both eclise and anti-eclipse were seen, the primary radius would only be $R\sim 0.8 \Rsol$, still much too low to match the spectral type. Therefore the target is a non-synchronised, fast-rotating F-star, which leaves three scenario open:

\begin{itemize}
\item{a transiting planet}
\item{a blend with a background eclipsing binary}
\item{a false positive of the transit detection}
\end{itemize}

The third possibility is discussed further at the end of this Section.

\ \\

OGLE-TR-124:
This object exhibits a linear drift in the radial velocity across eight nights of measurements, $\dot{V_r}=-73\pm 15 $m\,s$^{-1}$day$^{-1}$. The measured rotational velocity, \vsini$=6.3\pm$0.1 \kms, is not compatible with tidal synchronisation. That the radial velocity drift is indeed a long-term drift and not part of a short-period orbit was verified with a single later measurement on this objects with the ESO HARPS spectrometer on $JD=2453151.53$ with $V_r=-8.982\pm13$ \kms. The residuals around a linear relation (90 \ms, 29\ms without the discrepant point at JD=53083) put an upper limit of around 0.1 ${\mathrm M}_{\mathrm J}$ on a putative planet. 

\ \\

OGLE-TR-131:
This object shows no velocity variations, with residuals of 60 \ms. Ajusting an orbital solution gives $K$=16$\pm$17 \ms, which implies $m\leq 0.2 {\mathrm M}_{\mathrm J}$ for a possible planet. The shape of the transit is quite remote from a typical central-transit shape, and the lightcurve fit yields $b>0.79$ for the impact parameter.

\ \\

For both OGLE-TR-124 and OGLE-TR-131, the best fit to the light curve is given by a grazing transit with a large impact parameter. Therefore the explanation in terms of transiting planet is much less likely than in terms of blend with a background eclipsing binary contributing a few percent of the light and not seen in the CCF, or in terms of false positive of the transit detection procedure.

We note that all three objects in this section belong to a category where the existence of the transit signal itself is not beyond doubt -- see Section~\ref{freq} and Fig.~\ref{conf}, where we try to estimate the actual detection threshold of this OGLE transit survey in Carina.
The position of OGLE-TR-109, 124 and 131 in Figure~\ref{conf} shows that they stand out below the other confirmed candidates in terms of significance of the transit detection. The position of these objects without detected velocity variations near or below the detectability threshold is unlikely to be a coincidence. Some or all of them may simply be false positive of the transit detection procedure, a suspicion reinforced by their non-square shape for OGLE-TR-124 and OGLE-TR-131. 

The likelihood of OGLE-TR-109 being a false transit detection is difficult to evaluate because OGLE-TR-109 is situated in a magnitude range where the sensitivity limit of the OGLE transit search is not easy to define. The sensitivity limit has two regimes: for $m_I \geq 16$ mag, the photon noise dominates, which makes the errors uncorrelated and produces a detection limit dependent on the $S_d$ parameter. For brighter magnitudes, the systematic drifts in the photometry become the dominant obstacle to the detection. The covariance of the residuals on timescales of a few hours -- comparable to the duration of the transit signal -- make the actual transit detection threshold much higher than in the case of white noise. This effect is clearly reflected in Figure~\ref{conf} by the total absence of low-$S_d$ detections below $m_I\simeq 15$. In this second regime, the detection threshold can be modelled with an ``effective'' $\sigma_{phot}$ that does not reduce below a certain value, even when the nominal individual errors on the photometric data decrease. The limits resulting from this assumption are displayed on Fig.~\ref{conf} for floor values  $\sigma_{phot}=4.5$ and 6 mmag. 

The general behaviour of the transit detection threshold is clear from the position of the OGLE data relative to these two detectability regimes, but the transition from one regime to the other is difficult to locate. Indeed, the status of OGLE-TR-109 is critical to the localisation of this threshold, since it is the only object in the relevant zone. 

A finer analysis of the detectability threshold, which is beyond the scope of this discussion, would be necessary to ascertain whether OGLE-TR-109 is a likely false positive or bona fide transit system. Given that OGLE-TR-109 is in absolute terms the shallowest of all OGLE transit detections ($d\simeq 8$ mmag), we would tend to lean towards the first explanation. In the second case, higher accuracy photometric coverage of the transit would be useful to determine if the duration and shape is coherent with the planet scenario. That would make it the first planet detected around a fast-rotating F-star. Nevertheless, even with better photometry, radial velocity confirmation would be out of reach of present observational means.

Figure~\ref{conf} is discussed further in Section~\ref{concl}, where we reflect on the characteristics of the actual OGLE transit detection threshold.

\begin{table}[h!]
\begin{tabular}{l l r r r r r}

Name & $N$ & \vsini & $<\epsilon_{V_r}>$ & {\small RMS} & $m_{max}$ \\
& & [\kms] & [\kms]& [\kms] & [$\MJ$] \\
 \hline \hline
OGLE-TR-109 & 8 &  35.4 $\pm$ 1.8 & 1.0 & 1.9 & 45\\
OGLE-TR-124 & 8 &  6.3$\pm$0.1 & 0.055 & 0.090 & 0.1 \\
OGLE-TR-131 & 8 & $<5$  & 0.035 & 0.037 & 0.2\\
 \end{tabular}
 \caption{Data for objects with no significant orbital signal in phase with the transits. $N$: number of measurements,  \vsini: projected rotation velocity, $<\epsilon_{V_r}>$ mean radial velocity uncertainty,  {\small RMS}: dispersion of radial velocity residuals,  $m_{max}$: upper limit on the mass of a transiting companion. In the case of OGLE-TR-124, the secular drift has been substracted.}
 \label{table_novar}
 \end{table}

\subsection{Objects with no clear signal in the spectral CCF}

\label{snoccf}
Some objects showed no unambiguous signal in the CCF, namely OGLE-TR-68, 82, 84, 89, 107, 118 and 127. The absence of signal can be due to several causes:
(i) the target is an early-type star (earlier than about F2) and fast-rotating, with few narrow metallic lines.
(ii) synchronous rotation broadens the lines below the detection threshold
(iii) the target is a multiple-line spectroscopic binary, with the flux from each components damping the depth of the CCF of the other component.
(iv) the signal-to-noise ratio of the spetrum was too low even to detect a narrow CCF dip.

\begin{figure}
\resizebox{8cm}{!}{\includegraphics{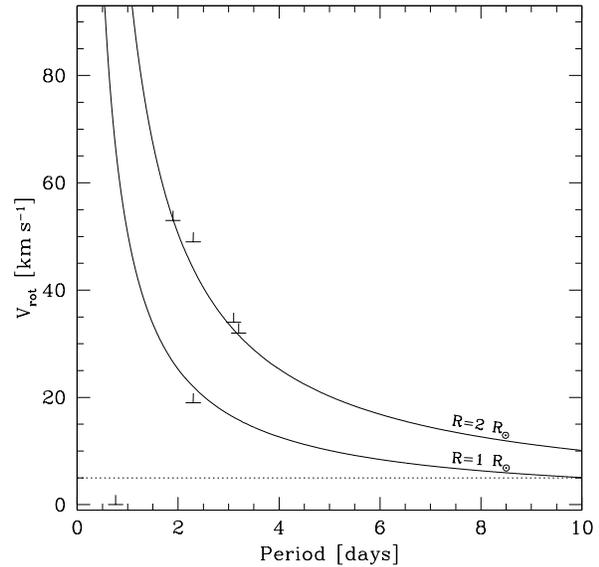}}
\caption{Lowe limit for the rotation velocity as a function of period for objects without detected signal in the CCF, if they are single stars of spectral type later than F2. Lines as in Fig.~\ref{rotvit}} 
\label{vrotmin}
\end{figure}

Only in case (iv) does the target remain a plausible planetary transit candidate. In case (i), the primary has a large radius, and therefore the eclipse depth implies a secondary radius larger than expected for planets. In case (ii),  the same arguments as in Section~\ref{select} indicate that the eclipsing companion is likely to be of stellar mass. In case (iii), the flux of the other components dilutes the transit, so that the minimum radius of the transiting body also becomes larger than expected for a planet. 

For each of the objects without a detected CCF signal, we calculated the lower limit of the rotation velocity compatible with the absence of detection in the case of a late-type primary.  We assume that the detection threshold of a CCF signal is five times the rms of the CCF continuum, and use a lower limit of 1.8 \kms for the surface of the CCF signal, suitable for stars later than F2 and not too metal-poor.
Table~\ref{vmin} gives the resulting rotation velocity lower limit for objects in this Section. Fig.~\ref{vrotmin} displays the results as a function of period, to be compared with Fig.~\ref{rotvit}. This calculation shows that all targets except OGLE-TR-82 cannot be isolated, slowly rotating late-type stars, and therefore the transiting object is most probably not a planet. All but OGLE-TR-127 have minimum rotational velocites compatible with an explanation in terms of single eclipsing binaires, albeit with $R\geq 2 \Rsol$. OGLE-TR-127 cannot be a single synchronised binary with a late-type primary, and may be either a multiple-lined binary or a hot star. Its spectrum shows strong $H_\beta$ compared to $H_\alpha$ which may indicate an A-type star. For OGLE-TR-82, our spectrum does not have high enough signal to detect any CCF (SNR$_{tot}\simeq 2.0$, SNR$_{sky}\simeq 1.7$).

The position of these objects in the detectability plot (Fig.~\ref{conf}) show that they quite neatly fall into two categories:

Three of the objects, OGLE-TR-89, 118 and 127, are also markedly lower than confirmed eclipse/transit cases in terms of reliability of the photometric transit detection.  Therefore they may well be false positives, and in that case the absence of CCF signal would indicate hot stars (stars earlier than about F2 do not produce a detectable CCF signal). They would be the early-type complement of the false detection group constituted by OGLE-TR-124 and 131. Indeed, very hot temperatures ($T>8000$ K) were found for OGLE-TR-89 and OGLE-TR-118 by \citet{Gal05}  from infrared photometry.

The other group, OGLE-TR-68, 82, 84 and 107, are objects situated above the detectability threshold, but near the faint end of the sample in terms of magnitudes. They are probably bona fide eclipsing binaries with large or early-type primaries, with synchronised rotation braodening the CCF below the detectability limit, with the exception of OGLE-TR-82. The spectrum of OGLE-TR-82 shows very little signal, 16 times less than OGLE-TR-84,  which was measured on the same exposure and is 0.4 magnitudes fainter in $I$. This may be due to an incorrect centering of the FLAMES fiber. If the fiber was correctly centered on the object, the weakness of the signal (the spectrum flux is centered in the $V$ filter) implies that OGLE-TR-82 is at least 3 magnitudes redder in $V-I$ than OGLE-TR-84. The best solution for the light curve of OGLE-TR-82 is of a $\sim 0.1 \Rsol$ body transiting across a $\sim\! 0.6 \Rsol$ star.  In this scenario the primary would be a K7/M0 dwarf. Such objects are extremely rare in a magnitude-selected sample, but the fact that the spectrum favors a very red object is intriguing. OGLE-TR-82 is therefore an interesting object that would deserve further observations. It may be a unique example of very small star or planet transiting in front of a late K dwarf on a very close orbit ($P=0.76$ days). Multi-colour photometry or low-resolution spectroscopy would indicate if the primary is indeed such a late-type object.

\begin{table}[h!]
\begin{tabular}{l l r r r }

Name & $N$ &  RMS$_{CCF}$& SNR & $v^{min}_{\sin i}$\\
& & [\%] &  & [\kms]\\  \hline \hline
OGLE-TR-68	& 1	& 0.9	& 3	& 19 \\
OGLE-TR-82	& 1	& 1.2	& $<1$	& - \\
OGLE-TR-84	& 1	& 0.6	& 4	& 34 \\
OGLE-TR-89	& 1	& 0.5	& 7	& 49 \\
OGLE-TR-107	& 1	& 0.6	& 4	& 32 \\
OGLE-TR-118	& 3	& 0.4	& 6	& 53 \\
OGLE-TR-127	& 1	& 0.2	& 15	& 119 \\

 \end{tabular}
 \caption{Data for objects with no detected signal in the CCF. $N$: number of measurements, $\sigma_{CCF}$: rms of the CCF continuum, SNR: signal-to-noise ratio of the spectrum, $v^{min}_{\sin i}$: minimum projected rotation velocity compatible with the absence of CCF signal for a single, late-type star.}
 \label{vmin}
 \end{table}

\section{Summary and individual notes}

Table~\ref{tsummary} summarizes the conclusions of the spectroscopic follow-up for our 42 targets in terms of mass and radius of the eclipsed and eclipsing body, and of the 
nature of the systems. Specific notes on some systems are included here.

\begin{table*}
\begin{tabular}{l l l l l l l l l }
Name & $R$ & $r$  & $M$ & $m$ & $i$ & Nature & Section\\ \hline \hline
Name &[$\Rsol$] & [$\Rsol$]&[$\Msol$]& [$\Msol$]& [$^0$] & & \\ \hline \hline

OGLE-TR-63      & (1.6)& (0.15) & & & & M eclipsing  binary? &[\ref{ssb1_3}]\\
OGLE-TR-64 &	      1.29$\pm$0.11& 0.97: &1.37$\pm$0.02& 0.87 $\pm$0.01 &83 & grazing eclipsing binary &[\ref{ssb2_1}]\\
OGLE-TR-65 &   1.58$\pm$0.07&1.59$\pm$0.05 &1.15$\pm$0.03&1.11$\pm$0.03 &70 & triple system & [\ref{ssb2_3}]\\
OGLE-TR-68	& & & & & & false transit detection?	&[\ref{snoccf}]\\
OGLE-TR-69 &  1.44 $\pm$ 0.10 & 1.06 $\pm$ 0.20 &   &   & 82 & grazing eclipsing binary &[\ref{ssb2_5}]\\
OGLE-TR-72  	    &1.49$\pm$ 0.12	    &0.31:   &             &0.26:   &86-87     & M eclipsing binary   &[\ref{ssb1_1}] \\
OGLE-TR-76 &   1.2$\pm$0.2 & & & 0.64: &$>$77 & triple system  &[\ref{ssb2_2}]\\
OGLE-TR-78  	    &1.84 $\pm$0.08&0.296 $\pm$0.017 &             &0.32$\pm$0.02 &$>$85 & M eclipsing binary &[\ref{ssb1_1}]\\
OGLE-TR-81    & (1.4)& & & & & triple system? &[\ref{ssb2_5}]\\
OGLE-TR-82	& & & & & & unsolved	&[\ref{snoccf}]	\\
OGLE-TR-84	& & & & & & eclipsing binary?	&[\ref{snoccf}]	\\
OGLE-TR-85 &   1.30$\pm$0.13  & &  & 0.39: &$>$77 & triple system &[\ref{ssb2_2}]\\
OGLE-TR-89	& & & & & & false transit detection?	&[\ref{snoccf}]	\\

OGLE-TR-93 &   (2.0):& &   & &  & triple system? &[\ref{ssb2_5}]\\
OGLE-TR-94       &  (1.7)& ($>$0.2) & & & & M eclipsing binary? &[\ref{ssb1_3}]\\
OGLE-TR-95    & (1.8)& & & & & multiple system &[\ref{ssb2_2}]\\
OGLE-TR-96 &   (1.4)&(1.2) &  &   &84 & grazing eclipsing binary &[\ref{ssb2_5}]\\
OGLE-TR-97    & & & & & & multiple system &[\ref{ssb2_5}]\\
OGLE-TR-98       & (2.2)& (0.34)& & & & M eclipsing binary? &[\ref{ssb1_3}]\\
OGLE-TR-99       & (1.1)& (0.19)& & & & M eclipsing binary? &[\ref{ssb1_3}]\\

OGLE-TR-105 	    &2.06$\pm$0.30 &  &             &1.05: &82 & G eclipsing binary &[\ref{ssb1_1}]\\
OGLE-TR-106 	    &1.31$\pm$0.09 &0.181$\pm$0.013 &             &0.116$\pm$0.021 &$>$85 & M eclipsing binary &[\ref{ssb1_1}]\\
OGLE-TR-107	& & & & & & eclipsing binary ?	&[\ref{snoccf}]	\\
OGLE-TR-109	& & & & & & unsolved &[\ref{snovar}]\\
OGLE-TR-110&   (1.32)&(1.28) &  & &84 & grazing eclipsing binary &[\ref{ssb2_5}]\\
OGLE-TR-111     & $0.88^{+0.10}_{-0.03}$  & $0.103^{+0.013}_{-0.06}$& $0.82^{+0.15}_{-0.02}$&  0.00050$\pm$0.00010& & planet &[\ref{planets}]\\
OGLE-TR-112 &  3.5::&  & &0.4:: & $>$70  & quadruple system &[\ref{ssb2_4}]\\
OGLE-TR-113       & 0.765 $\pm$ 0.025&  0.111 $\pm$ 0.006&  0.77 $\pm$ 0.06 & 0.00130 $\pm$ 0.00020 & & planet &[\ref{planets}]\\
OGLE-TR-114 &  0.73$\pm$0.09&0.72$\pm$0.09 &0.82$\pm$0.08&0.82$\pm$0.08 &83 & triple system  &[\ref{ssb2_3}]\\ 	
OGLE-TR-118	& & & & & & false transit detection?		&[\ref{snoccf}]\\
OGLE-TR-120 	    &1.75$\pm$0.16 &0.45$\pm$0.05 &             &0.50$\pm$0.06 &$>$88 & M eclipsing binary &[\ref{ssb1_1}]\\
OGLE-TR-121 	    &1.33$\pm$0.15 &0.33:$\pm$ 0.05 &             &0.35: &85 (fixed) & M eclipsing binary &[\ref{ssb1_1}]\\

OGLE-TR-122 	    &1.00$\pm$0.05 &0.114$\pm$0.009 &             &0.087$\pm$0.008 &$>$88 & M eclipsing binary &[\ref{ssb1_1}]\\
OGLE-TR-123	& & & & & & M eclipsing binary? &[\ref{ssb1_2}]\\
OGLE-TR-124	& & & & & & false detection? &[\ref{snovar}]\\
OGLE-TR-125 	    &1.94$\pm$0.18 &0.211$\pm$0.027 &             &0.209$\pm$0.033 &$>$86 &M eclipsing binary  &[\ref{ssb1_1}]\\
OGLE-TR-126      & (1.8)& (0.25)& & & & M eclipsing binary? &[\ref{ssb1_3}]\\
OGLE-TR-127	& & & & & & false detection?	&[\ref{snoccf}]	\\
OGLE-TR-129	& & & & & & M eclipsing binary? &[\ref{ssb1_2}]\\
OGLE-TR-130 	    &1.13$\pm$0.04 &0.25: &             &0.39: &85 (fixed)  &M eclipsing binary &[\ref{ssb1_1}] \\
OGLE-TR-131	& & & & & & false detection? &[\ref{snovar}]\\     
OGLE-TR-132       & 1.43 $\pm$ 0.10 & 0.116 $\pm$ 0.008  &  1.35 $\pm$ 0.06&  0.00113 $\pm$ 0.00012& & planet &[\ref{planets}]\\

\end{tabular}   

\caption{Results of the lightcurve+spectroscopy solution for all objects in our sample. $R,r,M,m$: radii and masses of the primary and secondary. $i$: orbital inclination. The last column summarizes the nature of the system, with a number referring to the Section where each case is treated. Question marks and brackets denote cases where the resolution is only tentative,  columns (":") the values with high uncertainties.}
\label{param}
\label{tsummary}
\end{table*}

\label{indiv}

\begin{description}

\item[OGLE-TR-64]

 In the first of the three measurements, the two sets of lines are blended and were deblended using information from the other two measurements. The period is double that of the OGLE estimate. With the new period, the even transits are clearly shallower than the odd transits (0.030 and 0.018 mag respectively), reflecting the luminosity difference between the two components.



\item[OGLE-TR-85]

The solution described in Section~\ref{ssb2} in terms of a triple system was obtained from the heavily mixed double-lined CCF by iteratively fixing the parameters of the different components, using the fact that the rotational broadenings are constant from one exposure to the next, and that the radial velocity of the second component does not change. The spectra consist of a broad component superimposed on an even broader component. We assumed that the wide-lined system had a constant radial velocity, which yields a much better orbital solution for the other components than allowing it to vary. The parameters of the second component somewhat depend on the choice of fixed parameters for the first, so that the real uncertainties are higher than indicated in Table~\ref{table_sb2}.

\item[OGLE-TR-89]

The CCF shows the possible presence of a very wide component, but the signal-to-noise ratio is not high enough to measure it with confidence.

\item[OGLE-TR-93]

Fitting the photometric transit curve with this constraint yields $r\sim 0.34 \Msol$ for the unseen eclipsing companion.  The light curve of this object shows a sinusoidal modulation with half the period of the transit \citep{Sir03}, with an amplitude compatible with the presence of an eclisping companion with $m\sim 0.34 \Msol$. While this tentative scenario explains all available data, other configurations cannot be eliminated entirely.



\item[OGLE-TR-99]

For this object, there is a slight indication of a second dip in the CCF, but the depth and width of this possible dip are not compatible with a simple explanation of the transit in terms of an eclipsing binary. We therefore prefer the explanation in terms of a single dip, that gives a coherent solution.





\item[OGLE-TR-105]

This object is an eclipsing binary constituted of a primary about twice larger than the secondary.  The temperature difference between the two, however, is close enough for both the eclipse and anti-eclipse to be visible in the lightcurve. Consequently the orbital period is double that given by the OGLE survey. The eclipse depth ratio and the derived radii are compatible with a solar-type G dwarf eclipsing an evolved late-F star, with both stars on the same isochrone.

\item[OGLE-112]

This  case  could serve as a textbook exercice for spectral CCF analysis. Two sets of narrow wide lines are superimposed on the very wide and shallow lines of the star undergoing the transit (see Fig.~\ref{ccf}), and must be resolved before its radial velocity can be determined even approximately. The flux ratios and the difference between the two systemic velocities are compatible with a gravitationally bound quadruple system consisting of two close binary systems, one of them eclipsing. 

\item[OGLE-TR-114]

The velocity of the third component of this triple system shows a slight drift over the measurement period ($\dot{V_r}=43\pm$12 m/s), which may indicate that it is gravitationally bound to the other two on a wider orbit. The surface of the three CCF dips are similar, which would be compatible with such a case. The orbit of OGLE-TR-114ab shows large velocity residuals around a circular orbit, that do not correspond to a Keplerian excentric orbit. 

\item[OGLE-TR-118]

The CCF in the two measured spectra is compatible with a very broad dip, with \vsini$\sim$40 \kms. This would imply $R\sim 1.4$ $\Rsol$ assuming tidal synchronisation, but the radial velocity change would then be of the order of 5 \kms, which would be too small for an orbital solution with a massive companion. This is the faintest star in our sample, and the signal-to-noise ratio of the spectra is not high enough to have confidence in the presence of a signal in the CCF.

\end{description}

\begin{figure*}[ht]
\resizebox{15cm}{!}{\includegraphics{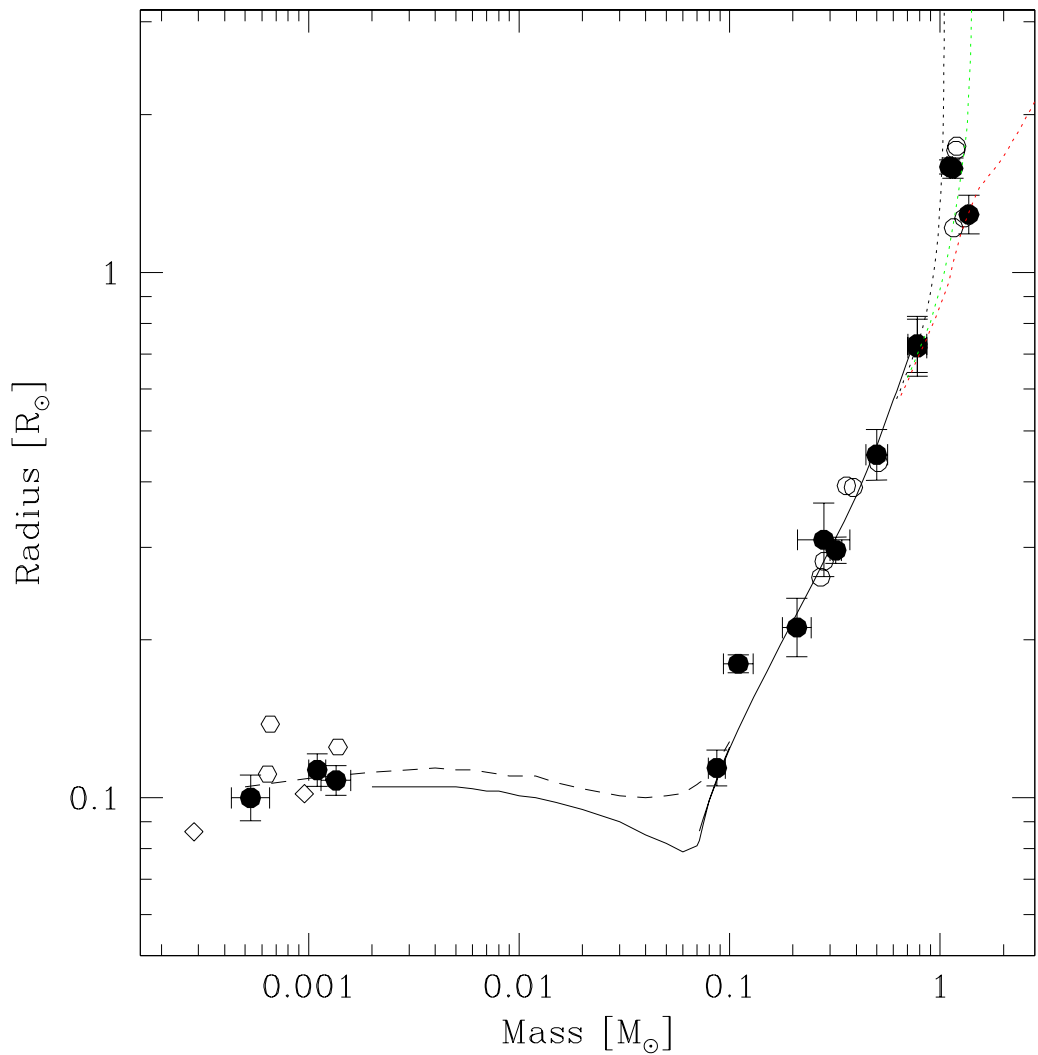}}
\caption{Mass-radius relation for low-mass stars and planets. Black dots show the objects in this study with well-determined mass and radius. Diamonds show Jupiter and Saturn, open circles the results of the OGLE bulge fields from Paper~I, the hexagons are the three other transiting planets HD~209458 \citep{Bro00}, OGLE-TR-56 \citep{Kon03} and TrES-1 \citep{Alo04}.
The lines show the models of \citet{Gir02} for Solar-type stars, for ages 0, 3 and 10 Gyr, and of \citet{Bar98} and \citet{Cha00b} for low-mass stars, brown dwarfs and planets for ages 0.5 and 5 Gyr.
 }
\label{mrr}
\end{figure*}

\section{ Discussion}

\label{concl}

\subsection{The mass-radius relation for planets and low-mass stars}

The detection and characterisation of transiting extrasolar planets is the main result of this study, but mass and radius measurements for very small stars is an important by-product. Stellar structure models predict a nearly proportional relation between mass and radius down to the stellar/brown dwarf transition, then a constant or slightly increasing radius throughout the brown dwarf domain down to the planet domain, with an age-dependent connection between the two domains \citep{Cha00b}. Few radii are known yet for stars lighter than 0.5 $\Msol$, some from eclipsing binaries (CM Dra, \citet{Met96}; CU Cnc, \citet{Rib03}), some from interferometric radius measurements \citep{Lan01, Seg03}, and some from the previous OGLE transiting candidates (Paper~I). 

Figure~\ref{mrr} shows the results of the present study in the mass-radius plot. It adds 6 new low-mass stars to the list of objects with accurate mass and radius determinations, together with three new planets. Moreover, two of our objects skirt the lower mass limit for stellar objects ($\sim 0.08 \Msol$) and their radius measurements are especially interesting because they explore the mass domain where the models predict a change of regime between solar-like behaviour and a degenerate equation of state.

\begin{figure}[ht!]
\resizebox{8cm}{!}{\includegraphics{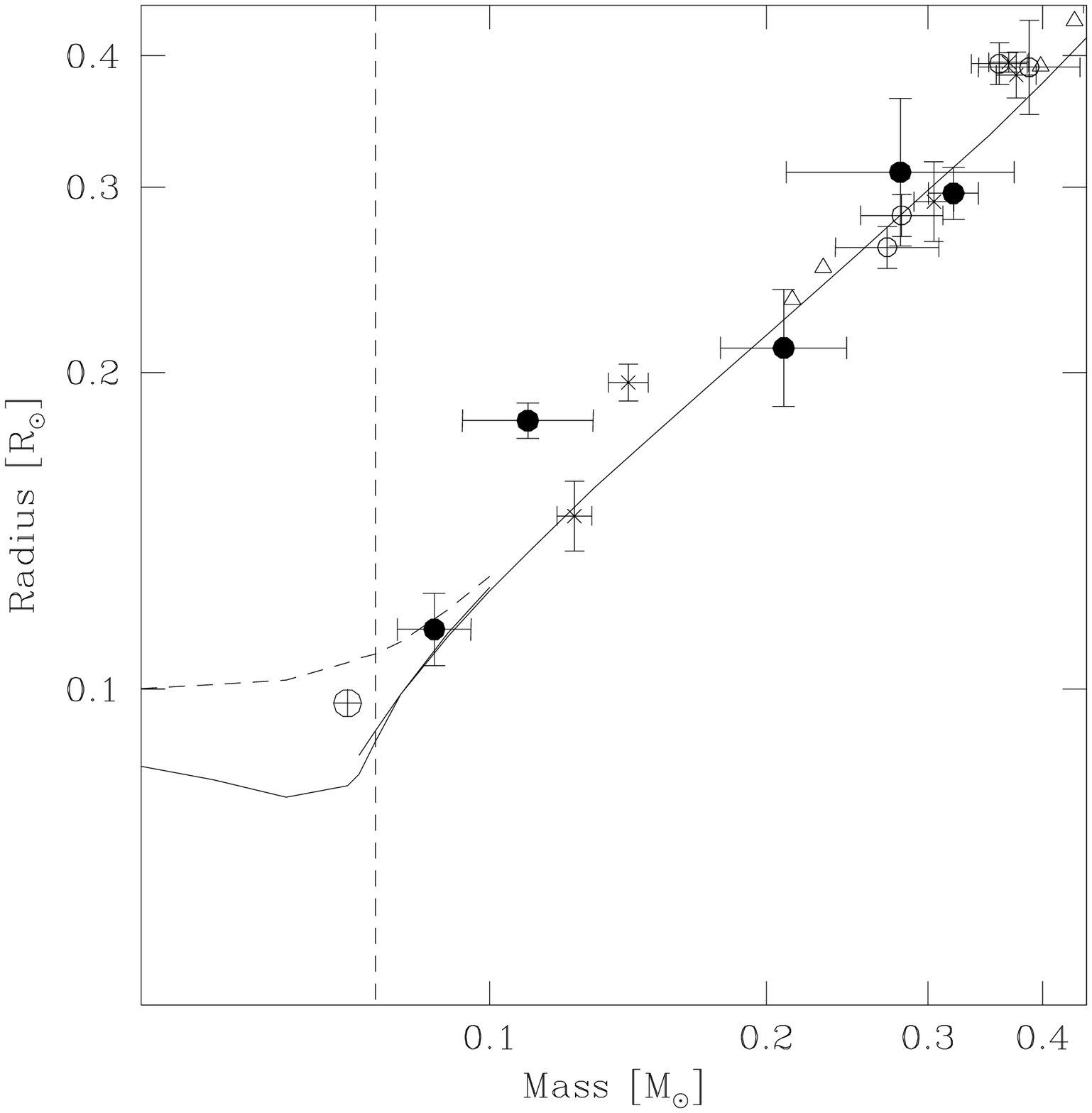}}
\caption{Mass-radius relation for M-dwarfs. Black dots show the objects in this study with well-determined mass and radius, open circles the results of the OGLE bulge fields from Paper~I. Triangles are data from eclipsing binaries \citep{Met96, Rib03} and crosses from interferometry \citep{Lan01, Seg03}. 
The lines show the models of \citet{Bar98} and \citet{Cha00b} for low-mass stars and brown dwarfs for ages 0.5 and 5 Gyr. The vertical line shows the brown dwarf limit. The crossed circle is the position of our tentative solution for OGLE-TR-123.}
\label{mrr2}
\end{figure}

Fig.~\ref{mrr2} zooms on the stellar region below 0.5 $\Msol$, and compares the OGLE data from this paper and Paper~I to other known stellar radii. The OGLE transit survey is seen to provide comparable constraints on the lower stellar mass-radius relation as all other studies combined, a very remarkable by-product of the planet search. Our mass-radius data for stellar objects confirm the general adequacy of the models. In fact the agreement of the results from independent methods is excellent and indicates that at the level of a few percent there is no ``second parameter'' in the mass-radius relation for M dwarfs down to the stellar limit, as predicted by the models. OGLE-TR-106 is about 2-sigma away from the theoretical relation and may indicate a departure from the models near $M=0.2 \Msol$, than needs to be confirmed with further data. OGLE-TR-122 is situated near the expected transition to a completely degenerate dependence of mass on radius. The uncertainty on its radius is still too high to constrain the models, but it does show that stars can reach smaller radii than planets, as predicted.


Fig.~\ref{mrr2} also plots the position of OGLE-TR-123 with the tentative solution derived from our three measurements. These would place it on the expected locus of young objects ($\tau<1$ Gyr), near or below the Hydrogen-burning limit and potentially within the brown-dwarf domain. This tantalizing result underlines the interest of collecting further spectroscopic data on this object, as well as a finer measurement of its photometric transit.

It is worth commenting on the absence of detection of transiting companions in four zones of the mass-radius plane:

\begin{enumerate}
\item{the brown-dwarf domain (0.01-0.07 $\Msol$)}
\item{heavy planets (2-10 $\MJ$)}
\item{high-radius hot Jupiters ($R\sim 1.4 \RJ$ like HD209458) }
\item{high radius lighter planets ($M<<M_J$, $R\sim 1 R_{\rm J}$)}
\end{enumerate}

The absence of detection in the brown-dwarf domain (pending the resolution of the case of OGLE-TR-123) is an expected consequence of the paucity of brown dwarfs compared to planets and H-burning stars \citep{Hal00}. However, theoretical radii for brown drawfs \citep[Fig.~\ref{mrr2}]{Cha00b} suggest that except in the first Gyr of their lifetime, they are expected to be significantly smaller than the detection threshold of the OGLE survey near 1~$\RJ$ (the smallest transiting objects detected all have $R\sim 1 \RJ$). Therefore, the lack of detection in the OGLE survey only indicates that brown dwarfs are either rare or smaller than 1 $\RJ$ (or both). 

The absence of heavy (2-10 $\MJ$) gas giants is more significant, since these are expected to have radii comparable or larger than Hot Jupiters. Their absence is in agreement with indications from Doppler surveys that heavy gas giants are not found on tight orbits \citep{Zuc02, Udr02, Pra02} except in binary systems \citep{Egg04}.

The absence of $R>1.3~\RJ$ Hot Jupiters like HD209458 in the OGLE Carina survey is a strong constraint, since such objects would have been much easier to detect in a transit survey than the three detected planets (see Section~\ref{freq} below). This indicates that cases such as HD209458 are relatively rare.

Finally, the absence of credible candidate for light transiting planets in our sample (OGLE-TR-109, 124 and 131 could in principle harbour such objects, but as exposed in Section~\ref{snovar} an explantion in terms of false transit positive is found to be more likely) indicates that ``hot Saturns'' generally have smaller radii, or are less numerous than hot Jupiters. Some models of pure H-He close-in gas giants \citep{Gui05} show that light close-in planets could be inflated to 1 $\RJ$ or higher by the incident flux of the star. Catastrophic evaporation scenarios like \citep{Bar04} also predict a light, expanded but very transient state. The non-detection of $R \sim 1 \RJ$ light planets in transit surveys needs to be confirmed with higher statistics but could indicate that hot Saturns have smaller radii than more massive gas giants, possibly indicating the presence of heavier material in their composition \citep{Gui05}.

There are several objects for which we obtained only one measurement indicating a probable small eclipsing companion. Further radial velocity monitoring of these objects would be useful in the context of the low-mass stellar mass-radius relation. These are OGLE-TR-98, 99, 126 (and to a lesser degree OGLE-TR-94 with a probable grazing eclipse). As mentioned above, OGLE-TR-123 is especially interesting as a potential transiting brown dwarf.

\subsection{Orbital circularisation timescale}

Binaries with periods smaller than about 10 days are observed to be circularised by tidal interaction, except when a third body can pump excentricity into the orbit \citep{Udr00}. In several cases our radial velocity measurements are numerous and precise enough compared to the orbital amplitude for a precise determination of the orbital excentricity, the value of the period being fixed by the transit signal. 
Simple models predict a $1/(1+q)q$ dependence of the orbital circularisation timescale \citep{Zah92}, where $q$ is the mass ratio, and therefore  the circularisation  is expected to take longer for lighter companions. This is indeed what we observed.  From this data we conclude that the typical circularisation period for small M-dwarf reach down to 5 days, i.e. as low as for hot Jupiters.

\label{ecc}




\subsection{Frequency of system types}

If we adopt for each system the resolution considered most likely in Section~\ref{results}, the bottom line of our follow-up is the following breakdown by type of system:

\begin{tabular}{l l}

Transit by a small stellar companion & 16 \\

Grazing binary & 3 \\

Multiple systems & 10\\

Transiting planets & 3 \\

False positives & 6 \\

Unspecified binary & 2 \\

Unsolved & 2 \\

\end{tabular}

Among the binary systems therefore, about half are transiting small stellar companions, a tenth are grazing binaries and the rest are triple or blended systems. The objects in this last category can be gravitationally bound systems or line-of-sight alignements. The observed velocity differences between the close binary and the third body are compatible with both scenarios.

We compared these results to the prediction of a Monte Carlo  simulation based on \citet{Bro03}. The simulation predicts, for a transit signal depth of 0.02, 40\% small eclipsing companions, 35\% grazing eclipses, 18\% physical triples and 7\% line-of-sight blends, in good agreement with the observed proportions. The only difference is the lower number of grazing binaries in the observed sample, which probably reflects the fact that many such cases were rejected by eye by \citet{Uda02b, Uda03} because of obviously V-shaped transit.

\subsection{Detection threshold and implications for planet properties}

\label{freq}

All three plantets discovered in the OGLE Carina survey are peculiar in some way compared to the Hot Jupiters found by radial velocity searches. OGLE-TR-113 and OGLE-TR-132 are "very hot Jupiters" with periods much shorter than any planet from Doppler surveys, and OGLE-TR-111 has a period of almost exactly 4 days. Here we try to understand the properties of the OGLE detection procedure and the statistical implications of the results for planet properties.

\citet{Uda02b, Uda03} have used a threshold of $\alpha=9$ for transit candidate selection, where $\alpha$ is the criteria defined by \citet{Kov02}, $\alpha \equiv d / \sigma_{phot} \sqrt {N q}$ ($d$ is the transit depth, $\sigma_{phot}$ the mean photometric uncertainty, $N$ the number of data points and $q$ the length of the transit in phase -- therefore $N\,q$ is the average number of points expected in the transit). With figures typical of the OGLE Carina data  ($N$=1150, $\sigma_{phot}=0.006$ mag), this implies that a  planet with $R\sim 1.3 \RJ$ with $P \sim 3$ days could be detected in front of a $R\sim 1.2 \Rsol$ star, and a $R\sim 1.1\ \RJ$ in front of a $R\sim 1 \Rsol$ star. The OGLE Carina fields contain several tens of thousands of objects with $\sigma_{phot}=0.006$ or better. Using a frequency of 1\% for hot Jupiters and a geometric transit probability of 10\%, several dozen detection of transiting hot Jupiters could be expected, while only three were detected up to now.

There are two explanations to this apparent paradox: first, planets are smaller and target stars are larger than usually assumed; second, the actual detection threshold is generally much higher than indicated by an $\alpha>9$ criteria.

All three transiting planets detected in Carina have radii comparable to Jupiter, contrasting with the high radius ($R\sim 1.4 \RJ$) and very low density of HD 209458. This shows that HD 209458 is not a typical Hot Jupiter in terms of size, and that most hot Jupiters probably have smaller radii. Moreover, in a magnitude-limited sample in a Galactic disc field, the median radius of the target stars is about 1.4 $\Rsol$ \citep[see e.g. Besan\c con Model of the Galaxy,][]{Rob03}. Therefore typical transiting hot Jupiter transits produce a signal of depth only $\sim 0.5$~\%, which is significantly below the detection capabilities of the OGLE survey. 

Secondly, although the OGLE detection used a  $\alpha>9$ threshold, \citet{Uda02b} also states that to avoid many obvious spurious detection, a further cut was also imposed in terms of the significance of the signal in the BLS periodogram  \citep[the "SDE" criteria of][]{Kov02}. This criteria expresses the significance of the detection of a {\em periodic} transit signal in the lightcurve. For a given $\alpha$, the SDE criteria is enhanced if 

\begin{itemize}
\item{The actual number of points in the transit is larger than $N\,q$}
\item{The points are distributed in a larger number of different transits\footnote{A given transit signal will be easier to detect if the data points are distributed among a larger number of different transits than if they are grouped in a low number of transits. The reason is that if the number of transits is low, systematic drifts in other parts of the lightcurve will produce some noise in the periodogram used to detect the signal, and result in a higher effective detection threshold.}}
\end{itemize}

\begin{figure}[ht!]
\resizebox{8cm}{!}{\includegraphics[angle=-90]{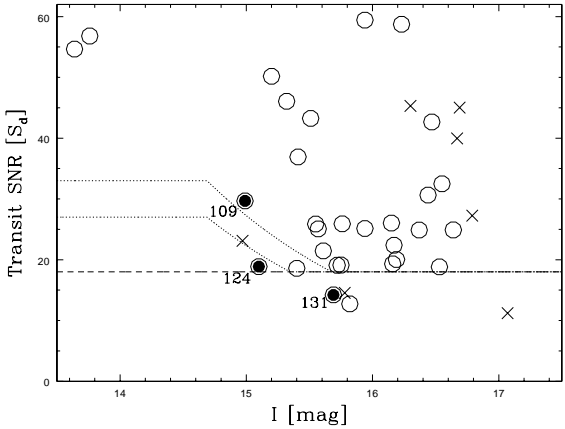}}
\caption{Transit signal-to-noise ratio ($d/\sigma_{phot} \sqrt{N_{tr}}$) as a function of magnitude, for the objects in our sample.   The crosses identify the objects without signal in the CCF, for which our measurements do not confirm the presence of a transiting companion. Black dots indicate the objects without significant radial velocity variations.  The lines show possible detectability limits according to the discussion in Section~\ref{detect}.}
\label{conf}
\end{figure}

The effect of the SDE criteria is difficult to model, especially in the presence of covariant residuals -- as is typical of ground-based photometric data. This criteria tends to favor shorter periods (more different transits sampled) and periods multiple of 1 day (number of points in the transit increased by the stroboscopic effect). We tried to approximate the effect of this selection using an $S_d$ criteria, defined in analogy to the $\alpha$ criteria, but with the actual number of points in the transit rather than the mean: $S_d\equiv d/\sigma_{phot} \sqrt{N_{tra}}$. Fig.~\ref{conf} illustrates our best attempt to model the transit detection threshold. A $S_d>18$ limit seems to define satisfactorily the limit of detectability of {\em confirmed} transits. Surprisingly, this is twice as high as the $\alpha=9$ threshold, indicating that for transits with low $\alpha$, the actual number of points in the transit ($N_{tra}$) must be enhanced by a large factor compared to the mean number ($N\,q$) in order for the signal to be clearly detectable.

Fig.~\ref{conf} shows that at magnitudes brighter than about $I=15.3$ mag, the detection threshold is even much higher. This can be expected because at these magnitude the systematic drifts in the photometry start dominating the photon noise \citep{Uda05}. These drifts (due to varying atmospheric conditions and airmass) operate on timescales similar to the transit duration, a few hours. They therefore produce an additional noise in the periodogram. We model this effect by assuming that below a certain value $\sigma_{min}$, the effective $\sigma_{phot}$ for transit detection no longer decreases, even if the nominal photometric uncertainty gets smaller. The detection limits for two possible values of $\sigma_{min}$, 4.5 mmag and 6 mmag, is plotted in Fig.~\ref{conf}. There is not enough data to decide which of these two values is more appropriate, but it does seem that our scheme provides a correct approximation of the selection limits.

\begin{figure}[ht!]
\resizebox{9cm}{!}{\includegraphics{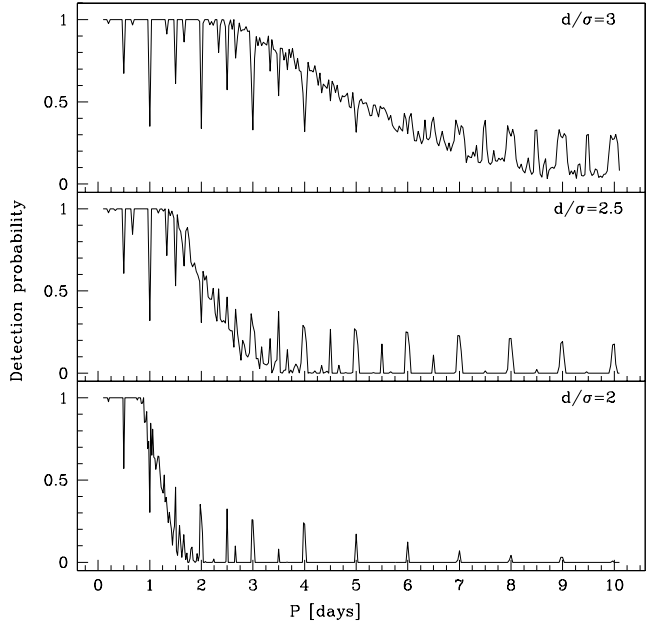}}
\caption{Detection probability of a transiting planet in the OGLE Carina survey as a function of orbital period, according to Monte Carlo simulations, for three transit depths: 3 times (top), 2.5 times (middle) and 2 times (bottom) the dispersion of the photometric data. The detection criteria is $d/\sigma_{phot} \sqrt{N_{tra}}>18$. A $R\sim \RJ$ planet in front of a solar-type star with $m_I<16$ would be in the lower panel, a $R\sim 1.3 \RJ$ planet would be in the upper panel.}
\label{probdec}
\end{figure}

With this caveat, we can model the behaviour of the detection threshold as a function of transit depth and period. We simulated sets of data with a time sampling typical of the OGLE Carina survey, and applied the $S_d>18$ selection criteria with various periods, and Monte Carlo realisations of a random phase shift. Fig.~\ref{probdec} plots, for three values of $d/\sigma_{phot}$ relevant for planetary transits, the proportion of detected transits in the simulations  as a function of periods. Near the detection threshold, a strong dependence of detectability on period is observed. This effect typical of ground-based, mono-site transit surveys was already discussed by \citet{Gau04} in the same context (cf their Fig. A4). The detection of transit signals with periods resonant with 1 day are slightly less favoured for deep transits, and greatly favoured for very shallow transits, compared with non-resonant periods. The lower panel of Fig.~\ref{probdec} shows that near the detection threshold, only transits with very short periods or periods very close to entire number of days can be detected.  

These elements make the peculiar characteristics of the OGLE planets look more reasonable. OGLE-TR-113 has a very short period ($P=1.43$ days), OGLE-TR-111 has a strongly resonant period ($P=4.02 $ days) and OGLE-TR-132 has both but to a weaker degree ($P=1.69\sim 5/3$ days). (Note that the same effects are visible in the OGLE bulge field, with OGLE-TR-56 at a very short period and OGLE-TR-10 near a resonance). This also provides a qualitative understanding of the surprising abundance of "very hot Jupiters" and scarcity of normal hot Jupiters in the OGLE survey: most $P<1.5$ day planets can be detected, but only a few percent of 3-4 day  hot Jupiters near resonances.  \citet{Gau04}, from scaling arguments only, estimate that the detection of P=1-2 days planets is enhanced by a factor 6 compared to 3-10 days planets. By revealing that some OGLE candidates were probable false positives, correspondingly increasing the effective detection threshold, our study indicates that this selection bias could even be higher, further increasing the marging for compatibility between the radial velocity surveys and the OGLE survey. Thus we can state that at this point there is no significant incompatibility between the results of the OGLE survey and the radial velocity survey in terms of planet frequencies. More quantitative estimates would require more detailed modelling of the detection threshold and a better knowledge of the actual selection procedure than described in \citet{Uda02b}.

The absence of detection of large transiting hot Jupiters in Carina is also a strong contraint on the radius distribution of hot Jupiters, because the transit depth has a square dependence on the planetary radius. Inflated Hot Jupiters with $R \sim 1.4 \RJ$ give rise to transits almost twice deeper and would be much easier to detect (compare the upper and lower panels of Fig.~\ref{probdec}). Given the fact that hot Jupiters are near the detectability threshold, the absence of detection indicates that inflated hot Jupiters do not constitute more than a minor fraction of the total number of hot Jupiters. The detection of OGLE-TR-10 in the OGLE bulge sample, on the other hand, shows that they are not totally absent.
On the small-radius side, it is possible that some hot Jupiters have substiantially smaller radii than 1.1 $\RJ$, and therefore the OGLE survey may have sampled only the high-radius end of the actual radius distribution.

\label{detect}

\subsection{Implications for spectroscopic follow-up of transiting candidates}

There are a number of things that were learned with this study concerning exoplanet transit surveys and the associated Doppler follow-up.
First, with FLAMES on the VLT we could push the limit of planet detection down to $I=16.5$ ($V\sim 17.5$). It is probably impractical at present to go much beyond this values, at least with fiber-fed instruments, because even if a CCF signal can be obtained, it is unlikely to be measured with the precision necessary to detect a planetary orbit, even with an 8-m telescope. A slit spectrograph could gather more signal, but at the loss of radial velocity accuracy because of the difficulty of controlling the position of the object in the slit with sufficient precision (see the UVES measurements for OGLE-TR-8, 10 and 12 in Paper I). This implies that the OGLE survey is near the upper limiting magnitude for spectroscopic follow-up with ground-based facilities, and that deeper photometric transit surveys - such as for instance with the HST - would run into the difficulty of having no possibility of confirmation with the velocity orbit for most of their candidates. This is a major difficulty, because Paper I and this study amply confirm that low-mass companions can perfectly mimic the signal of a transiting planet in photometry.

Another conclusion is that the list of possible configurations of the photometric and spectroscopic data is almost endless. The bestiary of cases encountered in our sample is even more varied than that of Paper I. It illustrates the range of contaminations expected in planetary transit survey, and the complexity of the subsequent spectroscopic follow-up. The systems fall in three broad categories: (i) bona fide planets (ii) eclipsing binaries in various configurations and (iii) false transit detections. Case (ii), in turn, divides in many subcases. For practical purposes there are two very different types of eclipsing binaries mimicking a planetary transit: either a grazing or blended eclipse of two large ($ R>0.5 \Rsol$) stars, or the transit of a small M-dwarf in from of a solar-type star. The duration and shape of the transit, as well as the presence of light-curve modulations outside the transit \citet{Sir03}, can be useful to reject these. Note that the main source of confusion (transiting M-dwarfs) do not produce a detectable colour signature in the transit signal. Therefore colour information \citep[see e.g.][]{Cha03} is not a very strong discriminant for ground-based transit surveys.

All transiting candidates with a good CCF signal and sufficiently high transit signal-to-noise ($S_d$) turned out to be eclipsing binaries or transiting planets, and it is worth noting that our follow-up did not reveal any intrinsic source of false positives, such as stellar spots or unusual variability patterns.


This study illustrated the very high potential of FLAMES/UVES on the VLT for follow-up of transit candidates in medium-deep fields. In only 8 half-nights (32 hours) of observations, we have followed 42 transit candidates including all the most promising candidates revealed by the photometric survey, detecting three planets among a much larger sample of eclipsing binaries, as well as collecting valuable data on very low-mass stars on the way. As a comparison, for the OGLE bulge field it has taken several observation runs for three different teams to cover only a fraction of the candidates (see references in the Introduction). Our strategy for the spectroscopic follow-up, detailed in Section~\ref{select}, has therefore proved its efficiency. Note however that it is designed to uncover planet cases among a large number of transiting candidates, and does not aim to be 100\% complete. Because fast rotation is used as a telltale signal of orbital synchronisation, planets orbiting a fast-rotating star would not be detected. However, in such a case the detection of a planet would be made very difficult by the decreased accuracy of the radial velocity determination due to rotational line broadening. Planets orbiting stars too hot for precise radial velocities would also be missed. 

Recently, \citet{Uda05} have announced a new list of 40 transiting candidates in six fields monitored during 2002 and 2003, using an improved candidate selection procedure taking into account the lessons of the high-resolution follow-up. If the resulting list provides comparable results to the Carina fields presented in this paper, the OGLE survey is on the way to securing its position as a very effective tool for ground-based transiting planet detection.


\acknowledgements{We thank the OGLE team to make their list of transiting candidates publicly available  for spectroscopic follow-up and accessible in a very user-friendly way.}

\bibliography{refCarina}

\begin{thebibliography}{63}
\expandafter\ifx\csname natexlab\endcsname\relax\def\natexlab#1{#1}\fi

\bibitem[{{Alonso} {et~al.}(2004){Alonso}, {Brown}, {Torres}, {Latham},
  {Sozzetti}, {Mandushev}, {Belmonte}, {Charbonneau}, {Deeg}, {Dunham},
  {O'Donovan}, \& {Stefanik}}]{Alo04}
{Alonso}, R., {Brown}, T.~M., {Torres}, G., {et~al.} 2004, \apjl, 613, L153

\bibitem[{{Baraffe} {et~al.}(1998){Baraffe}, {Chabrier}, {Allard}, \&
  {Hauschildt}}]{Bar98}
{Baraffe}, I., {Chabrier}, G., {Allard}, F., \& {Hauschildt}, P.~H. 1998, \aap,
  337, 403

\bibitem[{{Baraffe} {et~al.}(2004){Baraffe}, {Selsis}, {Chabrier}, {Barman},
  {Allard}, {Hauschildt}, \& {Lammer}}]{Bar04}
{Baraffe}, I., {Selsis}, F., {Chabrier}, G., {et~al.} 2004, \aap, 419, L13

\bibitem[{{Barban} {et~al.}(2003){Barban}, {Goupil}, {Van't Veer-Menneret},
  {Garrido}, {Kupka}, \& {Heiter}}]{Bar03}
{Barban}, C., {Goupil}, M.~J., {Van't Veer-Menneret}, C., {et~al.} 2003, \aap,
  405, 1095

\bibitem[{{Bouchy} {et~al.}(2005){Bouchy}, {Pont}, {Melo}, {Santos}, {Mayor},
  {Queloz}, \& {Udry}}]{Bou05}
{Bouchy}, F., {Pont}, F., {Melo}, C., {et~al.} 2005, to appear in A\& A

\bibitem[{{Bouchy} {et~al.}(2004){Bouchy}, {Pont}, {Santos}, {Melo}, {Mayor},
  {Queloz}, \& {Udry}}]{Bou04}
{Bouchy}, F., {Pont}, F., {Santos}, N.~C., {et~al.} 2004, \aap, 421, L13

\bibitem[{{Brown}(2003)}]{Bro03}
{Brown}, T.~M. 2003, \apjl, 593, L125

\bibitem[{{Brown} \& {Charbonneau}(2000)}]{Bro00}
{Brown}, T.~M. \& {Charbonneau}, D. 2000, in ASP Conf. Ser. 219: Disks,
  Planetesimals, and Planets, 584

\bibitem[{{Brown} {et~al.}(2001){Brown}, {Charbonneau}, {Gilliland}, {Noyes},
  \& {Burrows}}]{Bro01}
{Brown}, T.~M., {Charbonneau}, D., {Gilliland}, R.~L., {Noyes}, R.~W., \&
  {Burrows}, A. 2001, \apj, 552, 699

\bibitem[{{Brown} {et~al.}(2002){Brown}, {Libbrecht}, \& {Charbonneau}}]{Bro02}
{Brown}, T.~M., {Libbrecht}, K.~G., \& {Charbonneau}, D. 2002, \pasp, 114, 826

\bibitem[{{Burrows} {et~al.}(2004){Burrows}, {Hubeny}, {Hubbard}, {Sudarsky},
  \& {Fortney}}]{Bur04}
{Burrows}, A., {Hubeny}, I., {Hubbard}, W.~B., {Sudarsky}, D., \& {Fortney},
  J.~J. 2004, \apjl, 610, L53

\bibitem[{{Chabrier} {et~al.}(2000){Chabrier}, {Baraffe}, {Allard}, \&
  {Hauschildt}}]{Cha00b}
{Chabrier}, G., {Baraffe}, I., {Allard}, F., \& {Hauschildt}, P. 2000, \apj,
  542, 464

\bibitem[{{Chabrier} {et~al.}(2004){Chabrier}, {Barman}, {Baraffe}, {Allard},
  \& {Hauschildt}}]{Cha04}
{Chabrier}, G., {Barman}, T., {Baraffe}, I., {Allard}, F., \& {Hauschildt},
  P.~H. 2004, \apjl, 603, L53

\bibitem[{{Charbonneau} {et~al.}(2004){Charbonneau}, {Brown}, {Dunham},
  {Latham}, {Looper}, \& {Mandushev}}]{Cha03}
{Charbonneau}, D., {Brown}, T.~M., {Dunham}, E.~W., {et~al.} 2004, in AIP Conf.
  Proc. 713: The Search for Other Worlds, 151--160

\bibitem[{{Charbonneau} {et~al.}(2000){Charbonneau}, {Brown}, {Latham}, \&
  {Mayor}}]{Cha00}
{Charbonneau}, D., {Brown}, T.~M., {Latham}, D.~W., \& {Mayor}, M. 2000, \apjl,
  529, L45

\bibitem[{{Charbonneau} {et~al.}(2002){Charbonneau}, {Brown}, {Noyes}, \&
  {Gilliland}}]{Cha02}
{Charbonneau}, D., {Brown}, T.~M., {Noyes}, R.~W., \& {Gilliland}, R.~L. 2002,
  \apj, 568, 377

\bibitem[{{Dreizler} {et~al.}(2002){Dreizler}, {Rauch}, {Hauschildt}, {Schuh},
  {Kley}, \& {Werner}}]{Dre02}
{Dreizler}, S., {Rauch}, T., {Hauschildt}, P., {et~al.} 2002, \aap, 391, L17

\bibitem[{{Eggenberger} {et~al.}(2004){Eggenberger}, {Udry}, \&
  {Mayor}}]{Egg04}
{Eggenberger}, A., {Udry}, S., \& {Mayor}, M. 2004, \aap, 417, 353

\bibitem[{{Gallardo} {et~al.}(2005){Gallardo}, {Minniti}, {Valls-Gabaud}, \&
  {Rejkuba}}]{Gal05}
{Gallardo}, J., {Minniti}, D., {Valls-Gabaud}, D., \& {Rejkuba}, M. 2005, a \&
  A, in press

\bibitem[{{Gaudi} {et~al.}(2005){Gaudi}, {Seager}, \& {Mallen-Ornelas}}]{Gau04}
{Gaudi}, B., {Seager}, S., \& {Mallen-Ornelas}, G. 2005, \apj, 623

\bibitem[{{Gilliland} {et~al.}(2000){Gilliland}, {Brown}, {Guhathakurta},
  {Sarajedini}, {Milone}, {Albrow}, {Baliber}, {Bruntt}, {Burrows},
  {Charbonneau}, {Choi}, {Cochran}, {Edmonds}, {Frandsen}, {Howell}, {Lin},
  {Marcy}, {Mayor}, {Naef}, {Sigurdsson}, {Stagg}, {Vandenberg}, {Vogt}, \&
  {Williams}}]{Gil00}
{Gilliland}, R.~L., {Brown}, T.~M., {Guhathakurta}, P., {et~al.} 2000, \apjl,
  545, L47

\bibitem[{{Girardi} {et~al.}(2002){Girardi}, {Bertelli}, {Bressan}, {Chiosi},
  {Groenewegen}, {Marigo}, {Salasnich}, \& {Weiss}}]{Gir02}
{Girardi}, L., {Bertelli}, G., {Bressan}, A., {et~al.} 2002, \aap, 391, 195

\bibitem[{{Guillot}(2005)}]{Gui05}
{Guillot}, T. 2005, to appear in Ann. Rev. Earth \& Plan. Sciences

\bibitem[{{Halbwachs} {et~al.}(2000){Halbwachs}, {Arenou}, {Mayor}, {Udry}, \&
  {Queloz}}]{Hal00}
{Halbwachs}, J.~L., {Arenou}, F., {Mayor}, M., {Udry}, S., \& {Queloz}, D.
  2000, \aap, 355, 581

\bibitem[{{Henry} {et~al.}(2000){Henry}, {Marcy}, {Butler}, \& {Vogt}}]{Hen00}
{Henry}, G.~W., {Marcy}, G.~W., {Butler}, R.~P., \& {Vogt}, S.~S. 2000, \apjl,
  529, L41

\bibitem[{{Hut}(1981)}]{Hut81}
{Hut}, P. 1981, \aap, 99, 126

\bibitem[{{Kane} {et~al.}(2001){Kane}, {Horne}, {Street}, {Pollaco}, {James},
  {Tsapras}, \& {Collier-Cameron}}]{Kan01}
{Kane}, S., {Horne}, K., {Street}, R., {et~al.} 2001, in Techniques for the
  detection of planets and life beyond the solar system, 4th Annual ROE
  Workshop, held at Royal Observatory Edinburgh, Scotland, Nov 7-8, 2001.
  Edited by W.R.F. Dent. Edinburgh, Scotland: Royal Observatory, 2001, p.6, 6

\bibitem[{{Konacki} {et~al.}(2003{\natexlab{a}}){Konacki}, {Torres}, {Jha}, \&
  {Sasselov}}]{Kon03}
{Konacki}, M., {Torres}, G., {Jha}, S., \& {Sasselov}, D.~D.
  2003{\natexlab{a}}, \nat, 421, 507

\bibitem[{{Konacki} {et~al.}(2003{\natexlab{b}}){Konacki}, {Torres},
  {Sasselov}, \& {Jha}}]{Kon03b}
{Konacki}, M., {Torres}, G., {Sasselov}, D.~D., \& {Jha}, S.
  2003{\natexlab{b}}, \apj, 597, 1076

\bibitem[{{Konacki} {et~al.}(2005){Konacki}, {Torres}, {Sasselov}, \&
  {Jha}}]{Kon05}
{Konacki}, M., {Torres}, G., {Sasselov}, D.~D., \& {Jha}, S. 2005, submitted to
  ApJL

\bibitem[{{Konacki} {et~al.}(2004){Konacki}, {Torres}, {Sasselov}, {Pietrzy{\'
  n}ski}, {Udalski}, {Jha}, {Ruiz}, {Gieren}, \& {Minniti}}]{Kon04}
{Konacki}, M., {Torres}, G., {Sasselov}, D.~D., {et~al.} 2004, \apjl, 609, L37

\bibitem[{{Kov{\' a}cs} {et~al.}(2002){Kov{\' a}cs}, {Zucker}, \&
  {Mazeh}}]{Kov02}
{Kov{\' a}cs}, G., {Zucker}, S., \& {Mazeh}, T. 2002, \aap, 391, 369

\bibitem[{{Lane} {et~al.}(2001){Lane}, {Boden}, \& {Kulkarni}}]{Lan01}
{Lane}, B.~F., {Boden}, A.~F., \& {Kulkarni}, S.~R. 2001, \apjl, 551, L81

\bibitem[{{Lecavelier des Etangs} {et~al.}(2004){Lecavelier des Etangs},
  {Vidal-Madjar}, {McConnell}, \& {H{\' e}brard}}]{Lec04}
{Lecavelier des Etangs}, A., {Vidal-Madjar}, A., {McConnell}, J.~C., \& {H{\'
  e}brard}, G. 2004, \aap, 418, L1

\bibitem[{{Levato}(1976)}]{Lev76}
{Levato}, H. 1976, \apj, 203, 680

\bibitem[{{Mandel} \& {Agol}(2002)}]{Man02}
{Mandel}, K. \& {Agol}, E. 2002, \apjl, 580, L171

\bibitem[{{Mayor} \& {Queloz}(1995)}]{May95}
{Mayor}, M. \& {Queloz}, D. 1995, \nat, 378, 355

\bibitem[{{Mazeh} {et~al.}(2000){Mazeh}, {Naef}, {Torres}, {Latham}, {Mayor},
  {Beuzit}, {Brown}, {Buchhave}, {Burnet}, {Carney}, {Charbonneau}, {Drukier},
  {Laird}, {Pepe}, {Perrier}, {Queloz}, {Santos}, {Sivan}, {Udry}, \&
  {Zucker}}]{Maz00}
{Mazeh}, T., {Naef}, D., {Torres}, G., {et~al.} 2000, \apjl, 532, L55

\bibitem[{{Mazeh} {et~al.}(2005){Mazeh}, {Zucker}, \& {Pont}}]{Maz05}
{Mazeh}, T., {Zucker}, S., \& {Pont}, F. 2005, to appear in MNRAS

\bibitem[{{Melo} {et~al.}(2005){Melo}, {Bouchy}, {Pont}, {Santos}, {Mayor},
  {Queloz}, \& {Udry}}]{Mel05}
{Melo}, C., {Bouchy}, F., {Pont}, F., {et~al.} 2005, submitted to A\& A

\bibitem[{{Metcalfe} {et~al.}(1996){Metcalfe}, {Mathieu}, {Latham}, \&
  {Torres}}]{Met96}
{Metcalfe}, T.~S., {Mathieu}, R.~D., {Latham}, D.~W., \& {Torres}, G. 1996,
  \apj, 456, 356

\bibitem[{{Moutou} {et~al.}(2004){Moutou}, {Pont}, {Bouchy}, \&
  {Mayor}}]{Mou04}
{Moutou}, C., {Pont}, F., {Bouchy}, F., \& {Mayor}, M. 2004, \aap, 424, L31

\bibitem[{{P{\" a}tzold} \& {Rauer}(2002)}]{Pra02}
{P{\" a}tzold}, M. \& {Rauer}, H. 2002, \apjl, 568, L117

\bibitem[{{Pont} {et~al.}(2004){Pont}, {Bouchy}, {Queloz}, {Santos}, {Melo},
  {Mayor}, \& {Udry}}]{Pon04}
{Pont}, F., {Bouchy}, F., {Queloz}, D., {et~al.} 2004, \aap, 426, L15

\bibitem[{{Ribas}(2003)}]{Rib03}
{Ribas}, I. 2003, \aap, 398, 239

\bibitem[{{Robin} {et~al.}(2003){Robin}, {Reyl{\' e}}, {Derri{\` e}re}, \&
  {Picaud}}]{Rob03}
{Robin}, A.~C., {Reyl{\' e}}, C., {Derri{\` e}re}, S., \& {Picaud}, S. 2003,
  \aap, 409, 523

\bibitem[{{S{\' e}gransan} {et~al.}(2003){S{\' e}gransan}, {Kervella},
  {Forveille}, \& {Queloz}}]{Seg03}
{S{\' e}gransan}, D., {Kervella}, P., {Forveille}, T., \& {Queloz}, D. 2003,
  \aap, 397, L5

\bibitem[{{Sackett} {et~al.}(2004){Sackett}, {Albrow}, {Beaulieu}, {Caldwell},
  {Coutures}, {Dominik}, {Greenhill}, {Hill}, {Horne}, {Jorgensen}, {Kane},
  {Kubas}, {Martin}, {Menzies}, {Pollard}, {Sahu}, {Wambsganss}, {Watson}, \&
  {Williams}}]{Sak04}
{Sackett}, P.~D., {Albrow}, M.~D., {Beaulieu}, J.-P., {et~al.} 2004, in IAU
  Symposium, 35

\bibitem[{{Sackett} {et~al.}(2005){Sackett}, {Freeman}, \& {Bridges}}]{Sak05}
{Sackett}, P.~D., {Freeman}, K.~C., \& {Bridges}, T.~J. 2005, to appear in ApJ

\bibitem[{{Santos} {et~al.}(2004){Santos}, {Israelian}, \& {Mayor}}]{San04}
{Santos}, N.~C., {Israelian}, G., \& {Mayor}, M. 2004, \aap, 415, 1153

\bibitem[{{Sasselov}(2003)}]{Sas03}
{Sasselov}, D.~D. 2003, \apj, 596, 1327

\bibitem[{{Sirko} \& {Paczy{\' n}ski}(2003)}]{Sir03}
{Sirko}, E. \& {Paczy{\' n}ski}, B. 2003, \apj, 592, 1217

\bibitem[{{Udalski} {et~al.}(2002{\natexlab{a}}){Udalski}, {Paczynski},
  {Zebrun}, {Szymaski}, {Kubiak}, {Soszynski}, {Szewczyk}, {Wyrzykowski}, \&
  {Pietrzynski}}]{Uda02a}
{Udalski}, A., {Paczynski}, B., {Zebrun}, K., {et~al.} 2002{\natexlab{a}}, Acta
  Astronomica, 52, 1

\bibitem[{{Udalski} {et~al.}(2003){Udalski}, {Pietrzynski}, {Szymanski},
  {Kubiak}, {Zebrun}, {Soszynski}, {Szewczyk}, \& {Wyrzykowski}}]{Uda03}
{Udalski}, A., {Pietrzynski}, G., {Szymanski}, M., {et~al.} 2003, Acta
  Astronomica, 53, 133

\bibitem[{{Udalski} {et~al.}(2002{\natexlab{b}}){Udalski}, {Szewczyk},
  {Zebrun}, {Pietrzynski}, {Szymanski}, {Kubiak}, {Soszynski}, \&
  {Wyrzykowski}}]{Uda02c}
{Udalski}, A., {Szewczyk}, O., {Zebrun}, K., {et~al.} 2002{\natexlab{b}}, Acta
  Astronomica, 52, 317

\bibitem[{{Udalski} {et~al.}(2005){Udalski}, {Szymanski}, {Kubiak},
  {Pietrzynski}, {Soszynski}, {Zebrun}, {Szewczyk}, \& {Wyrzykowski}}]{Uda05}
{Udalski}, A., {Szymanski}, M., {Kubiak}, M., {et~al.} 2005, submitted to Acta
  Astronomica

\bibitem[{{Udalski} {et~al.}(2002{\natexlab{c}}){Udalski}, {Zebrun},
  {Szymanski}, {Kubiak}, {Soszynski}, {Szewczyk}, {Wyrzykowski}, \&
  {Pietrzynski}}]{Uda02b}
{Udalski}, A., {Zebrun}, K., {Szymanski}, M., {et~al.} 2002{\natexlab{c}}, Acta
  Astronomica, 52, 115

\bibitem[{{Udry} {et~al.}(2000){Udry}, {Mayor}, {Delfosse}, {Forveille}, \&
  {Perrier-Bellet}}]{Udr00}
{Udry}, S., {Mayor}, M., {Delfosse}, X., {Forveille}, T., \& {Perrier-Bellet},
  C. 2000, in IAU Symposium, 158

\bibitem[{{Udry} {et~al.}(2002){Udry}, {Mayor}, {Naef}, {Pepe}, {Queloz},
  {Santos}, \& {Burnet}}]{Udr02}
{Udry}, S., {Mayor}, M., {Naef}, D., {et~al.} 2002, \aap, 390, 267

\bibitem[{{Vidal-Madjar} {et~al.}(2004){Vidal-Madjar}, {D{\' e}sert},
  {Lecavelier des Etangs}, {H{\' e}brard}, {Ballester}, {Ehrenreich}, {Ferlet},
  {McConnell}, {Mayor}, \& {Parkinson}}]{Vid04}
{Vidal-Madjar}, A., {D{\' e}sert}, J.-M., {Lecavelier des Etangs}, A., {et~al.}
  2004, \apjl, 604, L69

\bibitem[{{Vidal-Madjar} {et~al.}(2003){Vidal-Madjar}, {Lecavelier des Etangs},
  {D{\' e}sert}, {Ballester}, {Ferlet}, {H{\' e}brard}, \& {Mayor}}]{Vid03}
{Vidal-Madjar}, A., {Lecavelier des Etangs}, A., {D{\' e}sert}, J.-M., {et~al.}
  2003, \nat, 422, 143

\bibitem[{{Zahn}(1989)}]{Zah92}
{Zahn}, J.-P. 1989, \aap, 220, 112

\bibitem[{{Zucker} \& {Mazeh}(2002)}]{Zuc02}
{Zucker}, S. \& {Mazeh}, T. 2002, \apjl, 568, L113

\end{thebibliography}

\clearpage

\begin{table}[p]
\caption{Radial velocity measurements (in the barycentric frame) and CCF parameters. 
Labels $a$, $b$ and $c$ indicate that several components are present in the CCF.}
\label{tablevr}
\begin{tabular}{r r r r r r} \hline \hline
BJD & RV & depth & FWHM & SNR & $\sigma_{\rm RV}$  \\ 
{[$-2453000\,$d]} & [{\kms}] & [\%] & [{\kms}] & & [{\kms}] \\ \hline 
OGLE-TR-63 & & & & & \\ \hline
78.63476 & -29.141 & 2.23 & 82.4 & 3.9 & 3.131 \\ 
79.57336 & -24.663 & 2.61 & 90.8 & 6.3 & 1.739 \\ 
80.58924 & -25.147 & 2.68 & 86.4 & 6.3 & 1.652 \\ \hline 
OGLE-TR-64a & & & & & \\ \hline
78.63475 & 4.373  & 10.00 & 18.0 & 4.0 & 0.320 \\
79.57336 & 56.232 & 11.78 & 18.4 & 6.5 & 0.172 \\
80.58923 & 43.547 & 11.79 & 18.5 & 6.1 & 0.183 \\ \hline
OGLE-TR-64b & & & & & \\ \hline
78.63475 & -10.012 & 2.40 & 13.9 & 4.0 & 1.166 \\
79.57336 & -89.975 & 2.62 & 17.3 & 6.5 & 0.734 \\
80.58923 & -70.125 & 2.31 & 14.7 & 6.1 & 0.817 \\ \hline
OGLE-TR-65a & & & & & \\ \hline
78.63475 & 72.265  & 1.60 & 53.6&  4.5&  3.051 \\
79.57336 & -99.104 & 2.98 & 49.1&  7.6&  0.929 \\
80.58923 & 111.614 & 2.73 & 43.9&  7.3&  0.998 \\
81.68999 & -76.895 & 2.73 & 52.0&  6.8&  1.166 \\
84.57980 & -45.244 & 2.60 & 66.9&  9.7&  0.974 \\
85.67894 & 107.523 & 2.33 & 40.2&  7.8&  1.047 \\ \hline
OGLE-TR-65b & & & & & \\ \hline
78.63475 & -82.835  & 2.98 & 47.8&  4.5&  1.547 \\
79.57336 & 97.622   & 2.66 & 38.2&  7.6&  0.918 \\
80.58923 & -119.194 & 2.79 & 53.2&  7.3&  1.075 \\
81.68999 & 75.493   & 1.70 & 37.9&  6.8&  1.598 \\
84.57980 & 44.270   & 2.26 & 41.5&  9.7&  0.882 \\
85.67894 & -116.179 & 2.74 & 51.3&  7.8&  1.006 \\ \hline
OGLE-TR-65c & & & & & \\ \hline
78.63475 & -7.327&  2.22 & 12.2 & 4.5&  1.049 \\
79.57336 & -5.425&  2.63 & 12.1 & 7.6&  0.523 \\
80.58923 & -5.002&  2.29 & 10.0 & 7.3&  0.569 \\
81.68999 & -5.827&  2.69 & 10.6 & 6.8&  0.535 \\
84.57980 & -5.676&  2.22 & 10.9 & 9.7&  0.461 \\
85.67894 & -6.383&  2.14 & 13.7 & 7.8&  0.666 \\ \hline
OGLE-TR-68 & & & & & \\ \hline
81.69000 & - & - & - & 3.1 & - \\ \hline
OGLE-TR-69a & & & & & \\ \hline
84.579811 & -14.988 & 3.14 & 27.1 & 4.1 & 1.214 \\  
85.678947 & 87.732 & 2.76 & 20.5 & 3.7 & 1.331 \\ \hline
OGLE-TR-69b & & & & & \\ \hline
84.579811 & 51.167 & 1.73 & 14.5 & 4.1 & 1.611 \\ 
85.678947 & -68.270 & 2.82 & 20.6 & 3.7 & 1.305 \\ \hline
OGLE-TR-72 & & & & & \\ \hline
81.68998 & 9.599 & 15.73 & 17.8 & 3.6 & 0.226 \\ 
84.57979 & 41.057 & 21.28 & 16.5 & 5.5 & 0.110 \\ 
85.67893 & 16.490 & 19.60 & 17.2 & 4.7 & 0.140 \\ \hline 
OGLE-TR-76a & & & & & \\ \hline 
78.659480 & -11.0 & 1.80&  28.0&  18.6&  0.475 \\
79.607748 & -11.0 & 1.80&  28.0&  19.6&  0.451 \\
80.624408 & -11.0 & 1.80&  28.0&  18.5&  0.478 \\
81.631519 & -11.0 & 1.80&  28.0&  19.6&  0.451 \\
82.676072 & -11.0 & 1.80&  28.0&  19.2&  0.461 \\
83.593052 & -11.0 & 1.80&  28.0&  15.6&  0.566 \\
84.686852 & -11.0 & 1.80&  28.0&  18.9&  0.468 \\
85.571295 & -11.0 & 1.80&  28.0&  20.6&  0.430 \\ \hline
\end{tabular}
\end{table}
\begin{table}[p]
\begin{tabular}{r r r r r r} \hline 
OGLE-TR-76b & & & & & \\ \hline 
78.659480 &  18.294 & 0.60&  34.6&  18.6& 1.582 \\
79.607748 & -30.235 & 0.70&  33.6&  19.6& 1.268 \\
80.624408 &  53.437 & 0.30&  36.2&  18.5& 3.252 \\
81.631519 & -38.871 & 0.60&  43.3&  19.6& 1.679 \\
82.676072 &  76.459 & 0.40&  31.0&  19.2& 2.175 \\
83.593052 & -57.686 & 0.60&  52.7&  15.6& 2.327 \\
84.686852 &  83.450 & 0.60&  28.2&  18.9& 1.405 \\
85.571295 & -55.996 & 0.60&  44.9&  20.6& 1.627 \\ \hline
OGLE-TR-78 & & & & & \\ \hline
78.65948 & 6.133 & 8.69 & 27.2 & 8.8 & 0.208 \\ 
79.60775 & -21.430 & 9.51 & 24.6 & 10.4 & 0.154 \\ 
80.62441 & -34.631 & 8.81 & 26.6 & 11.9 & 0.152 \\ 
81.63152 & -22.449 & 9.98 & 23.9 & 11.9 & 0.128 \\ 
82.67607 & 10.708 & 8.58 & 25.9 & 9.8 & 0.185 \\ 
83.59305 & 16.285 & 7.15 & 24.5 & 5.7 & 0.366 \\ 
84.68685 & -15.866 & 10.03 & 24.7 & 12.3 & 0.126 \\ 
85.57130 & -32.386 & 9.05 & 25.9 & 10.0 & 0.172 \\ \hline 
OGLE-TR-81a & & & & & \\ \hline
78.684263 & -17.175 & 2.24 & 9.3 & 9.4 & 0.436 \\ \hline 
OGLE-TR-81b & & & & & \\ \hline
78.684263 & 14.555 & 1.30 & 31.6 & 9.4 & 1.380 \\ \hline 
OGLE-TR-82 & & & & & \\ \hline
78.684264 & - & - & - & 2.0 & - \\ \hline
OGLE-TR-84 & & & & & \\ \hline
78.684266  & - & - & - &  4.1 & - \\ \hline
OGLE-TR-85a & & & & & \\ \hline
78.659504 & -1.0 & 1.00&  90.0 & 6.2&  4.591 \\
79.607771 & -1.0 & 1.00&  90.0 & 9.1&  3.128 \\
80.624432 & -1.0 & 1.00&  90.0 & 8.5&  3.348 \\
81.631542 & -1.0 & 1.00&  90.0 & 9.5&  2.996 \\
82.676095 & -1.0 & 1.00&  90.0 & 6.7&  4.248 \\
83.593075 & -1.0 & 1.00&  90.0 & 4.6&  6.187 \\
84.686875 & -1.0 & 1.00&  90.0 & 8.0&  3.558 \\
85.571317 & -1.0 & 1.00&  90.0 & 6.9&  4.125 \\ \hline
OGLE-TR-85b & & & & & \\ \hline
78.659504 &  42.924 & 2.00 & 39.9 & 6.2 & 1.529 \\
79.607771 & -20.555 & 2.90 & 49.3 & 9.1 & 0.799 \\
80.624432 &  23.797 & 2.80 & 44.5 & 8.5 & 0.842 \\
81.631542 &  -7.581 & 3.10 & 43.1 & 9.5 & 0.670 \\
82.676095 &  14.049 & 2.30 & 49.4 & 6.7 & 1.369 \\
83.593075 &   9.384 & 1.60 & 45.4 & 4.6 & 2.747 \\
84.686875 &   0.020 & 2.60 & 50.1 & 8.0 & 1.021 \\
85.571317 &  32.024 & 2.50 & 36.0 & 6.9 & 1.044 \\ \hline
OGLE-TR-89 & & & & & \\ \hline
82.603122 & - & - & - & 7.4 & - \\ \hline 
OGLE-TR-93a & & & & & \\ \hline
82.603127 & 41.872 & 7.63 & 12.1 & 8.6 & 0.163 \\ \hline 
OGLE-TR-93b & & & & & \\ \hline
82.603127 & 9.492 & 3.05 & 55.6 & 8.6 & 0.854 \\ \hline 
OGLE-TR-94 & & & & & \\ \hline
82.60312 & -28.875 & 6.06 & 36.8 & 15.9 & 0.192 \\ \hline 
OGLE-TR-95a & & & & & \\ \hline
82.603116 & -0.332 & 2.93 & 18.1 & 5.4 & 0.807 \\ \hline 
OGLE-TR-95b & & & & & \\ \hline
82.603116 & -90.494 & 1.25 & 127.0 & 5.4 & 5.009 \\ \hline
OGLE-TR-96a & & & & & \\ \hline
82.603113 & -85.557 & 2.37 & 16.3 & 11.2 & 0.458 \\ \hline 
OGLE-TR-96b & & & & & \\ \hline
82.603113 & -3.783 & 1.80 & 94.6 & 11.2 & 1.448 \\ \hline 
OGLE-TR-96c & & & & & \\ \hline
82.603113 & 55.246 & 3.73 & 15.9 & 11.2 & 0.288 \\ \hline 
\end{tabular}
\end{table}
\begin{table}[p]
\begin{tabular}{r r r r r r} \hline 
OGLE-TR-97a & & & & & \\ \hline
83.628839 & 56.177 & 7.81 & 17.7 & 10.0 & 0.165 \\ \hline 
OGLE-TR-97b & & & & & \\ \hline
83.628839 & -48.176 & 3.55 & 16.1 & 10.0 & 0.341 \\ \hline 
OGLE-TR-97c & & & & & \\ \hline
83.628839 & 13.911 & 1.73 & 11.9 & 10.0 & 0.599 \\ \hline 
OGLE-TR-98 & & & & & \\ \hline
83.62884 & -0.210 & 7.71 & 24.6 & 4.8 & 0.404 \\ \hline 
OGLE-TR-99 & & & & & \\ \hline
83.62884 & -24.145 & 3.27 & 70.2 & 5.7 & 1.349 \\ \hline 
OGLE-TR-105 & & & & & \\ \hline
78.60418 & -8.707 & 2.02 & 22.0 & 8.3 & 0.840 \\ 
79.64234 & 50.273 & 1.85 & 26.2 & 9.3 & 0.893 \\ 
81.59491 & 117.739 & 1.63 & 17.5 & 7.3 & 1.055 \\ 
82.71279 & 57.870 & 2.37 & 22.5 & 8.7 & 0.691 \\ 
83.66467 & 2.705 & 1.77 & 23.9 & 8.2 & 1.011 \\ 
84.65148 & -9.314 & 2.13 & 20.9 & 10.6 & 0.608 \\ 
85.60720 & 37.973 & 1.75 & 20.3 & 9.2 & 0.840 \\ \hline 
OGLE-TR-106 & & & & & \\ \hline
78.60418 & -24.828 & 4.59 & 42.1 & 3.2 & 1.326 \\ 
79.64234 & 9.532 & 5.89 & 33.9 & 3.8 & 0.781 \\ 
80.65956 & -13.450 & 4.93 & 35.1 & 3.0 & 1.202 \\ 
81.59491 & -12.040 & 5.22 & 33.5 & 2.9 & 1.148 \\ 
82.71279 & 6.894 & 5.20 & 35.5 & 3.4 & 1.012 \\ 
83.66467 & -23.908 & 5.97 & 39.0 & 3.4 & 0.924 \\ 
84.65148 & 9.035 & 5.66 & 39.2 & 3.8 & 0.874 \\ 
85.60720 & -6.853 & 6.22 & 34.2 & 4.0 & 0.706 \\ \hline 
OGLE-TR-107 & & & & & \\ \hline
82.603106 & - & - & - &  4.1 & - \\ \hline
OGLE-TR-109 & & & & & \\ \hline
78.60420 & -11.963 & 1.77 & 53.5 & 12.2 & 1.017 \\ 
79.64235 & -12.455 & 1.82 & 59.3 & 14.2 & 0.895 \\ 
80.65958 & -13.858 & 1.65 & 60.9 & 11.4 & 1.245 \\ 
81.59492 & -15.263 & 1.69 & 56.5 & 10.9 & 1.225 \\ 
82.71280 & -11.850 & 1.62 & 56.5 & 13.7 & 1.017 \\ 
83.66469 & -13.758 & 1.75 & 61.1 & 13.4 & 1.001 \\ 
84.65149 & -12.460 & 1.72 & 52.9 & 14.6 & 0.870 \\ 
85.60721 & -8.818 & 1.74 & 54.8 & 14.6 & 0.875 \\ \hline 
OGLE-TR-110a & & & & & \\ \hline
80.659567 & -39.425 & 5.45 & 19.7 & 6.7 & 0.366 \\ \hline 
OGLE-TR-110b & & & & & \\ \hline
80.659567 & 56.957 & 5.00 & 16.0 & 6.7 & 0.360 \\ \hline
OGLE-TR-111 & & & & & \\ \hline
78.60420 & 25.111 & 31.14 & 10.4 & 6.9 & 0.045 \\ 
79.64235 & 25.163 & 30.56 & 10.4 & 6.9 & 0.046 \\ 
80.65958 & 25.220 & 27.78 & 10.1 & 5.0 & 0.069 \\ 
81.59492 & 25.106 & 25.23 & 10.8 & 4.3 & 0.091 \\ 
82.71280 & 25.029 & 31.08 & 10.4 & 7.4 & 0.042 \\ 
83.66469 & 25.175 & 29.02 & 10.3 & 5.9 & 0.056 \\ 
84.65149 & 25.219 & 33.48 & 10.2 & 8.6 & 0.033 \\ 
85.60721 & 25.149 & 32.78 & 10.2 & 7.8 & 0.037 \\ \hline 
OGLE-TR-112a & & & & & \\ \hline
78.604197  &-49.156 & 3.29 & 12.1 & 22.5 & 0.145 \\
79.642351  &-41.946 & 3.45 & 12.8 & 25.4 & 0.127 \\
80.659577  & -8.420 & 3.72 & 14.5 & 21.8 & 0.145 \\
81.594925  & 86.548 & 3.23 & 11.3 & 19.3 & 0.166 \\
82.712800  & 58.124 & 3.15 & 12.6 & 25.9 & 0.135 \\
83.664686  &  9.111 & 3.56 & 12.4 & 24.0 & 0.129 \\
84.651492  &-14.857 & 3.50 & 12.5 & 24.8 & 0.127 \\
85.607208  &-30.259 & 3.30 & 11.8 & 24.4 & 0.133 \\ \hline
\end{tabular}
\end{table}
\begin{table}[p]
\begin{tabular}{r r r r r r} \hline 
OGLE-TR-112b & & & & & \\ \hline
78.604197 &   31.991 & 2.16 & 11.3 & 22.5 & 0.210 \\
79.642351 &   22.937 & 2.53 & 11.9 & 25.4 & 0.165 \\
80.659577 &  -15.744 & 2.70 & 10.6 & 21.8 & 0.170 \\
81.594925 & -119.516 & 2.36 & 14.2 & 19.3 & 0.251 \\
82.712800 &  -89.284 & 2.29 & 11.7 & 25.9 & 0.177 \\
83.664686 &  -33.716 & 2.35 & 11.9 & 24.0 & 0.187 \\
84.651492 &   -4.470 & 2.48 & 13.2 & 24.8 & 0.181 \\
85.607208 &   10.060 & 2.35 & 11.3 & 24.4 & 0.179 \\ \hline
OGLE-TR-112c & & & & & \\ \hline
78.604197 &  22.192 & 1.55 & 59.3 & 22.5 & 0.663 \\
79.642351 & -11.342 & 1.40 & 59.0 & 25.4 & 0.649 \\
80.659577 & -37.596 & 1.23 & 53.9 & 21.8 & 0.822 \\
81.594925 &  20.250 & 1.06 & 54.5 & 19.3 & 1.083 \\
82.712800 &  13.662 & 1.14 & 51.4 & 25.9 & 0.729 \\
83.664686 & -17.880 & 1.40 & 54.0 & 24.0 & 0.657 \\
84.651492 & -28.478 & 1.44 & 49.2 & 24.8 & 0.590 \\
85.607208 &  18.729 & 1.38 & 59.1 & 24.4 & 0.686 \\ \hline
OGLE-TR-113 & & & & & \\ \hline
78.60419 & -7.883 & 39.95 & 11.7 & 14.1 & 0.039 \\ 
79.64235 & -8.302 & 39.77 & 11.8 & 14.4 & 0.039 \\ 
80.65957 & -7.972 & 39.39 & 11.6 & 13.0 & 0.040 \\ 
81.59492 & -7.887 & 38.73 & 11.6 & 11.5 & 0.042 \\ 
82.71279 & -8.120 & 39.89 & 11.7 & 14.7 & 0.039 \\ 
83.66468 & -8.120 & 39.29 & 11.6 & 13.1 & 0.040 \\ 
84.65149 & -7.641 & 40.60 & 11.6 & 16.1 & 0.038 \\ 
85.60720 & -8.082 & 40.12 & 11.6 & 15.0 & 0.039 \\ \hline 
OGLE-TR-114a & & & & & \\ \hline
78.604187 &  -72.952 &  6.50 &  16.7 &  9.6  & 0.200 \\
79.642340 &   34.714 &  5.70 &  18.7 &  8.1  & 0.283 \\
80.659566 &   69.960 &  6.20 &  15.9 &  7.5  & 0.260 \\
81.594914 &  -61.504 &  5.90 &  16.1 &  6.6  & 0.311 \\
82.712790 &  -32.185 &  6.43 &  17.0 & 10.0  & 0.196 \\
83.664676 &   84.788 &  5.66 &  17.3 &  8.0  & 0.278 \\
84.651481 &   -4.038 &  7.19 &  18.8 & 11.5  & 0.161 \\
85.607198 &  -68.153 &  6.22 &  16.8 & 10.1  & 0.199 \\ \hline
OGLE-TR-114b & & & & & \\ \hline
78.604187 &   84.598 &  5.94  & 16.3  &  9.6  & 0.215 \\
79.642340 &  -24.713 &  6.38  & 17.8  &  8.1  & 0.247 \\
80.659566 &  -59.671 &  6.15  & 16.6  &  7.5  & 0.267 \\
81.594914 &   71.205 &  5.84  & 14.9  &  6.6  & 0.302 \\
82.712790 &   42.987 &  6.31  & 17.2  & 10.0  & 0.200 \\
83.664676 &  -73.867 &  6.10  & 17.9  &  8.0  & 0.262 \\
84.651481 &   13.530 &  7.44  & 19.5  & 11.5  & 0.159 \\
85.607198 &   77.476 &  6.55  & 16.8  & 10.1  & 0.189 \\ \hline
OGLE-TR-114c & & & & & \\ \hline
78.604187  & 7.166  & 8.57  &  9.4 &  9.6  & 0.117 \\
79.642340  & 7.358  & 7.81  & 10.3 &  8.1  & 0.156 \\
80.659566  & 7.183  & 7.84  &  9.4 &  7.5  & 0.160 \\
81.594914  & 7.232  & 7.49  &  9.0 &  6.6  & 0.185 \\
82.712790  & 7.110  & 8.49  & 10.0 & 10.0  & 0.117 \\
83.664676  & 6.915  & 7.72  &  9.5 &  8.0  & 0.154 \\
84.651481  & 6.943  & 8.20  &  9.5 & 11.5  & 0.104 \\
85.607198  & 6.897  & 8.17  &  9.3 & 10.1  & 0.116 \\ \hline
OGLE-TR-118 & & & & & \\ \hline
78.574215 & - & - & - &  3.3 & - \\
79.692810 & - & - & - &  6.7 & - \\
80.695892 & - & - & - &  5.1 & - \\ \hline
\end{tabular}
\end{table}
\begin{table}[p]
\begin{tabular}{r r r r r r} \hline 
OGLE-TR-120 & & & & & \\ \hline
78.57422 & 29.764 & 10.28 & 16.6 & 4.4 & 0.272 \\ 
79.69282 & 42.857 & 15.48 & 16.2 & 10.7 & 0.081 \\ 
80.69590 & 56.690 & 14.24 & 15.9 & 8.1 & 0.109 \\ 
81.72915 & 69.946 & 14.28 & 15.9 & 8.2 & 0.108 \\ 
82.64057 & 60.727 & 14.11 & 15.8 & 7.2 & 0.122 \\ 
83.70041 & 15.286 & 14.89 & 16.1 & 9.1 & 0.095 \\ 
84.61587 & 5.387 & 13.36 & 16.0 & 6.0 & 0.154 \\ 
85.64256 & 9.553 & 13.55 & 16.2 & 6.6 & 0.139 \\ \hline 
OGLE-TR-121 & & & & & \\ \hline
78.57422 & 22.508 & 7.85 & 28.3 & 5.5 & 0.371 \\ 
79.69282 & -30.078 & 9.90 & 30.7 & 12.1 & 0.143 \\ 
80.69590 & 29.621 & 9.37 & 29.7 & 10.1 & 0.176 \\ 
81.72916 & 28.150 & 9.41 & 29.2 & 9.8 & 0.179 \\ 
82.64057 & -28.573 & 9.43 & 31.4 & 9.1 & 0.199 \\ 
83.70041 & 13.803 & 9.64 & 30.2 & 10.4 & 0.168 \\ 
84.61587 & 45.009 & 8.42 & 28.7 & 6.9 & 0.279 \\ 
85.64256 & -19.195 & 9.26 & 29.9 & 8.1 & 0.221 \\ \hline 
OGLE-TR-122 & & & & & \\ \hline
78.63477 & -7.720 & 17.54 & 11.2 & 3.7 & 0.159 \\ 
79.57337 & -2.620 & 21.85 & 11.4 & 5.3 & 0.094 \\ 
80.58925 & 2.930 & 19.22 & 11.8 & 4.7 & 0.119 \\ 
81.69001 & 7.820 & 21.48 & 11.9 & 5.3 & 0.097 \\ 
84.57982 & -9.535 & 26.78 & 12.2 & 7.5 & 0.063 \\ 
85.67896 & -8.809 & 25.92 & 12.1 & 7.3 & 0.065 \\ \hline 
OGLE-TR-123 & & & & & \\ \hline
81.68998 & 0.234 & 5.97 & 44.9 & 10.7 & 0.317 \\ 
84.57979 & -6.208 & 6.13 & 43.4 & 15.2 & 0.215 \\ 
85.67893 & 12.276 & 6.14 & 43.9 & 11.9 & 0.274 \\ \hline 
OGLE-TR-124 & & & & & \\ \hline
78.65948 & -5.308 & 23.50 & 12.0 & 10.0 & 0.056 \\ 
79.60775 & -5.385 & 24.00 & 12.1 & 12.0 & 0.050 \\ 
80.62441 & -5.458 & 24.15 & 12.0 & 13.4 & 0.048 \\ 
81.63152 & -5.545 & 24.35 & 12.2 & 13.2 & 0.048 \\ 
82.67608 & -5.594 & 23.78 & 12.1 & 10.7 & 0.054 \\ 
83.59306 & -5.446 & 20.55 & 11.9 & 6.4 & 0.086 \\ 
84.68686 & -5.810 & 24.59 & 12.0 & 14.0 & 0.046 \\ 
85.57130 & -5.880 & 23.48 & 12.1 & 10.8 & 0.054 \\ \hline 
OGLE-TR-125 & & & & & \\ \hline
78.65949 & 32.731 & 3.68 & 22.7 & 7.5 & 0.519 \\ 
79.60776 & 46.168 & 3.72 & 23.8 & 8.8 & 0.448 \\ 
80.62442 & 37.267 & 3.56 & 26.3 & 9.1 & 0.476 \\ 
81.63153 & 16.405 & 3.64 & 25.2 & 10.3 & 0.403 \\ 
82.67608 & 9.185 & 3.33 & 26.6 & 7.1 & 0.655 \\ 
83.59306 & 25.120 & 2.19 & 19.4 & 5.1 & 1.184 \\ 
84.68686 & 45.041 & 3.62 & 23.5 & 8.9 & 0.453 \\ 
85.57130 & 42.347 & 3.35 & 21.3 & 7.5 & 0.552 \\ \hline 
OGLE-TR-126 & & & & & \\ \hline
78.68427 & 5.519 & 4.40 & 24.9 & 7.4 & 0.461 \\ \hline 
OGLE-TR-127 & & & & & \\ \hline
82.603117  & - & - & - & 16.8 & - \\ \hline
OGLE-TR-129 & & & & & \\ \hline
78.57423 & 28.954 & 1.18 & 29.1 & 4.2 & 3.266 \\ 
79.69283 & -1.838 & 2.37 & 37.3 & 10.2 & 0.759 \\ 
80.69591 & -3.341 & 2.09 & 34.2 & 7.6 & 1.105 \\ 
81.72916 & 1.110 & 2.14 & 33.8 & 8.3 & 0.983 \\ 
82.64058 & 5.068 & 1.91 & 35.5 & 7.4 & 1.265 \\ 
83.70042 & 14.841 & 1.83 & 42.1 & 8.7 & 1.223 \\ 
84.61588 & 15.003 & 1.00 & 51.9 & 5.6 & 3.860 \\ 
85.64257 & -3.476 & 1.83 & 39.9 & 6.4 & 1.618 \\ \hline 
\end{tabular}
\end{table}
\begin{table}[p]
\begin{tabular}{r r r r r r} \hline
OGLE-TR-130 & & & & & \\ \hline
78.57422 & 7.894 & 12.60 & 17.6 & 6.8 & 0.151 \\ 
79.69281 & 43.016 & 14.62 & 17.9 & 13.3 & 0.074 \\ 
80.69589 & 15.732 & 14.47 & 17.3 & 10.6 & 0.089 \\ 
81.72915 & -28.779 & 13.47 & 17.9 & 10.1 & 0.100 \\ 
82.64056 & -25.896 & 13.49 & 17.5 & 8.7 & 0.113 \\ 
83.70040 & 21.937 & 13.93 & 17.6 & 10.6 & 0.092 \\ 
84.61586 & 42.805 & 14.05 & 17.5 & 9.1 & 0.104 \\ 
85.64255 & 10.091 & 13.50 & 17.6 & 9.3 & 0.106 \\ \hline 
OGLE-TR-131 & & & & & \\ \hline
78.57421 & 18.879 & 19.59 & 10.7 & 3.2 & 0.160 \\ 
79.69280 & 18.937 & 35.15 & 10.4 & 7.4 & 0.051 \\ 
80.69588 & 18.937 & 32.13 & 10.7 & 5.9 & 0.062 \\ 
81.72914 & 18.997 & 32.19 & 10.4 & 6.0 & 0.061 \\ 
82.64055 & 18.988 & 29.93 & 10.2 & 4.8 & 0.075 \\ 
83.70039 & 18.935 & 32.65 & 10.4 & 6.3 & 0.059 \\ 
84.61585 & 18.934 & 27.08 & 10.6 & 4.5 & 0.087 \\ 
85.64254 & 18.918 & 27.46 & 10.7 & 4.7 & 0.084 \\ \hline 
OGLE-TR-132 & & & & & \\ \hline
81.72913 & 39.690 & 30.61 & 10.5 & 9.3 & 0.049 \\ 
82.64054 & 39.676 & 28.80 & 10.6 & 7.9 & 0.055 \\ 
83.70038 & 39.517 & 30.88 & 10.6 & 10.0 & 0.047 \\ 
84.61585 & 39.760 & 29.89 & 10.4 & 8.5 & 0.052 \\ 
85.64253 & 39.491 & 30.32 & 10.5 & 9.2 & 0.049 \\ \hline 
\end{tabular}
\end{table}

\end{document}